\documentclass[11pt,a4paper]{article}
\usepackage{amsmath,amsfonts,epsfig}
\usepackage{amssymb}
\usepackage{mathrsfs}
\usepackage{hyperref}
\begin{document}
\numberwithin{equation}{section}
\setlength{\unitlength}{.8mm}

\begin{titlepage} 
%\begin{flushright}
%version1\\
%\today
%\end{flushright}
\vspace*{0.5cm}
\begin{center}
{\Large\bf On the finite volume expectation values of local operators in the sine-Gordon model}
\end{center}
\vspace{1.5cm}
\begin{center}
{\large \'Arp\'ad Heged\H us}
\end{center}
\bigskip

\vspace{0.1cm}

\begin{center}
Wigner Research Centre for Physics,\\
H-1525 Budapest 114, P.O.B. 49, Hungary\\ 
\end{center}
\vspace{1.5cm}
\begin{abstract}
In this paper we present sets of linear integral equations which make possible to compute the 
finite volume expectation values of the trace of the stress energy tensor ($\Theta$) and the 
$U(1)$ current ($J_\mu$) in any eigenstate of the Hamiltonian of the sine-Gordon model. 
 The solution of these equations in the large volume 
limit allows one to get exact analytical formulas for the expectation values in the Bethe-Yang limit. 
These analytical formulas are used to test an earlier conjecture for the Bethe-Yang limit of 
expectation values in non-diagonally scattering theories. 
The analytical tests have been carried out upto three particle states and gave agreement with the conjectured 
formula, provided the definition of polarized symmetric diagonal form-factors is modified appropriately. 
Nevertheless, we point out that our results provide only a partial confirmation 
of the conjecture and further investigations 
are necessary to fully determine its validity. 
The most important missing piece in the confirmation is the mathematical proof of the finiteness 
of the symmetric diagonal limit of form-factors in a non-diagonally scattering theory. 
\end{abstract}

\end{titlepage}

%%%%%%%%%%%%%%%%%%%%%%%%%%%%%%%%%%%%%%%%%%%%%%%%%%%%%%%%%%%%%%
\section{Introduction} \label{intro}

Finite volume form-factors of integrable quantum field theories play an important role in the 
$AdS_5/CFT_4$ correspondence \cite{BJsftv,BJhhl} and in condensed matter applications \cite{KoEss}, as well. 
In AdS/CFT, their knowledge is indispensable for the computation of the string field 
theory vertex \cite{BJsftv} and of the heavy-heavy-light 3-point functions \cite{BJhhl} of the theory. 
In condensed matter systems finite volume form-factors can be used to compute quantum 
correlation-functions describing quasi 1-dimensional quantum-magnets, Mott insulators and carbon nanotubes \cite{KoEss}.

The systematic study of finite volume form-factors of integrable quantum field theories was initiated in \cite{PT08a,PT08b}, where 
the finite volume matrix-elements of local operators are sought in the form of a systematic large volume series. 
From the investigation of finite volume 2-point functions it turned out, that 
upto exponentially small finite volume corrections, but including all corrections in the inverse of the volume, 
the non-diagonal finite volume form-factors are equal to their infinite volume counterparts taken at the 
positions of the solutions of the  Bethe-Yang equations and normalized by the square roots of the densities of the sandwiching states. 

As a consequence of the Dirac-delta contact terms in the crossing relations of the form-factor axioms, the diagonal form-factors 
cannot be obtained from the non-diagonal ones by taking their straightforward diagonal limit. Thus, diagonal form-factors are related 
to the infinite volume form-factors in a bit more indirect way. According to the conjectures \cite{PT08b},
 they can be 
represented as density weighted linear combinations of the so-called "connected" or "symmetric" diagonal limits of the infinite volume form-factors of the theory.
In \cite{BWu}, it has been shown, that the conjecture  \cite{PT08b} being valid for purely elastic scattering-theories, can be derived from the 
leading order formula for the non-diagonal finite volume form-factors by considering such non-diagonal matrix elements, in which there is one particle 
more in the "bra" sandwiching state and the rapidity of this additional particle is taken to infinity appropriately.

The conjectures for purely elastic scattering theories \cite{PT08b} went through extensive analytical and numerical tests \cite{PST14} providing convincing amount of evidence 
for their validity. So far the conjecture for the more subtle non-diagonally scattering theories \cite{Palmai13} has not gone through convincing amount of tests. 
It has been tested in the sine-Gordon model, where it was checked  analytically in the whole pure soliton sector against exact \cite{En,En1} and numerical \cite{FPT11} results 
and numerically  against TCSA data for mixed soliton-antisoliton  two particles states \cite{Palmai13}. 
Thus, analytical tests of this conjecture is still missing in the soliton-antisoliton mixed sector. In this paper we would like to fill this gap and we 
check the conjecture of \cite{Palmai13} upto 3-particle soliton-antisoliton  mixed states.

Beyond the leading polynomial in the inverse of the volume terms, the exponentially small in volume corrections are also necessary. Their determination is still an open problem in general. 
Nevertheless, some progress has been reached in this direction, as well. For the non-diagonal form-factors in purely elastic scattering-theories there is some knowledge about the 
leading order exponentially small in volume corrections termed the L\"uscher corrections. The so-called $\mu$-term L\"uscher corrections were determined in \cite{Pmu} and the F-term corrections 
for vacuum-1-particle form factors has been determined in \cite{BBCL}. Unfortunately, the L\"uscher corrections to form-factors in non-diagonally scattering theories and 
higher order exponentially small in volume corrections in any integrable quantum field theory are presently out of reach. 

Nevertheless, much is known about the exact finite volume behavior of the diagonal form-factors both in purely elastic and non-diagonally scattering theories.

In \cite{Pozsg13,PST14} a LeClair-Mussardo type \cite{LM99} series representation was conjectured to describe exactly the finite volume dependence 
of diagonal matrix elements of local operators in purely elastic scattering theories. In non-diagonally scattering theories the description of 
finite volume diagonal matrix elements is less complete. So far only the sine-Gordon model has been studied in this class of theories. 
There, based on computations done in the framework of its integrable lattice regularization \cite{ddvlc}, a LeClair-Mussardo type series representation was proposed 
to describe the finite volume dependence of the expectation values of local operators in pure soliton states \cite{En}. Nevertheless, soliton-antisoliton mixed states 
have not been investigated so far. In this paper we partly fill this gap and derive integral equations
 to get any finite volume diagonal matrix elements of two important 
operators of the theory. These are the trace of the stress energy tensor ($\Theta$) and the $U(1)$ current ($J_\mu$). %of the theory.
Our formulas are valid to any value of the volume and to any eigenstate of the Hamiltonian of the model. 

In the paper the Bethe-Yang limit of the diagonal form-factors will play an important role.  
In our terminology this limit means, that the exponentially small in volume corrections are 
neglected from the large volume expansion  of the exact result{\footnote{Only the terms being polynomials 
in the inverse of the volume remain.}}.

In the repulsive regime, where there are no breathers in the spectrum, we solve our equations 
%for the expectation values 
in the Bethe-Yang limit and 
give exact formulas for the expectation values of our operators in this limit. The formulas depend on the rapidities 
of the physical particles and on the magnonic Bethe-roots of the Bethe-Yang equations.

With the help of these exact  formulas we check the conjecture of \cite{Palmai13} for the 
%leading order large volume behavior of 
Bethe-Yang limit of the diagonal matrix elements of local operators in non-diagonally scattering theories. The conjectured formula in \cite{Palmai13} contains the symmetric 
diagonal limit of the infinite volume form-factors.  The determination of these symmetric diagonal form-factors becomes more and more complicated 
as the number of particles increases. This is why we complete the test upto three particle states. 
Upto 3-particle states our exact formulas give perfect 
agreement 
with the conjectured formula of \cite{Palmai13} for the operators $\Theta$ and $J_\mu,$ 
provided the sandwiching color wave function $\Psi$ is replaced by its complex conjugate in the original 
formulas of \cite{Palmai13}.

However despite the success of our checks, the details of the computations shed light on some subtle points 
of the conjecture, which require further work to be confirmed. The most important of them is to prove that 
the symmetric diagonal limit of form-factors is finite for a generic sandwiching state in a non-diagonally 
scattering theory. Though, this statement looks intuitively quite trivial, in section 
\ref{sect9}, where we comment the computation of the symmetric diagonal limit of form-factors, 
we argue that this statement is not trivial at all. 

% of  $\Theta$ and $J_\mu.$ 

The outline of the paper is as follows.

In section \ref{sect2}. we summarize the most important facts about the models and about the operators of our interest. 
In section \ref{sect3}. the equations governing the exact finite volume dependence of diagonal matrix elements of the operators $\Theta$ and $J_\mu$ are derived. 
The solution of the equations in the large volume limit is given in section \ref{sect4}. 

The basic ingredients of the form-factor bootstrap program for the sine-Gordon model can be found in section \ref{sect5}.  
In section \ref{PTsejt}. we summarize the conjecture of \cite{Palmai13} for the Bethe-Yang limit of the 
diagonal matrix elements of local operators in non-diagonally scattering theories. 
In section \ref{sect7}. we compute the symmetric diagonal form-factors of the 
operators under consideration upto 3-particle states. In section \ref{sect8}. we perform the analytical checks of the conjecture \cite{Palmai13} upto 3-particle states.
In section \ref{sect9}. we comment on some subtle points of the conjecture of \cite{Palmai13}.
The body of the paper is closed by our summary and conclusions in section \ref{sect10}.
 
The paper contains two appendices, as well. Appendix \ref{appA} contains the detailed form of the linear integral equations governing the finite volume dependence of the expectation values of the operators $\Theta$ and $J_\mu.$ 
In appendix \ref{appB} the diagonalization of the soliton transfer-matrix is performed by means of algebraic Bethe-Ansatz. This appendix contains the classification of Bethe-roots, as well.
%In this appendix the classification of the Bethe-roots can also be found.

\section{The models and operators} \label{sect2}

In this paper we investigate the sine-Gordon  and  the massive Thirring models.
They are given by the well known Lagrangians:
\begin{equation}
\label{sG_Lagrangian}
{\cal L}_{SG}= \displaystyle\frac{1}{2}\partial _{\nu }\Phi \partial ^{\nu }\Phi +\displaystyle \alpha_0  \left( \cos \left( \beta \Phi \right)-1 \right), \,  \qquad 0<\beta^2<8 \pi,
\end{equation}
and
\begin{equation}
\label{mTh_Lagrangian}
{\cal L}_{MT}= \bar{\Psi }(i\gamma _{\nu }\partial ^{\nu }-m_{0})\Psi -\displaystyle\frac{g}{2}\bar{\Psi }\gamma^{\nu }\Psi \bar{\Psi }\gamma _{\nu }\Psi \,,
\end{equation}
where $m_0$ and $g$ denote the bare mass and the coupling constant of the massive Thirring model, respectively.
In (\ref{mTh_Lagrangian}) $\gamma_\mu$ stand for the $\gamma$-matrices, which satisfy 
the algebraic relations: $\{\gamma^\mu,\gamma^\nu\}=2 \eta^{\mu \nu}$ with $\eta^{\mu \nu}=\text{diag}(1,-1)$. 

The two models are equivalent in their even $U(1)$ charge sector \cite{s-coleman,klassme}, 
provided the coupling constants of the two theories are related by the formula:
\begin{equation} \label{gbeta}
1+\frac{g}{4 \pi}=\frac{4 \pi}{\beta^2}.
\end{equation}
In the sequel we will prefer the following parameterization of the coupling constant $\beta:$
\begin{equation}
\frac{\beta^2}{4 \pi}=\frac{2 \, p}{p+1}, \qquad 0<p\in \mathbb{R}.
\end{equation}
The ranges $0<p<1$ and $1<p$ correspond to the attractive and repulsive regimes of the theory respectively. 

The fundamental particles in the theory are the soliton ($+$) and the antisoliton ($-$)
 of mass ${\cal M}.$
Their exact S-matrix is well known \cite{ZamZam} and in terms of the coupling constant $p$ 
it can be written in the form as follows:
\begin{equation} \label{Smatr}
\begin{split}
{\cal S}_{ab}^{cd}(\theta)=S_0(\theta) \, S_{ab}^{cd}(\theta), \qquad a,b,c,d, \in \{\pm\},
\end{split}
\end{equation}
where $\theta$ is the relative rapidity of the scattering particles,
$S_0(\theta)$ is the soliton-soliton scattering amplitude:
\begin{equation} \label{CHI}
\begin{split}
S_0(\theta)=-e^{i \chi(\theta)}, \qquad \chi(\theta)=\!  \int\limits_{0}^{\infty} \! d\omega \, 
\frac{\sin(\omega \, \theta)}{\omega} \, \frac{\sinh(\tfrac{(p-1) \,\pi \omega}{2})}{ \cosh( \tfrac{\pi \omega}{2}) 
\,\sinh(\tfrac{p \,  \pi \, \omega}{2})}.
\end{split}
\end{equation}
The nonzero matrix elements of $S_{ab}^{cd}(\theta)$ in (\ref{Smatr}) 
can be expressed in terms of elementary functions as follows:
\begin{equation} \label{Selem}
\begin{split}
S_{++}^{++}(\theta)&=S_{--}^{--}(\theta)=1, \\
S_{+-}^{+-}(\theta)&=S_{-+}^{-+}(\theta)=B_0(\theta), \\
S_{+-}^{-+}(\theta)&=S_{-+}^{+-}(\theta)=C_0(\theta), 
\end{split}
\end{equation}
where 
\begin{equation} \label{B0}
\begin{split}
B_0(\theta)=\frac{\sinh \tfrac{\theta}{p}}{\sinh \tfrac{i \, \pi-\theta}{p}},
\end{split}
\end{equation}
\begin{equation} \label{C0}
\begin{split}
C_0(\theta)=\frac{\sinh \tfrac{i \, \pi}{p}}{\sinh \tfrac{i \, \pi-\theta}{p}}.
\end{split}
\end{equation}
The S-matrix (\ref{Smatr}) obeys the Yang-Baxter equation{\footnote{The matrix part $S_{ab}^{cd}(\theta)$ of the 
S-matrix also satisfies the Yang-Baxter equation.}}: 
\begin{equation} \label{YBE}
\begin{split}
{\cal S}_{k_2 k_3}^{j_2 j_3}(\theta_{23}) \, {\cal S}_{k_1 i_3}^{j_1 k_3}(\theta_{13}) \,
{\cal S}_{i_1 i_2}^{k_1 k_2}(\theta_{12})=
{\cal S}_{k_1 k_2}^{j_1 j_2}(\theta_{12}) \, 
{\cal S}_{i_1 k_3}^{k_1 j_3}(\theta_{13}) \,
{\cal S}_{i_2 i_3}^{k_2 k_3}(\theta_{23}), 
\end{split}
\end{equation}
with $ \theta_{ij}=\theta_i-\theta_j,$ for $i,j\in\{1,2,3\},$ 
and it satisfies the properties as follows:
%\begin{equation} \bullet \text{Parity-symmetry:} \qquad 
%{\cal S}_{ab}^{cd}(\theta)={\cal S}_{ba}^{dc}(\theta), \label{Bose}\end{equation}
%\begin{equation} \bullet \text{Time-reversal symmetry:} \qquad 
%{\cal S}_{ab}^{cd}(\theta)={\cal S}_{cd}^{ab}(\theta), \label{Time}\end{equation}
%\begin{equation} \bullet \text{Crossing-symmetry:} \qquad 
%{\cal S}_{ab}^{cd}(\theta)={\cal S}_{a\bar{d}}^{c\bar{b}}(\theta), \label{Cross}\end{equation}
%\begin{equation} \bullet \text{Unitarity:} \qquad 
%{\cal S}_{ab}^{ef}(\theta)\, {\cal S}_{ef}^{cd}(\theta)=\delta_a^c \, \delta_b^d, \label{Unit}\end{equation}
%\begin{equation} \bullet \text{Real analyticity:} \qquad 
%{\cal S}_{ab}^{* cd}(\theta)={\cal S}_{ab}^{cd}(-\theta^*), \label{Real}\end{equation}
\begin{eqnarray} & \bullet & \text{Parity-symmetry:} \qquad  \qquad \qquad
{\cal S}_{ab}^{cd}(\theta)={\cal S}_{ba}^{dc}(\theta), \label{Bose} \\
 & \bullet & \text{Time-reversal symmetry:} \qquad  \quad \!
{\cal S}_{ab}^{cd}(\theta)={\cal S}_{cd}^{ab}(\theta), \label{Time} \\
 & \bullet & \text{Crossing-symmetry:} \qquad  \qquad \quad
{\cal S}_{ab}^{cd}(\theta)={\cal S}_{a\bar{d}}^{c\bar{b}}(i \, \pi-\theta), \label{Cross} \\
 & \bullet & \text{Unitarity:} \qquad \qquad \qquad \, \,
{\cal S}_{ab}^{ef}(\theta)\, {\cal S}_{ef}^{cd}(-\theta)=\delta_a^c \, \delta_b^d, \label{Unit} \\
 & \bullet & \text{Real analyticity:} \qquad  \qquad \quad \quad \!
{\cal S}_{ab}^{cd}(\theta)^*={\cal S}_{ab}^{cd}(-\theta^*),  \label{Real} 
\end{eqnarray}
where for any index $a,$ $\bar{a}$ denotes the charge conjugated particle ($\bar{a}=-a$). 
The charge conjugate of a soliton is an antisoliton and vice versa, thus the 
charge conjugation matrix acting on the two dimensional vector space spanned by the soliton and the 
antisoliton, is equal to the first Pauli-matrix:
\begin{equation} \label{Cmatr}
C=\sigma_x=\begin{pmatrix}
0 & 1 \\
1 & 0
\end{pmatrix}, \quad \text{or equivalently:} \quad C_{ab}=\delta_{a \bar{b}}.
\end{equation} 

In this paper we determine the finite volume expectation values of the operators as follows;
the trace of the stress energy tensor:
\begin{equation}\label{THETA}
\Theta=2 \, \alpha_0 \, (1-\tfrac{\beta^2}{8 \pi})\, \cos(\beta \, \Phi).
\end{equation}
and the $U(1)$ current of the theory:
\begin{equation}\label{Jmu}
J_\mu=\tfrac{\beta}{2 \pi}\, \epsilon_{\mu \nu} \partial^\nu \Phi, \qquad \mu=0,1,
\end{equation}
where $\epsilon_{\mu \nu}$ denotes the antisymmetric matrix with nonzero entries: 
$\epsilon_{10}=-\epsilon_{01}=1.$ 

Both operators correspond to some conserved quantity of the theory and  in the 
subsequent sections their finite volume expectation values will be expressed in terms of the 
counting-function governing the finite volume spectrum of the theory. 

The two operators have different parities under charge conjugation; $\Theta$ is positive, while $J_\mu$ is negative. 
This property proves to be an important difference between the two operators, when the symmetric 
diagonal limit of their form-factors are computed.

\section{Finite volume expectation values of $\Theta$ and $J_\mu$} \label{sect3}

 In this section we give the equations, which govern the finite volume dependence of all diagonal form-factors of the trace of the stress energy tensor ($\Theta$) and of the $U(1)$ 
 current ($J_\mu$) of the sine-Gordon theory. The equations for pure solitonic expectation values have been derived in \cite{En,En1}. The derivations were based on an integrable 
 lattice regularization of the model, on the so-called light-cone lattice regularization \cite{ddvlc}. In this section we extend the results of \cite{En,En1} from the pure soliton 
 sector to all excited states of the model. To keep the paper within reasonable size, instead of repeating the lattice regularization based derivations we will derive the equations 
 in a more pragmatic way. From \cite{zamiz} it is well known, that the expectation values of the trace of the stress energy tensor can be computed from the finite volume dependence of the 
 energy of the sandwiching state by the formula as follows:
 \begin{equation} \label{Th0}
\begin{split}
\langle \Theta \rangle_L={\cal{M}} \left( \frac{E(\ell)}{\ell}+\frac{d}{d\ell}E(\ell) \right),
\end{split}
\end{equation}
where $\ell={\cal M} L$ with ${\cal M}$ and $L$ being the soliton mass and the finite volume respectively.
  This implies that the diagonal form-factors of $\Theta$ can be expressed in terms of certain derivatives of the counting-function of the model \cite{En1}. In the case of $\Theta,$ 
 the derivatives entering the equations are the derivative with respect to the spectral parameter and the derivative with respect to the dimensionless volume of the model. 
 The computations achieved in the pure soliton sector \cite{En} imply, that the same derivatives describe the finite volume dependence of the expectation values of the $U(1)$ current, too.

In order to formulate the equations describing the finite volume diagonal form-factors of our interest, we have to recall how the finite volume spectrum of the theory 
is described in terms of  the nonlinear integral equations (NLIE) \cite{KP1,ddv92} 
satisfied by the counting-function. Since we know that the expectation values of our interest can be 
expressed in terms of certain derivatives of the counting function, we can skip the intermediate lattice versions of the equations, and we can formulate the problem 
directly in the continuum limit.

\subsection{Nonlinear integral equations for the counting-function}

The counting-function $Z(\theta)$ is a periodic function on the complex plane with period $i {\pi}(1+p).$
To describe general excited states of the model one needs to know how to determine $Z(\theta)$ for any $\theta$ lying in the whole strip 
$\left[ -i \frac{\pi}{2}(1+p),i \frac{\pi}{2}(1+p)\right].$ The counting-function satisfies different equations in the different domains of the 
periodicity strip. In the fundamental domain defined by the strip $|\mbox{Im} \theta|\leq \mbox{min}(p \pi,\pi)$  the continuum limit of the counting-function 
satisfies the nonlinear-integral equations as follows \cite{ddv97,FRT1,FRT2,FRT3}:
%The continuum limit of the nonlinear-integral equations satisfied by the  counting-function take the form as follows \cite{}: 
\begin{equation} \label{DDV}
\begin{split}
Z(\theta)=\ell \sinh \theta \! +\! \sum\limits_{k=1}^{m_H} \chi(\theta-h_k) \! -\! \sum\limits_{k=1}^{m_C} \chi(\theta-c_k) \!
-\!\sum\limits_{k=1}^{m_S} \left(\chi(\theta-y_k+i \eta)\!+\!\chi(\theta-y_k-i \eta) \right)
\\
-\!\sum\limits_{k=1}^{m_W}  \chi_{II}(\theta-w_k)\!+ \! \int\limits_{-\infty}^{\infty} \frac{d\theta'}{2 \pi i}  G(\theta-\theta'-i\eta)  L_{+}(\theta'+i \eta) 
- \int\limits_{-\infty}^{\infty} \frac{d\theta'}{2 \pi i} G(\theta-\theta'+i\eta) L_-(\theta'-i \eta),
\end{split}
\end{equation}
 where 
 \begin{equation} \label{Lpm}
L_{\pm}(\theta)=\ln\left(1+(-1)^\delta \, e^{\pm i \, Z(\theta)} \right),
\end{equation}
such that the parameter $\delta$ can take values $0$ or $1.$ 
Its value affect the quantization equations of the objects entering the source terms of the integral equation. 
In (\ref{DDV}), $\chi(\theta)$ is the soliton-soliton scattering 
phase given by (\ref{CHI}) and $G(\theta)$ denotes its derivative. It can be given 
by the Fourier-integral as follows:
\begin{equation} \label{G}
G(\theta)= \frac{d}{d\theta} \chi(\theta)=\! \int\limits_{-\infty}^{\infty} \! d\omega \, 
e^{-i \, \omega \theta} \, \frac{\sinh(\tfrac{(p-1) \,\pi \omega}{2})}{2 \cosh( \tfrac{\pi \omega}{2}) 
\,\sinh(\tfrac{p \,  \pi \, \omega}{2})}.
\end{equation}
The equations contain the so-called second determination \cite{ddv97} of $\chi(\theta),$ as well. 
For any function $f,$ the definition of second determination is different in the attractive ($0<p<1$) 
and repulsive ($1<p$) regimes of the model: 
\begin{equation} \label{fII}
\begin{split}
f_{II}(\theta)=\left\{ 
\begin{array}{r}
f( \theta)+f(\theta-i \, \pi \,\text{sign}(\text{Im} \theta)), \qquad
\qquad 1<p, \\
f( \theta)-f(\theta-i \, \pi \,p \, \text{sign}(\text{Im} \theta)),
\qquad 0<p<1.
\end{array}\right.
\end{split}
\end{equation}
For the function $\chi(\theta)$ we provide the concrete functional forms as well \cite{Fevphd}:
\begin{equation} \label{CHI2}
\begin{split}
\chi_{II}(\theta)=\left\{ 
\begin{array}{r}
i \, \, \text{sign}(\text{Im} \, \theta)\left( \log\sinh\frac{\theta}{p}-\log \sinh\frac{\theta-i \, \pi \, \text{sign}\,  \text{Im} \, \theta}{p} \right), \qquad
\qquad 1<p, \\
i \, \, \text{sign}(\text{Im} \, \theta)\left( \log\left(-\tanh\frac{\theta}{p}\right)+\log \tanh\frac{\theta-i \, \pi \, p \, \text{sign}\,  \text{Im} \, \theta}{p} \right),
\qquad 0<p<1.
\end{array}\right.
\end{split}
\end{equation}
In (\ref{DDV}), $\eta$ is an arbitrary positive contour-deformation parameter, which should be in the range 
$[0,\text{min}(p \pi,\pi,|\mbox{Im} \, c_j|)].$  
As we have already mentioned, $\ell$ denotes the 
dimensionless volume made out of the  the soliton mass ${\cal M}$ and of the volume $L$ of the theory  
 by the formula $\ell={\cal M} L.$ 
All objects entering the source terms in (\ref{DDV}) satisfy the equation: 
\begin{equation} \label{qo}
\begin{split}
1+(-1)^{\delta}e^{i Z({\frak O})}=0, \qquad {\frak O} \in \{h_k\}_{k=1}^{m_H} \cup \{c_k\}_{k=1}^{m_C} \cup \{w_k\}_{k=1}^{m_W}\cup \{y_k\}_{k=1}^{m_S}.
\end{split}
\end{equation}
It is useful to classify them as follows \cite{ddv97}:
\begin{itemize}
\item holes: $h_k \in \mathbb{R}, \quad k=1,...,m_H$
\item close roots: $c_k \quad k=1,...,m_C$, with $|\mbox{Im} c_k|\leq \mbox{min}(\pi,p \pi)$, 
\item wide roots: $w_k \quad k=1,...,m_W$, with $\mbox{min}(\pi,p \pi)<|\mbox{Im} w_k|\leq \tfrac{\pi}{2}(1+p)$,
\item special objects{\footnote{With this interpretation of special objects the contour deformation parameter 
$\eta$ should be considered to be a positive infinitesimal number. %Thus in the sequel, when special objects 
%enter the equations, one should consider $\eta$ to be infinitesimal.
}}
: $y_k \in \mathbb{R}, \quad k=1,...,m_S \,\,$ defined by the equations $1\!+\!(-1)^{\delta}e^{i Z(y_k)}\!=\!0$ with $Z'(y_k)<0.$
\end{itemize}
Their numbers determine the topological charge $Q$ of the state  by the so-called counting-equation:
\begin{equation} \label{counteq}
\begin{split}
Q=m_H-2 m_S-m_C-2 H(p-1) m_W,
\end{split}
\end{equation}
where here $H(x)$ denotes the Heaviside-function.
As a consequence of (\ref{qo}) the source objects satisfy the quantization equations as follows:
%\begin{itemize}
%\item holes: $Z(h_k)=2\pi \, I_{h_k}, \qquad I_{h_k} \in \mathbb{Z}+\tfrac{1+\delta}{2}, \qquad k=1,..,m_H,$ \begin{equation}\label{Hkvant}\end{equation}
%\item close roots: $Z(c_k)=2\pi \, I_{c_k}, \qquad I_{c_k} \in \mathbb{Z}+\tfrac{1+\delta}{2}, \qquad k=1,..,m_C,$
%\item wide roots: $Z(w_k)=2\pi \, I_{w_k}, \qquad I_{w_k} \in \mathbb{Z}+\tfrac{1+\delta}{2}, \qquad k=1,..,m_W,$
%\item special objects: $Z(y_k)=2\pi \, I_{y_k}, \qquad I_{y_k} \in \mathbb{Z}+\tfrac{1+\delta}{2}, \qquad k=1,..,m_S.$
%\end{itemize}
%
%jkdsgggfdjghfjkdgjfgjhgfghjgjh
\begin{eqnarray}
&\bullet& \, \text{holes:}  \qquad Z(h_k)=2\pi \, I_{h_k}, \qquad I_{h_k} \in \mathbb{Z}+\tfrac{1+\delta}{2}, \qquad k=1,..,m_H, \label{Hkvant}\\
&\bullet& \, \text{close roots:}  \qquad Z(c_k)=2\pi \, I_{c_k}, \qquad I_{c_k} \in \mathbb{Z}+\tfrac{1+\delta}{2}, \qquad k=1,..,m_C, \label{Ckvant} \\
&\bullet& \, \text{wide roots:}  \qquad Z(w_k)=2\pi \, I_{w_k}, \qquad I_{w_k} \in \mathbb{Z}+\tfrac{1+\delta}{2}, \qquad k=1,..,m_W, \label{Wkvant} \\
&\bullet& \, \text{special objects:}  \qquad Z(y_k)=2\pi \, I_{y_k}, \qquad I_{y_k} \in \mathbb{Z}+\tfrac{1+\delta}{2}, \qquad k=1,..,m_S. \label{Skvant}
\end{eqnarray}
From this list one can see that the actual value of the parameter $\delta \in \{0,1\}$ determines whether the source objects are quantized by integer or half integer quantum numbers. 
It was shown in \cite{FRT1,FRT2,FRT3}, that not all choices of $\delta$ are possible to describe properly the states of the sine-Gordon or of the Massive Thirring model. To describe the proper 
states of these quantum field theories the following selection rules have to be satisfied by the parameter $\delta:$
 \begin{eqnarray} 
&\bullet& \quad \frac{Q+\delta+M_{sc}}{2} \in \mathbb{Z}, \qquad \text{sine-Gordon,} \qquad \qquad\qquad \qquad \qquad\qquad \label{SGkvant} \\
%\label{MTkvant}
&\bullet& \quad \frac{\delta+M_{sc}}{2} \in \mathbb{Z}, \qquad  \text{massive Thirring,} \qquad \qquad\qquad \qquad \qquad\qquad \label{MTkvant}
\end{eqnarray}
where here $M_{sc}$ stands for the number of self-conjugate roots, which are such wide roots, whose imaginary parts are fixed by the periodicity of $Z(\theta)$ to 
$i\, \tfrac{\pi}{2}(1+p).$

In order to be able to impose the quantization equations (\ref{Wkvant}) for the wide roots, the integral representation of $Z(\theta)$ must be known in the strip 
$\text{min}(p \, \pi,\pi)<\text{Im} \, \theta\leq \tfrac{\pi}{2}(1+p),$ as well. 
In this "wide-root domain" $Z(\theta)$ is given by the equations as follows \cite{ddv97,FRT1,FRT2,FRT3}:
\begin{equation} \label{DDVII}
\begin{split}
Z(\theta)=\ell \sinh_{II} (\theta) \! +{\cal D}_{II}(\theta)+\sum\limits_{\alpha=\pm} \, \alpha 
\! \int\limits_{-\infty}^{\infty} \frac{d\theta'}{2 \pi i}  G_{II}(\theta-\theta'-i \, \alpha \, \eta)  L_{\alpha}(\theta'+i \, \alpha \,  \eta),
\end{split}
\end{equation}
where ${\cal D}_{II}(\theta)$ is the second determination (\ref{fII}) of the source term function of (\ref{DDV}):
\begin{equation} \label{calD}
\begin{split}
{\cal D}(\theta)\!\!=\!\!\sum\limits_{k=1}^{m_H} \chi(\theta\!-\!h_k) \! -\! \sum\limits_{k=1}^{m_C} \!\chi(\theta\!-\!c_k) \!
-\!\!\sum\limits_{k=1}^{m_S} \left(\chi(\theta\!-\!y_k\!+\!i \eta)\!+\!\chi(\theta\!-\!y_k\!-\!i \eta) \right)\!-\!\sum\limits_{k=1}^{m_W}  \chi_{II}(\theta\!-\!w_k).
\end{split}
\end{equation}
The energy and momentum of the model can be expressed in terms of the solution of the nonlinear integral equations by the following formulas \cite{ddv97,FRT1,FRT2,FRT3}:
\begin{equation} \label{EL}
\begin{split}
E(L)\!\!=\!\!{\cal M} \left( \sum\limits_{k=1}^{m_H} \cosh(h_k) \! -\! \sum\limits_{k=1}^{m_C} \!\cosh(c_k) \!
-\!\!\sum\limits_{k=1}^{m_S} \left(\cosh(y_k\!+\!i \eta)\!+\!\cosh(\!y_k\!-\!i \eta) \right)\!-\!\!\sum\limits_{k=1}^{m_W}  \cosh_{II}(w_k) - \right.\\
\left. \int\limits_{-\infty}^{\infty} \! \frac{d\theta}{2 \pi i}  \sinh(\theta+i\eta)  L_{+}(\theta+i \eta) 
+ \! \int\limits_{-\infty}^{\infty} \! \frac{d\theta}{2 \pi i} \sinh(\theta-i\eta) L_-(\theta-i \eta)
 \right),
\end{split}
\end{equation}
\begin{equation} \label{PL}
\begin{split}
P(L)\!\!=\!\!{\cal M} \left( \sum\limits_{k=1}^{m_H} \sinh(h_k) \! -\! \sum\limits_{k=1}^{m_C} \!\sinh(c_k) \!
-\!\!\sum\limits_{k=1}^{m_S} \left(\sinh(y_k\!+\!i \eta)\!+\!\sinh(\!y_k\!-\!i \eta) \right)\!-\!\!\sum\limits_{k=1}^{m_W}  \sinh_{II}(w_k) - \right.\\
\left. \int\limits_{-\infty}^{\infty} \! \frac{d\theta}{2 \pi i}  \cosh(\theta+i\eta)  L_{+}(\theta+i \eta) 
+ \! \int\limits_{-\infty}^{\infty} \! \frac{d\theta}{2 \pi i} \cosh(\theta-i\eta) L_-(\theta-i \eta)
 \right).
\end{split}
\end{equation}

\subsection{Expectation values of $\Theta$}

The computations of finite volume expectation values of the trace of the stress energy tensor goes 
analogously to the former computations done in purely elastic scattering theories \cite{PST14}. 
Formula (\ref{Th0}) implies that the finite volume expectation values of 
the trace of the stress energy tensor 
$\Theta$ can be expressed in terms of the $\theta$ and $\ell$ 
derivatives of $Z(\theta).$ By differentiating the equations (\ref{DDV})-(\ref{DDVII}) it is easy to show that these derivatives satisfy linear integral equations with 
kernels containing the counting-equation itself \cite{En,En1}. 

We introduce two functions with related sets of discrete variables by the definitions as follows:
\begin{equation} \label{Gd}
\begin{split}
{\cal G}_d(\theta)&=Z'(\theta), \qquad \\
X_{d,k}^{(h)}&=\frac{{\cal G}_d(h_k)}{Z'(h_k)}=1, \qquad k=1,...,m_H, \\
X_{d,k}^{(c)}&=\frac{{\cal G}_d(c_k)}{Z'(c_k)}=1, \qquad k=1,...,m_C, \\
X_{d,k}^{(y)}&=\frac{{\cal G}_d(y_k)}{Z'(y_k)}=1, \qquad k=1,...,m_S, \\
X_{d,k}^{(w)}&=\frac{{\cal G}_d(w_k)}{Z'(w_k)}=1, \qquad k=1,...,m_W.
\end{split}
\end{equation}
and
\begin{equation} \label{Gl}
\begin{split}
{\cal G}_\ell(\theta)&=\frac{d}{d\ell}Z(\theta|\ell), \qquad \\
X_{\ell,k}^{(h)}&=\frac{{\cal G}_\ell(h_k)}{Z'(h_k)}=-h_k'(\ell), \qquad k=1,...,m_H, \\
X_{\ell,k}^{(c)}&=\frac{{\cal G}_\ell(c_k)}{Z'(c_k)}=-c_k'(\ell), \qquad k=1,...,m_C, \\
X_{\ell,k}^{(y)}&=\frac{{\cal G}_\ell(y_k)}{Z'(y_k)}=-y_k'(\ell), \qquad k=1,...,m_S, \\
X_{\ell,k}^{(w)}&=\frac{{\cal G}_\ell(w_k)}{Z'(w_k)}=-w_k'(\ell), \qquad k=1,...,m_W.
\end{split}
\end{equation}
%From the NLIE (\ref{DDV})-(\ref{DDVII}) it can be shown, the variables (\ref{Gd}) and (\ref{Gl}) satisfy the set of linear integral equations (\ref{ujset})-(\ref{bfGdef}).
Taking the appropriate derivatives of the NLIE (\ref{DDV})-(\ref{DDVII}) it can be shown, that the variables in (\ref{Gd}) and in (\ref{Gl}) satisfy  sets of linear integral equations. We relegated these equations to appendix \ref{appA}, 
where their explicit form is given by the formulas (\ref{ujset})-(\ref{bfGdef}).

With the help of (\ref{Th0}) it can be shown, that he finite volume expectation value of $\Theta$ in a state described by the NLIE (\ref{DDV})-(\ref{DDVII}) can be expressed in terms of the 
variables (\ref{Gd}) and (\ref{Gl}) by the following formula:
\begin{equation} \label{THexp}
\begin{split}
\langle \Theta \rangle_L=\langle \Theta \rangle_\infty+{\cal M}^2 \, \Theta_{rest}(\ell), 
\end{split}
\end{equation}
where $\langle \Theta \rangle_\infty$ stands for the infinite volume "bulk" vacuum expectation value \cite{ddvaft,ddv95}:
\begin{equation} \label{THinf}
\begin{split}
\langle \Theta \rangle_\infty=-\frac{{\cal M}^2}{4} \tan\left(\tfrac{p \pi}{4}\right),
\end{split}
\end{equation}
and $\Theta_{rest}(\ell)$ denotes the dimensionless part of the rest of the expectation value. 
It is given by the formula:
\begin{equation} \label{THrest}
\begin{split}
\Theta_{rest}(\ell)\!=\!\sum\limits_{k=1}^{m_H} \!\! \left(\cosh h_k \frac{X_{d,k}^{(h)}}{\ell}\!-\!\sinh h_k X_{\ell,k}^{(h)}\right)
\!-\!\sum\limits_{k=1}^{m_C} \!\! \left(\cosh c_k \frac{X_{d,k}^{(c)}}{\ell}\!-\!\sinh c_k X_{\ell,k}^{(c)}\right)- \\
\!-\!\sum\limits_{k=1}^{m_S} \!\! \left(\left(\cosh (y_k+i \eta)+\cosh (y_k-i \eta)\right) \frac{X_{d,k}^{(y)}}{\ell}\!-\!\left(\sinh (y_k+i \eta)+\sinh (y_k-i \eta)\right) X_{\ell,k}^{(y)}\right) \\
\!-\!\sum\limits_{k=1}^{m_W} \!\! \left(\cosh_{II}(w_k) \frac{X_{d,k}^{(w)}}{\ell}\!-\!\sinh_{II} (w_k) X_{\ell,k}^{(w)}\right)+ \\
\sum\limits_{\alpha=\pm} \! \int\limits_{-\infty}^{\infty} \! \frac{d\theta}{2 \pi} \! 
\left[\cosh(\theta+i \, \alpha \, \eta) \, \frac{{\cal G}_d(\theta+i \, \alpha \, \eta)}{\ell}-\sinh(\theta+i \, \alpha \, \eta) \, {\cal G}_\ell(\theta+i \, \alpha \, \eta) \right] 
{\cal F}_{\alpha}(\theta+i \, \alpha \, \eta),
\end{split}
\end{equation}
where ${\cal F}_\pm(\theta)$ stands for the nonlinear combinations:
\begin{equation} \label{calF}
\begin{split}
{\cal F}_{\pm}(\theta)=\frac{(-1)^\delta \, e^{\pm i \, Z(\theta)}}
{1+(-1)^\delta \, e^{\pm i \, Z(\theta)}}. 
\end{split}
\end{equation}

\subsection{Expectation values of $J_\mu$}

The finite volume expectation values of the $U(1)$ current can be derived from the light-cone lattice 
regularization \cite{ddvlc} of the model. In this way the expectation values of $J_\mu$  between pure soliton states 
have been determined in \cite{En}. Nevertheless, the computations of \cite{En} can be easily extended to all excited 
states of the model. Here, we skip the lengthy, but quite straightforward computations and present only the final 
result. As the pure soliton results of \cite{En} suggest, the expectation values of $J_0$ and $J_1$ can be expressed 
in terms of the set of variables of (\ref{Gd}) and of (\ref{Gl}), respectively:
\begin{equation} \label{J0exp}
\begin{split}
\langle J_0 \rangle_L=\frac{1}{L}\left\{ \sum\limits_{j=1}^{m_H} X_{d,j}^{(h)} -
2 \sum\limits_{j=1}^{m_S} X_{d,j}^{(y)}-\sum\limits_{j=1}^{m_C} X_{d,j}^{(c)}-
2 \, H(p-1) \, \sum\limits_{j=1}^{m_W} X_{d,j}^{(w)}- \right. \\
\left. \sum\limits_{\alpha=\pm} \int\limits_{-\infty}^{\infty} \! \frac{d\theta}{2 \pi} \,
{\cal G}_d(\theta+i \, \alpha \, \eta) \, {\cal F}_\alpha(\theta+i \, \alpha \,\eta)
\right\},
\end{split}
\end{equation}
\begin{equation} \label{J1exp}
\begin{split}
\langle J_1 \rangle_L={\cal M}\left\{ \sum\limits_{j=1}^{m_H} X_{\ell,j}^{(h)} -
2 \sum\limits_{j=1}^{m_S} X_{\ell,j}^{(y)}-\sum\limits_{j=1}^{m_C} X_{\ell,j}^{(c)}-
2 \, H(p-1) \, \sum\limits_{j=1}^{m_W} X_{\ell,j}^{(w)}- \right. \\
\left. \sum\limits_{\alpha=\pm} \int\limits_{-\infty}^{\infty} \! \frac{d\theta}{2 \pi} \,
{\cal G}_\ell(\theta+i \, \alpha \, \eta) \, {\cal F}_\alpha(\theta+i \, \alpha \,\eta)
\right\}.
\end{split}
\end{equation}
Here $H(x)$ is the Heaviside function. Using the definitions (\ref{Gd}) and 
the counting equation (\ref{counteq}), it is easy to show that formula (\ref{J0exp}) gives the 
 correct result $\langle J_0 \rangle_L=\tfrac{Q}{L}$ for the finite volume expectation values of $J_0$ in 
sandwiching states with topological charge $Q.$

\section{Large volume solution} \label{sect4}

In this section we provide exact formulas for the Bethe-Yang limit of the  
expectation values of the trace of the stress energy tensor and 
of the $U(1)$ current in the repulsive ($1<p$) regime of the sine-Gordon model{\footnote{Our formulas are valid for the massive Thirring model, as well. 
Only the value of the parameter $\delta$ should be set in accordance with (\ref{MTkvant}).}}. 
The reason why we restrict ourselves to the repulsive regime is that in this 
regime the correspondence between the Bethe-roots entering the NLIE (\ref{DDV}) and magnonic Bethe-roots of 
the  Bethe-Yang equations (\ref{ABAE}) is quite direct. In the attractive regime the correspondence 
is much more complicated and more indirect. 

The first step to make a correspondence between the source objects or the Bethe-roots of the NLIE and the roots of 
(\ref{ABAE}), is to find the relation between the holes of the NLIE (\ref{DDV}) and the rapidity of physical particles entering the  magnonic part of the Bethe-Yang equations (\ref{ABAE}). 
It is well known in the literature \cite{ddv97} that the holes in the NLIE description describe the rapidities of the physical particles in the large volume limit 
($\{h_j\}=\{\theta_j\}$). 
Then one has to know what kind of complexes the roots of the NLIE fall into, when the infinite volume limit is taken. 
In the repulsive regime these complexes are as follows \cite{Viallet}:
%\begin{itemize}
%\item 2-strings: $\quad s^{(2)}_j=s_j\pm i \, \tfrac{\pi}{2}, \qquad j=1,...,n_2,$
%\item quartets: $\quad q_j=\{q^{(1)}_j \!\pm\! i  \tfrac{\pi}{2}, q^{(2)}_j \! \pm\! i  \tfrac{\pi}{2}\}, \qquad \text{with } \quad 
%|\text{Im}\, q^{(a)}_j|\leq \tfrac{\pi}{2} \qquad a\!=\!1,2 \quad j\!=\!1,...,n_4,$
%\end{itemize}
\begin{eqnarray}
&\bullet& \text{2-strings:} \quad s^{(2)}_j=s_j\pm i \, \tfrac{\pi}{2}, \qquad \text{with:} \quad s_j\in{\mathbb R}, \quad j=1,...,n_2, \nonumber \\
%&\bullet& \text{quartets:} \quad q_j=\{q^{(1)}_j \!\pm\! i  \tfrac{\pi}{2}, q^{(2)}_j \! \pm\! i  \tfrac{\pi}{2}\}, \quad \text{with:} \quad 
%|\text{Im}\, q^{(a)}_j|\leq \tfrac{\pi}{2}, \qquad a\!=\!1,2, \quad j\!=\!1,...,n_4, \nonumber \\
&\bullet& \text{quartets:} \quad q_j=\{q^{(\pm)}_j \!\pm\! i  \tfrac{\pi}{2}\}, \quad \text{with:} \quad q^{(+)}_j=(q^{(-)}_j)^*, \quad
|\text{Im}\, q^{(\pm)}_j|\leq \tfrac{\pi}{2}, \quad  \, j\!=\!1,...,n_4, \nonumber \\
&\bullet& \text{wide-roots:} \quad w_j, \quad \text{with:} \quad \pi<|\text{Im}\, w_j|\leq\tfrac{(1+p)\, \pi}{2}, \quad j\!=\!1,...,m_w, 
\label{quartett}
\end{eqnarray}
such that wide-roots either form complex conjugate pairs or they are self-conjugate roots with 
fixed imaginary part: $\text{Im} \, w^{(sc)}_j=\frac{(1+p) \, \pi}{2}.$
From this classification, one can see that only the close-roots fall into special configurations in the infinite volume limit
 ($m_C=2n_2+4n_4$). Namely, they form either 
quartets or 2-strings, where the latter can be thought of as degenerate quartets. Here we note that in the $\ell \to \infty$ limit 
there are no special objects, so they do not enter the expressions in this limit.

The counting equation (\ref{counteq}) tells us how these root configurations act on the topological charge of a state:
\begin{itemize}
\item a 2-string decreases the charge by 2,
\item a quartet decreases the charge by 4,
\item a wide-root decreases the charge by 2.
\end{itemize}

On the other hand each root of the magnonic part of the Bethe-Yang equations decrease the topological charge of a state by 2 units and as it was mentioned in appendix 
\ref{appABA2} the roots of these equations form conjugate pairs with respect to the line $\text{Im} \, z=\tfrac{\pi}{2}.$
These suggest the following identification between the $\ell \to \infty$ complexes  in (\ref{quartett}) and the different types of roots of the 
magnonic part of the Bethe-Yang equations given in (\ref{rootclassR}):

\begin{itemize}
\item Real-roots of (\ref{rootclassR}) correspond to 2-strings in (\ref{quartett}), such that the real part of a real-root 
is equal to the center of the corresponding 2-string: $\lambda_j-i \tfrac{\pi}{2}=s_j.$
\item a close-pair $\lambda^{(\pm)}_j$ in (\ref{rootclassR}) corresponds to a quartet in (\ref{quartett}), such that the positions of the close-pair are given by
complex conjugate pair describing the quartet: 
\newline $\{\lambda^{(\pm)}_j-i \tfrac{\pi}{2}\}=\{q_j^{(\pm)}\}.$
\item wide-roots in (\ref{rootclassR}) correspond to wide roots of (\ref{quartett}), such that:
\newline
 $\lambda^{wide}_j-i \tfrac{\pi}{2}=w_j-i \,\text{sign}(\text{Im}\, w_j) \,\tfrac{\pi}{2}.$
\end{itemize}

%In addition to the above correspondence between the Bethe-roots entering the NLIE (\ref{DDV}) and the Asymptotic Bethe equations (\ref{ABAE})

With this correspondence 
%between the Bethe-roots of the NLIE (\ref{DDV}) and of the Asymptotic Bethe equations (\ref{ABAE}),
 the $\ell \to \infty$ limit of the NLIE can be mapped to the equations (\ref{ABAE}). 
This proves to be important in finding an exact
formula expressed in terms of the roots of (\ref{ABAE}) 
for the Bethe-Yang limit of the diagonal form-factors of $\Theta$ and $J_\mu.$

To find the leading order large volume solution of the diagonal form-factors of $\Theta$ and $J_\mu,$ one should recognize that the integral terms in 
(\ref{THexp}), (\ref{THrest}), (\ref{J0exp}) and (\ref{J1exp}) are exponentially small in the volume, and so negligible at leading order. 
Consequently, the only task is to determine the Bethe-Yang limit of the discrete variables $X_{d,j}$ and $X_{\ell,j}.$
They are solutions of the equations (\ref{ujset})-(\ref{Qyk}). For the first sight, it does not seem to be easy to find the general solutions of these 
equations in the large volume limit, but the relations of these $X$-variables to the $\theta$ and $\ell$ derivatives of the counting-function 
given in (\ref{Gd}) and in  (\ref{Gl}), makes it quite easy to find the required solutions. 

First let us consider the variables related to the $\ell$ derivative of $Z(\theta)$ in (\ref{Gl}). 
Then using (\ref{Gl}), for a complex root{\footnote{For more precise notation see (\ref{ujset}) in appendix \ref{appA}.}} $u_j$  the corresponding $X$-variable can be written as:
\begin{equation} \label{Xul}
\begin{split}
X^{(u)}_{\ell,j}=-u_j'(\ell)=-\sum\limits_{k=1}^{m_H} \frac{\partial u_j}{\partial h_k} \, h_k'(\ell)
\stackrel{\ell \to \infty}{\approx} \sum\limits_{k=1}^{m_H} \frac{\partial \lambda_j}{\partial h_k} \, X^{(h)}_{\ell,j},
\end{split}
\end{equation}
where we used (\ref{Gl}) and exploited the large volume correspondence between the Bethe-roots of the
 NLIE (\ref{DDV}) and  magnonic the Bethe-roots of (\ref{ABAE}). 
The formula (\ref{Xul}) expresses the complex root's $X$-variables in terms of those of the holes 
in the large volume limit. 
Then the large $\ell$ solution goes as follows:
One should insert (\ref{Xul}) into (\ref{Qhk}) taken at $\nu=\ell,$ such that the integral terms are neglected because they are exponentially small in volume.
 This way one gets a closed discrete set of linear equations for the variables $X^{(h)}_{\ell,j}.$ 

The equations through (\ref{Xul}) contain the derivative matrix $\frac{\partial \lambda_j}{\partial h_k},$ 
which can be computed by differentiating the logarithm of the equations (\ref{ABAE}).
The final result takes the form:
\begin{equation} \label{dldh}
\begin{split}
\frac{\partial \lambda_j}{\partial h_k}=\sum\limits_{s=1}^r \psi_{js}^{-1} \, V_{s k}, \qquad 
V_{s  k}=(\ln B_0)'(\lambda_s-h_k), \qquad k=1,...,m_H, \quad
s=1,..,r,  
\end{split}
\end{equation}
where we exploited the infinite volume correspondence between the holes and the rapidities of the physical particles  
$\{h_j\}_{j=1}^{m_H} \leftrightarrow \{\theta_j\}_{j=1}^n$
and we introduced $\psi,$ a symmetric $r \times r$ matrix with the definition as follows:
\begin{equation} \label{kispsi}
\begin{split}
\psi_{jk}=z'(\lambda_j) \, \delta_{jk}+(\ln E_0)'(\lambda_j-\lambda_k), \qquad j,k=1,..,r,
\end{split}
\end{equation}
with
\begin{equation} \label{kisz}
\begin{split}
z(\lambda)=\sum\limits_{k=1}^{m_H} \ln B_0(\lambda-h_k)-\sum\limits_{k=1}^r \ln E_0(\lambda-\lambda_k), \qquad 
E_0(\lambda)=\frac{B_0(\lambda)}{B_0(-\lambda)}.
\end{split}
\end{equation}
Then  equations (\ref{Qhk}) with $\nu=\ell$ for $X_{\ell,j}^{(h)}$ can be written in the repulsive regime as follows:
\begin{equation} \label{Xhdet1}
\begin{split}
\sum\limits_{k=1}^{m_H} \Phi_{jk} \,X_{\ell,k}^{(h)}=\sinh h_j, \qquad j=1,...,m_H,
\end{split}
\end{equation}
where $\Phi_{jk}$ is the Gaudin-matrix of the physical particles in the state described by the magnonic roots $\{\lambda_j\}_{j=1}^{r}:$
\begin{equation} \label{Phi}
\begin{split}
\Phi_{jk}=\left\{ \begin{array}{l}\ell \cosh h_j+\sum\limits_{s=1 \atop s\neq j}^{m_H} {\tilde G}_{js}, \qquad j=k, \\
-{\tilde G}_{jk}, \qquad \qquad \qquad \qquad \! \! \! j\neq k, 
\end{array} \right.
\end{split}
\end{equation}
\begin{equation} \label{tildeG}
\begin{split}
{\tilde G}_{jk}=G_{jk}+\tfrac{1}{i} \sum\limits_{s,q=1}^{r} V_{sj}\, \psi_{sq}^{-1} \, V_{qk}, \qquad 
j,k=1,..,m_H.
\end{split}
\end{equation}
Here $V_{jk}$ is defined in (\ref{dldh}) and we introduced the short notation: $G_{jk}=G(h_j-h_k).$
Now it is easy to solve (\ref{Xhdet1}) for $X_{\ell,j}^{(h)}:$ 
\begin{equation} \label{Xhres}
\begin{split}
X_{\ell,j}^{(h)}=\sum\limits_{k=1}^{m_H} \Phi_{jk}^{-1} \, \sinh h_k, \qquad j=1,..,m_H.
\end{split}
\end{equation}
Then using (\ref{Xul}) and (\ref{dldh}), the $X$-variables of the complex roots can also be obtained from (\ref{Xhres}):
\begin{equation} \label{Xures1}
\begin{split}
X^{(u)}_{\ell,j}=\sum\limits_{k=1}^{m_H} \sum\limits_{s=1}^r \psi_{js}^{-1} V_{sk} \sum\limits_{k'=1}^{m_H} \Phi_{kk'}^{-1} \, \sinh h_{k'}, \qquad j=1,..,r.
\end{split}
\end{equation}

As for the $d$-type of $X$-variables,  from (\ref{Gd}) we know exactly that the value of each of them is exactly 1. 
Nevertheless, with the help of the large $\ell$ solution of (\ref{Qhk}), this value can be expressed in a more complicated way, as well: 
\begin{equation} \label{Xdhres}
\begin{split}
X_{d,j}^{(h)}=\sum\limits_{k=1}^{m_H} \Phi_{jk}^{-1} \,\ell  \cosh h_k, \qquad j=1,..,m_H,
\end{split}
\end{equation}
\begin{equation} \label{Xdures1}
\begin{split}
X^{(u)}_{d,j}=\sum\limits_{k=1}^{m_H} \sum\limits_{s=1}^r \psi_{js}^{-1} V_{sk} \sum\limits_{k'=1}^{m_H} \Phi_{kk'}^{-1} \, \ell \, \cosh h_{k'}, \qquad j=1,..,r.
\end{split}
\end{equation}
For the derivation of the second expression the following discrete set of equations 
should have been used as well:
\begin{equation} \label{identity}
\begin{split}
\sum\limits_{j=1}^{r} \sum\limits_{k=1}^{m_H} V_{jk} \sum\limits_{q=1}^r \psi_{jq}^{-1} V_{qs}=
\sum\limits_{j=1}^r V_{js}, \qquad s=1,...,m_H,
\end{split}
\end{equation}
which can be derived from the logarithmic derivative of the (\ref{ABAE}).
The point in making the simple to complicated is that in this way the solutions for both subscript $\nu=\ell,d$ 
can be written on equal footing:
\begin{equation} \label{Xnuhres}
\begin{split}
X_{\nu,j}^{(h)}=\sum\limits_{k=1}^{m_H} \Phi_{jk}^{-1} \,  f_{\nu} (h_k), \qquad 
\nu\in\{d,\ell\},\qquad j=1,..,m_H,
\end{split}
\end{equation}
\begin{equation} \label{Xnuures}
\begin{split}
X^{(u)}_{d,j}=\sum\limits_{k=1}^{m_H} \sum\limits_{s=1}^r \psi_{js}^{-1} V_{sk} \sum\limits_{k'=1}^{m_H} \Phi_{kk'}^{-1} \, f_\nu( h_{k'}), \quad \nu\in\{d,\ell\},\qquad j=1,..,r,
\end{split}
\end{equation}
where $f_\nu(\theta)$ is the source term of the linear problem (\ref{Gnu}). It is given in (\ref{fd}) and (\ref{fl}).
Inserting the large volume solutions (\ref{Xnuhres}), (\ref{Xnuures}) into the expectation value formulas: 
(\ref{THrest}), (\ref{J0exp}) and (\ref{J1exp}), one ends up with the large volume solutions as follows:
\begin{equation} \label{THrest8}
\begin{split}
\Theta_{rest}(\ell)_{BY}=\sum\limits_{j,k=1}^{m_H} \left( 
\cosh h_j \, \Phi_{jk}^{-1} \cosh h_k -\sinh h_j \, \Phi_{jk}^{-1} \sinh h_k\right),
\end{split}
\end{equation}
\begin{equation} \label{J08}
\begin{split}
\langle J_0 \rangle_{BY}={\cal M} \sum\limits_{k,s=1}^{m_H} \Phi_{sk}^{-1} \, \cosh h_k \, 
\left( 1-2 \, \sum\limits_{j,q=1}^r \psi_{jq}^{-1} V_{qs}\right),
\end{split}
\end{equation}
\begin{equation} \label{J18}
\begin{split}
\langle J_1 \rangle_{BY}={\cal M} \sum\limits_{k,s=1}^{m_H} \Phi_{sk}^{-1} \, \sinh h_k \, 
\left( 1-2 \, \sum\limits_{j,q=1}^r \psi_{jq}^{-1} V_{qs}\right).
\end{split}
\end{equation}

In the computations we have done so far, it was not necessary to impose the quantization 
equations (\ref{Hkvant}) for the holes. Thus the formulas above can be considered as 
 analytical functions of the $m_H$ pieces of holes (rapidities). Nevertheless, if one would like to 
get the Bethe-Yang limit of the expectation values, formulas (\ref{THrest8}), (\ref{J08}) and (\ref{J18}) 
must be taken at the solutions of Bethe-Yang limit of the quantization equations (\ref{Hkvant}), which takes 
the well-known form:
\begin{equation} \label{BYhole}
\begin{split}
e^{i \ell \sinh \tilde{h}_j} \, \Lambda(\tilde{h}_j|\vec{\tilde{h}})=1, \qquad j=1,..,m_H,
\end{split}
\end{equation}
where $\tilde{h}_j$ denotes the solutions of the Bethe-Yang equations and $\Lambda(\theta|\vec{\tilde{h}})$
denotes that eigenvalue (\ref{eival}) of the soliton transfer matrix (\ref{trans}), which corresponds to the sandwiching state.

%Finally we just emphasize that in all the formulas of this subsection from (\ref{dldh}) to (\ref{J18}) it is assumed, 
%that the Bethe-roots $\{\lambda_j\}_{j=1}^r$ are solutions of the ABA equations (\ref{ABAE}) and so they are 
%continuous functions of the rapidities or holes. Since we are in the infinite volume limit these latter are not 
%subjected to any quantization conditions.

\section{Form-factors in the sine-Gordon theory} \label{sect5}

Having the exact formulas (\ref{THrest8}), (\ref{J08}) and (\ref{J18}), for the Bethe-Yang limit of the expectation values of our operators, 
we would like to check analytically the conjecture of \cite{Palmai13}  
for the Bethe-Yang limit of the diagonal matrix elements of local operators in a non-diagonally 
scattering theory. 
To do so, we need the infinite volume form-factors of the theory. There are several ways to determine these form-factors. 
The earliest construction is written in the seminal work of Smirnov \cite{Smirnov}. Later other constructions 
arose in the literature, like Lukyanov's free-field representation \cite{LUK1,LUK2} and 
the off-shell Bethe-Ansatz based method of \cite{BABK1,BABK2}.
In this section we summarize the axiomatic equations satisfied by the form-factors of 
local operators in an integrable quantum field theory.

Let ${\cal O}(x,t)$ a local operator of the theory. Then its matrix elements between asymptotic multiparticle 
states is given by \cite{Smirnov}:
\begin{equation} \label{FFxt}
\begin{split}
 {}^{(in)}\langle \gamma_1,b_1;...;\gamma_m,b_m|{\cal O}(x,t)|\beta_1,a_1;...;\beta_n,a_n \rangle^{(in)}=
e^{i \, t \, (E_\gamma-E_\beta)-i\, x\, (P_\gamma-P_\beta) }\,\times \\
F^{\cal O}_{\bar{b}_m...\bar{b}_1\, a_1...a_n}(\gamma_m+i \, \pi-i \epsilon_m,...,\gamma_1+i \, \pi-i\, \epsilon_1,
\beta_1,...,\beta_n)+\text{Dirac-delta terms},
\end{split}
\end{equation}
where the form-factors of the local operator ${\cal O}$ is denoted by $F^{\cal O},$ 
$\epsilon_j$s are positive infinitesimal numbers and the orderings 
$\beta_n<...<\beta_2<\beta_1,$  $\gamma_1<\gamma_2<...<\gamma_m$
 are meant in the $in$ states.
 The Latin and Greek letters denote the polarizations and rapidities of the sandwiching multisoliton states, 
respectively. Thus $a_j,b_j\in\{\pm\},$ and $E_\gamma, E_\beta, P_\gamma, P_\beta$ denote the energies and the 
momenta of the corresponding states: 
\begin{equation} \label{EPgb}
\begin{split}
E_\gamma&=\sum\limits_{j=1}^m {\cal M} \cosh \gamma_j, \qquad E_\beta=\sum\limits_{j=1}^n {\cal M} \cosh \beta_j, \\
P_\gamma&=\sum\limits_{j=1}^m {\cal M} \sinh \gamma_j, \qquad P_\beta=\sum\limits_{j=1}^n {\cal M} \sinh \beta_j.
\end{split}
\end{equation}
We choose the normalization  for the scalar product of states in infinite volume as follows:
\begin{equation} \label{scalar8}
\begin{split}
 {}^{(in)}\langle \gamma_1,b_1;...;\gamma_n,b_n|\beta_1,a_1;...;\beta_n,a_n \rangle^{(in)}=(2 \pi)^n \, \prod\limits_{j=1}^n 
\delta_{b_j a_j} \, \delta(\gamma_j-\beta_j). 
\end{split}
\end{equation}
In this convention the form-factors $F^{{\cal O}}$ of the operator ${\cal O}(x,t)$ satisfy the the following 
axioms \cite{Smirnov}:
\newline{I. Lorentz-invariance:}
\begin{equation} \label{ax1}
\begin{split}
F^{\cal O}_{a_1...a_n}(\theta_1+\theta,...,\theta_n+\theta)=e^{s({\cal O}) \theta} \, F^{\cal O}_{a_1...a_n}(\theta_1,...,\theta_n),
\end{split}
\end{equation}
where $s({\cal O})$ is the Lorentz-spin of ${\cal O}.$
\newline{II. Exchange:}
\begin{equation} \label{ax2}
\begin{split}
F^{\cal O}_{...a_j a_{j+1}...}(...,\theta_j,\theta_{j+1},...)={\cal S}_{a_j a_{j+1}}^{b_{j+1} b_j}(\theta_j-\theta_{j+1}) \, F^{\cal O}_{...b_{j} b_{j+1}...}(...,\theta_{j+1},\theta_{j},...),
\end{split}
\end{equation}
\newline{III. Cyclic permutation:}
\begin{equation} \label{ax3}
\begin{split}
F^{\cal O}_{a_1 a_2...a_n}(\theta_1+2 \pi \, i,...,\theta_n)=e^{2 \pi \, i \,\omega({\cal O})}\, 
F^{\cal O}_{a_2...a_n a_1}(\theta_2,...,\theta_n,\theta_1),
\end{split}
\end{equation}
where $\omega({\cal O})$ denotes the mutual locality index between 
${\cal O}$ and the asymptotic field which creates the solitons.
\newline{IV. Kinematical singularity:}
\begin{equation} \label{ax4}
\begin{split}
F^{\cal O}_{ab u_1...u_n}(\theta+i\, \pi+\epsilon,\theta, \theta_1,...,\theta_n)\stackrel{\epsilon \to 0}{\simeq} 
\frac{i}{\epsilon} \bigg\{ 
C_{ab} \, F^{\cal O}_{u_1...u_n}( \theta_1,...,\theta_n)-\\
 e^{2 \pi \, i \,\omega({\cal O})} \! 
\sum\limits_{v_1,..,v_n=\pm}{\cal T}_{b}^{\bar{a}}(\theta|\theta_1,..,\theta_n)_{u_1...u_n}^{v_1...v_n} \,
F^{\cal O}_{v_1...v_n}(\theta_1,...,\theta_n)
\bigg\},
\end{split}
\end{equation}
where ${\cal T}$ denotes the soliton monodromy matrix defined in (\ref{monodr}), and $C_{ab}$ is the 
charge conjugation matrix (\ref{Cmatr}). 
In this paper we will focus on the repulsive regime of the sine-Gordon theory, where there are no 
soliton-antisoliton bound states in the spectrum. 
This is why we skipped to present the dynamical singularity axiom, which relates the form-factors of bound states to 
those of its constituents. 
 
We just remark that for the operators of our interest 
%$\Theta$ and $J_\mu$ 
the mutual locality index is zero: 
$\omega(\Theta)=\omega(J_\mu)=0.$

\section{The P\'almai-Tak\'acs conjecture} \label{PTsejt}

In this section we summarize the conjecture of P\'almai and Tak\'acs \cite{Palmai13} 
for the Bethe-Yang limit of the diagonal matrix elements of local operators in a non-diagonally 
scattering theory. By Bethe-Yang limit we mean those terms of the large volume expansion which 
are polynomials in the inverse of the volume. Namely, the exponentially small in volume corrections are 
neglected from the exact result. 

In finite volume the particle rapidities become quantized. The quantization equations 
which account for the polynomial in the inverse of the large volume correctons are called 
the Bethe-Yang equations. In a non-diagonally scattering theory at large volume, the amplitudes 
describing the "color part" of the multisoliton eigenstates of the Hamiltonian are 
eigenvectors of the soliton transfer matrix (\ref{trans}). 
They form a complete normalized basis on the space of "color" degrees of freedom of the wave function. 
On an $n$-particle state it can be formulated as follows: 
\begin{equation} \label{bazis}
\begin{split}
\tau(\theta|\vec{\theta})_{a_1...a_n}^{b_1...b_n} \, \Psi^{(t)}(\vec{\theta})_{b_1...b_n}\,
=\Lambda^{(t)}(\theta|\vec{\theta}) \,
\Psi^{(t)}(\vec{\theta})_{a_1...a_n} \qquad t=1,..,2^n,
\end{split}
\end{equation}
\begin{equation} \label{ortog}
\begin{split}
\sum\limits_{a_1,...,a_n=\pm}\Psi^{(s)}(\vec{\theta})_{a_1...a_n} \Psi^{(t)*}(\vec{\theta})_{a_1...a_n}&=\delta_{st}, \\
\sum\limits_{t=1}^{2^n}\Psi^{(t)}(\vec{\theta})_{a_1...a_n} \Psi^{(t)*}(\vec{\theta})_{b_1...b_n}&=
\prod\limits_{j=1}^n \delta_{a_j b_j},
\end{split}
\end{equation}
where for short we use the notation $\vec{\theta}=\{\theta_1,...,\theta_n\}$ and $\Lambda^{(t)}(\theta|\vec{\theta})$ 
stands for the eigenvalue of the $t$th eigenstate. This eigenvalue can be obtained by the Algebraic Bethe method \cite{FST79} summarized in appendix \ref{appB}. Its expression in terms of the magnonic Bethe-roots 
is given by (\ref{eival}). 

In the language of the soliton transfer matrix (\ref{trans}), 
for an $n$-particle state
the Bethe-Yang quantization equations take the form:
\begin{equation} \label{BY1}
\begin{split}
e^{i \ell \sinh \tilde{\theta}_j} \, \Lambda^{(t)}(\tilde{\theta_j}|\vec{\tilde{\theta}})=1, \qquad j=1,..,n,
\end{split}
\end{equation}
where we introduced the notation that the set $\{\theta_j\}_{j=1}^n,$ means an arbitrary unquantized set of rapidities, while the set with tilde $\{\tilde{\theta}_j\}_{j=1}^n,$  denotes the set of rapidities satisfying the Bethe-Yang equations (\ref{BY1}).
%if the set of rapidities are generic then they are denoted by 
%$\{\theta_j\}_{j=1}^n,$  and when they satisfy the 
%Bethe-Yang equations, then they are written with a tilde $\{\theta_j\}_{j=1}^n,$.
It is more common to rephrase (\ref{BY1}) in its logarithmic form. To do so, first one has to define the 
function:
\begin{equation} \label{Qt}
\begin{split}
Q^{(t)}(\theta|\theta_1,..,\theta_n)=\ell \sinh \theta+\tfrac{1}{i} \Lambda^{(t)}({\theta}|\{\theta_1,..,\theta_n\}).
\end{split}
\end{equation}
Then the logarithmic form of the Bethe-Yang equations take the form:
\begin{equation} \label{BYln}
\begin{split}
Q^{(t)}(\tilde{\theta}_j|\tilde{\theta}_1,..,\tilde{\theta}_n)=2 \pi \, I^{(t)}_j, \qquad j=1,...,n, \quad 
t=1,..,2^n,
\end{split}
\end{equation}
where $I_j^{(t)} \in{\mathbb Z}$ are the quantum numbers characterizing the individual rapidities of the $t\,$th 
eigenstate. The function $Q^{(t)}$ in (\ref{Qt}) allows one to define the density of states in the 
$t\,$th eigenstate $\Psi^{(t)}$ by the Jacobi determinant as follows: %of the left hand side of (\ref{BYln}):
%Bethe-Yang equations:
\begin{equation} \label{Rhot}
\begin{split}
\rho^{(t)}(\theta_1,...,\theta_n)=\text{det}\left\{ \frac{\partial Q^{(t)}(\theta_j|\theta_1,..,\theta_n)}{\partial \theta_k} \right\}_{j,k=1,..,n} \qquad t=1,...,2^n.
\end{split}
\end{equation}
With the help of the basis (\ref{bazis}), (\ref{ortog}) one can define form-factors being polarized 
with respect to the eigenvectors (\ref{bazis}). In \cite{Palmai13} this quantity was defined by the formula as follows:
\begin{equation} \label{polform}
\begin{split}
&F^{{\cal O}}_{(s,t)}(\theta_m',...,\theta'_1|\theta_1,...,\theta_n)=
\sum\limits_{b_1,..,b_m=\pm} \, \sum\limits_{a_1,..,a_n=\pm} \Psi_{b_1...b_m}^{(s)*}(\theta'_1,..,\theta'_m) \times \\
&F^{{\cal O}}_{\bar{b}_m...\bar{b}_1 a_1...a_n}(\theta'_m+i \, \pi,...,\theta'_1+i \, \pi,\theta_1,...,\theta_n) \,
\Psi_{a_1...a_n}^{(t)}(\theta_1,..,\theta_n), \quad s=1,...,2^m, \quad t=1,..,2^n.
\end{split}
\end{equation}
Based on our computations described in the forthcoming sections, we suggest the following slightly modified 
definition:
\begin{equation} \label{polformen}
\begin{split}
&F^{{\cal O}}_{(s,t)}(\theta_m',...,\theta'_1|\theta_1,...,\theta_n)=
\sum\limits_{b_1,..,b_m=\pm} \, \sum\limits_{a_1,..,a_n=\pm} \Psi_{b_1...b_m}^{(s)}(\theta'_1,..,\theta'_m) \times \\
&F^{{\cal O}}_{\bar{b}_m...\bar{b}_1 a_1...a_n}(\theta'_m+i \, \pi,...,\theta'_1+i \, \pi,\theta_1,...,\theta_n) \,
\Psi_{a_1...a_n}^{(t)*}(\theta_1,..,\theta_n), \quad s=1,...,2^m, \quad t=1,..,2^n.
\end{split}
\end{equation}
The only difference between the two definitions is a $\Psi \to \Psi^*$ exchange.
Or equivalently, as a consequence of the hermiticity property of the soliton transfer matrix (\ref{tauH}),
one can maintain the original form (\ref{polform}) for the definition 
of polarized form-factors, but in this case the vector $\Psi$ should not be considered as 
a right eigenvector of the $\tau(\theta|\vec{\theta}),$ but it should be considered as 
a left eigenvector of the soliton transfer matrix (\ref{trans}).
In the rest of the paper we will keep the form of the original definition (\ref{polform}), but we will 
consider $\Psi$ as a left eigenvector of $\tau(\theta|\vec{\theta}).$

Now we are in the position to formulate the conjecture of P\'almai and Tak\'acs for the expectation values 
of local operators in non-diagonally scattering theories.
Let
\begin{equation} \label{state}
\begin{split}
|\bar{\theta}_1,..,\bar{\theta}_n \rangle^{(s)}_L, 
\end{split}
\end{equation}
that eigenstate of the  Hamiltonian defined in finite volume $L$ of the system, which 
is described by the eigenstate $\Psi^{(s)}$ of the soliton transfer matrix in the large volume limit.
Here $\{\bar{\theta}\}_{j=1}^n$ denote the exact finite volume rapidities, which become $\{\tilde{\theta}\}_{j=1}^n$ 
if the exponentially small in volume corrections are neglected in the large volume limit. 
Then the conjecture of \cite{Palmai13} states that the finite volume expectation value of a local operator in an 
$n$-particle state can be written as follows: 
%that the Bethe-Yang limit of an $n$-particle 
%finite volume expectation value of a local operator of the theory can be computed by taking a function  
%as follows:
\begin{equation} \label{PT1}
\begin{split}
{}^{(s)}\langle \bar{\theta}_1,..,\bar{\theta}_n |{\cal O}(0,0)|\bar{\theta}_1,..,\bar{\theta}_n \rangle^{(s)}_L=
{\sc F}_n^{{\cal O},(s)}(\tilde{\theta}_1,...,\tilde{\theta}_n)+O(e^{-\ell}), \qquad s=1,..,2^n,
\end{split}
\end{equation}
where according to the conjecture, the function ${\sc F}_n^{{\cal O},(s)}$, 
which should be taken at the positions of the roots of the Bethe-Yang equations (\ref{BYln})  
can be constructed from the infinite volume form-factors of the theory by the following 
formula:
%way. 
%take the form as follows:
\begin{equation} \label{PTfv}
\begin{split}
{\sc F}_n^{{\cal O},(s)}({\theta}_1,...,{\theta}_n)=\frac{1}{\rho^{(s)}_n(1,...,n)} \sum\limits_{A\subset\{1,..,n\}} 
\sum\limits_{q,t} |C_{qt}^{(s)}\left(\{\theta_k\}|A\right)|^2 \,
{\sc F}_{2|A|,symm}^{{\cal O},(q)}(A) \, \rho^{(t)}_{|\bar{A}|}(\bar{A}),  
\end{split}
\end{equation}
where the first sum runs for all bipartite partitions of the set of indexes $A^{(n)}=\{1,2,...,n\}.$ Namely,  
 $A \cup \bar{A}=A^{(n)}.$ 
The number of elements of $A$ is denoted by $|A|,$ then the number of elements of $\bar{A}$ is $|\bar{A}|=n-|A|.$ 
In the sequel we denote the elements of the sets $A$ and $\bar{A}$ as follows{\footnote{Though it was not specified clearly in \cite{Palmai13}, we assume the following orderings within these sets: 
$A_i<A_j$ and $\bar{A}_i<\bar{A}_j$ if $i<j.$ }}:
\begin{equation} \label{AAbar}
\begin{split}
A=\{A_1,A_2,...,A_{|A|}\}, \\
\bar{A}=\{\bar{A}_1,\bar{A}_2,...,\bar{A}_{|\bar{A}|}\}.
\end{split}
\end{equation}
The second sum in (\ref{PTfv}) runs for all decompositions of the $n$-particle color wave function with respect to 
the normalized eigenvectors{\footnote{Normalized eigenvectors mean that they fullfill the conditions (\ref{bazis}) and (\ref{ortog}).}} of the transfer matrices acting only on the index sets $A$ and $\bar{A}:$ 
\begin{equation} \label{decomp}
\begin{split}
\Psi^{(t)}_{a_1...a_n}(\theta_1,...,\theta_n)\!\!=\!\!\sum\limits_{q=1}^{2^{|A|}} \sum\limits_{s=1}^{2^{|\bar{A}|}} 
C_{qs}^{(t)}(\{\theta_k\}|A) \, 
\Psi^{(q)}_{a_{A_1}...a_{A_{|A|}}}\!(\theta_{A_1},...,\theta_{A_{|A|}}) \,
\Psi^{(s)}_{a_{\bar{A}_1}...a_{\bar{A}_{|\bar{A}|}}}\!(\theta_{\bar{A}_1},...,\theta_{\bar{A}_{|\bar{A}|}}),
\end{split}
\end{equation}
where as a consequence of (\ref{ortog}) the branching coefficients $C_{qs}^{(t)}(\{\theta_k\}|A)$ satisfy the 
normalization condition:
\begin{equation} \label{branchnorm}
\begin{split}
\sum\limits_{q,s}|C_{qs}^{(t)}(\{\theta_k\}|A)|^2=1.
\end{split}
\end{equation}
Here we note, that the earlier discussed $\Psi \to \Psi^*$ exchange in the formulation of the problem, doesnot 
cause problem in the determination of these branching coefficients, since it corresponds to a simple 
complex conjugation. This is irrelevant from the conjecture's point of view, since the final formula depends 
only on the absolute value square of these branching coefficients.

Now we have two further missing definitions in (\ref{PTfv}). 
In accordance with \cite{Palmai13} we introduced some more compact notations for the densities:
\begin{equation} \label{densrov}
\begin{split}
\rho^{(s)}_{n}(1,2,..,n)=\rho^{(s)}(\theta_1,\theta_2,..,\theta_n), \\
\rho^{(t)}_{|\bar{A}|}(\bar{A})=\rho^{(t)}(\theta_{\bar{A}_1},\theta_{\bar{A}_2},..,\theta_{\bar{A}_{|\bar{A}|}}),
\end{split}
\end{equation}
with $\rho^{(s)}$ functions in the right hand side given by (\ref{Rhot}).
The last so far undefined object in (\ref{PTfv}) is ${\sc F}_{2|A|,symm}^{{\cal O},(q)}(A).$  
It is defined as the uniform diagonal limit of a $(q,q)$ polarized form-factor of ${\cal O},$ such that 
the indexes of the rapidities of the sandwiching states run the set $A:$ 
\begin{equation} \label{szimFA}
\begin{split}
{\sc F}_{2|A|,symm}^{{\cal O},(q)}(A)=\lim\limits_{\epsilon \to 0} F^{{\cal O}}_{(q,q)}(\theta_{A_{|A|}}+\epsilon,...,\theta_{A_1}+\epsilon|\theta_{A_1},...,\theta_{A_{|A|}}),
\end{split}
\end{equation}
with $F^{{\cal O}}_{(q,q)}$ defined in (\ref{polform}). 
In analogy with the terminology in purely elastic scattering theories  the function 
${\sc F}_{2n,symm}^{{\cal O},(q)}$ is called the the $2n$-particle $q$-polarized symmetric diagonal form-factor of the operator ${\cal O}.$

For the operators $\Theta$ and $J_\mu$ the functions ${\sc F}_n^{{\cal O},(s)}(\theta_1,...,\theta_n)$ were computed 
in the previous sections. Their form taken at the positions of the holes $\{h_j\}_{j=1}^{m_H}$
are given by the formulas (\ref{THrest8}), (\ref{J08}) and (\ref{J18}). In the rest of the paper we will 
compare these formulas with the conjecture (\ref{PT1}), (\ref{PTfv}) applied to the operators $\Theta$ and $J_\mu.$ 
In the forthcoming sections we will do the comparison upto 3-particle states. 
The only missing piece to this comparison is the knowledge of the symmetric diagonal form-factors. 
Thus our next task is to compute them upto the required particle numbers.

\section{Symmetric diagonal form-factors for $\Theta$ and $J_\mu$} \label{sect7}

Both the trace of the stress energy tensor and the $U(1)$ current  are related to some 
conserved quantities of the theory. In purely elastic scattering theories the symmetric diagonal 
from-factors of such operators can be computed in a simple way \cite{LM99,Mussbook}. 
The key point in the computation is that by exploiting of the corresponding conservation law,  
 it is not necessary to find the explicit solutions of the axioms (\ref{ax1})-(\ref{ax4}).
%The sought symmetric diagonal form-factors can be obtained by appropriate use the axioms 
%(\ref{ax2})-(\ref{ax4}). 
%without finding the explicit solutions of the axioms (\ref{ax1})-(\ref{ax4}). %such that 
%One should only exploit the conservation law corresponding to the operator under consideration and 
%use the axioms (\ref{ax2})-(\ref{ax4}) appropriately.
%exchange (\ref{ax2}) and kinematical singularity (\ref{ax4}) axioms appropriately.
% by exploiting the 
%related conservation law and by applying the the cyclic (\ref{ax3}) and the kinematical 
%singularity (\ref{ax4}) axioms. 

In this paper we use the same method to compute the symmetric diagonal 
form-factors upto 3-particle states. It turns out that this simple method allows one to compute 
the symmetric diagonal form-factors for any number of particles in the pure soliton sector, but 
for soliton-antisoliton mixed states it works only upto 3-particle states. For higher number of 
particles the explicit solution of the axioms (\ref{ax1})-(\ref{ax4}) is required.

The form-factor axioms allow one to compute form-factors of higher number of particles from 
those of lower number of particles. Thus, we should start with the computation of 
the 2-particle symmetric diagonal form-factors of the operators of our interest.

\subsection{2-particle symmetric diagonal form-factors}

{\bf The case of $\Theta$:}
\newline

The stress energy tensor $T_{\mu \nu}$ is a conserved quantity, which implies that it can be 
written as appropriate derivative of some Lorentz scalar field $\phi:$ 
\begin{equation} \label{Tmn}
\begin{split}
T_{\mu\nu}=(\partial_\mu \partial_\nu-\eta_{\mu\nu} \partial^{\tau} \partial_{\tau})\, \phi,
\end{split}
\end{equation}
where $\eta_{\mu\nu}$ is the 2-dimensional Minkowski metric. In this representation the 
trace of the stress energy tensor take the form:
\begin{equation} \label{THfi}
\begin{split}
\Theta=T_{\, \mu}^{\mu}=(\partial^2_1-\partial^2_0)\, \phi.
\end{split}
\end{equation}
%From the analysis of Wightman-functions, 
It can be shown \cite{BMN97}, that the Lorentz scalar field $\phi$ is not a 
local quantum field. Consequently, not all of its form-factors satisfy the axioms (\ref{ax1})-(\ref{ax4}). 
To be more precise from the representation (\ref{Tmn}), it can be shown, that the 3- or more particle 
form-factors of $\phi$ satisfy the axioms (\ref{ax1})-(\ref{ax4}), but the 2-particle ones 
become more singular, than it is expected from (\ref{ax4}). (See (\ref{TH2A}).)

Using the space-time structure of the form-factors (\ref{FFxt}), the form-factors of $\Theta$ 
being close to the diagonal limit can be written as follows:   
\begin{equation} \label{THff1}
\begin{split}
F^{\Theta}(\hat{\theta}_n,...,\hat{\theta}_1,\theta_1,..,\theta_n)=-{\cal M}^2 
\left[ \sum\limits_{j,k=1}^n 
\epsilon_j \epsilon_k \cosh(\theta_j-\theta_k)+O(\epsilon^3) \right]  F^{\phi}(\hat{\theta}_n,...,\hat{\theta}_1,\theta_1,..,\theta_n), 
\end{split}
\end{equation}
where $F^{\phi}$ denotes the form-factors of the  scalar operator
%{\footnote{They satisfy the usual form-factor axioms (\ref{ax2})-(\ref{ax4}).}}
 $\phi$ in (\ref{THfi}) and 
we introduced the notation $\hat{\theta}_j=\theta_j+i\, \pi+\epsilon_j$ for all values of the index $j.$
In (\ref{THff1}) the symbol $O(\epsilon^3)$ means at least cubic in $\epsilon$ terms when the uniform 
$\epsilon_1=...=\epsilon_n=\epsilon \to 0$ 
%$\lim\limits_{\epsilon \to 0}$ 
limit is taken. For the sake of simplicity we did not write out the subscripts of the form-factors.
 
The basic idea of computing the 2-particle form-factors near their diagonal limit is that the 
near diagonal matrix elements of the Hamiltonian ${\cal H}=\int dx \, T_{00}$ can be computed in two different ways. 
First, it can be computed directly by acting with ${\cal H}$ on the eigenstates:
\begin{equation} \label{HszamTH}
\begin{split}
\langle \theta+\epsilon,a|{\cal H}|\theta,b\rangle=2\, \pi \, {\cal M} \, \cosh \theta \, 
\delta_{ab} \, \delta(\epsilon), \qquad a,b \in\{\pm\}.
\end{split}
\end{equation}
Second, it can be computed by using the representation $\int dx \, T_{00}$ for the Hamiltonian, and 
 the matrix element is computed by integrating the space-time dependence of the corresponding form-factor:
\begin{equation} \label{FszamTH1}
\begin{split}
\langle \theta\!+\!\epsilon,a|{\cal H}|\theta,b\rangle\!=\!\!\!\int \!\!dx \langle \theta\!+\!\epsilon,a|T_{00}(x,0)|\theta,b\rangle\!=\!
- 2\pi (\epsilon^2\!+\!O(\epsilon^3)) {\cal M} \cosh \theta \, \delta(\epsilon) \, 
F^{\phi}_{\bar{a}b}(\theta\!+\!i\, \pi\!+\!\epsilon,\theta),
\end{split}
\end{equation}
where we used (\ref{FFxt}) and (\ref{THff1}). Comparing the results (\ref{HszamTH}) and (\ref{FszamTH1}) 
of the two different computations allows one to compute the near diagonal limit of the scalarized form-factor:
\begin{equation} \label{TH2A}
\begin{split}
F^{\phi}_{ab}(\theta\!+\!i\, \pi\!+\!\epsilon,\theta)=-\frac{1}{\epsilon^2} \delta_{\bar{a}b}+O(\tfrac{1}{\epsilon})
, \qquad a,b\in\{\pm\}.
\end{split}
\end{equation}
Combining (\ref{TH2A}) with (\ref{THff1}), the symmetric diagonal 2-particle form-factor of $\Theta$ can also be determined:  
\begin{equation} \label{TH2F}
\begin{split}
F^{\Theta}_{ab}(\theta\!+\!i\, \pi,\theta)={\cal M}^2 \delta_{\bar{a}b}, \qquad a,b\in\{\pm\}.
\end{split}
\end{equation} 
The matrix structure $\delta_{\bar{a}b}$ in (\ref{TH2A}) and (\ref{TH2F}) accounts for the charge conjugation 
invariance of the operator $\Theta.$ 
\begin{equation}\nonumber
\end{equation}

{\bf The $J_\mu$ case:}
\newline

The computation of the near diagonal limit of the 2-particle form-factors of the $U(1)$ current goes 
analogously to that of the operator $\Theta.$ The conservation law for the current implies the following 
representation:
\begin{equation} \label{Jmuff1}
\begin{split}
J_0=-i\, \partial_1 \psi, \qquad J_1=-i \, \partial_0 \psi,
\end{split}
\end{equation}
with $\psi$ being a (non-local) Lorentz scalar operator. The form-factors of $\psi$ satisfy the same 
form-factor axioms as the form-factors of $\phi$ do. 
This together with (\ref{FFxt}) 
gives the following representation for the near diagonal 
form-factors:  
\begin{equation} \label{Jmuff2}
\begin{split}
F^{J_0}(\hat{\theta}_n,...,\hat{\theta}_1,\theta_1,..,\theta_n)=
-{\cal M} \left[ \sum\limits_{j=1}^n \cosh \theta_j \, \epsilon_j+O(\epsilon^2)\right]  
F^{\psi}(\hat{\theta}_n,...,\hat{\theta}_1,\theta_1,..,\theta_n), \\
F^{J_1}(\hat{\theta}_n,...,\hat{\theta}_1,\theta_1,..,\theta_n)=
{\cal M} \left[ \sum\limits_{j=1}^n \sinh \theta_j \, \epsilon_j +O(\epsilon^2)\right]  
F^{\psi}(\hat{\theta}_n,...,\hat{\theta}_1,\theta_1,..,\theta_n).
\end{split}
\end{equation}
The topological charge $Q\!=\!\int \! dx J_0$ acts on one-particle states as follows:
\begin{equation} \label{Qact1}
\begin{split}
Q|\theta,a\rangle\!=\!\sum\limits_{b=\pm} \! q_{ab}|\theta,b\rangle, \quad a=\pm,\quad 
\text{with} \quad
q_{++}\!=\!1, \quad q_{--}\!=\!-1, \quad q_{+-}\!=\!q_{-+}\!=\!0.
\end{split}
\end{equation}
Using this action and the scalar product formula (\ref{scalar8}) 
the near diagonal limit of the matrix elements of the charge can be computed directly:
\begin{equation} \label{Qdirect}
\begin{split}
\langle\theta+\epsilon,a|Q|\theta,b\rangle=2\pi\, q_{ba} \, \delta(\epsilon).
\end{split}
\end{equation}
On the other hand this matrix element can also be computed by integrating the space-time dependence of the form-factor of $J_0:$ 
%in the representation (\ref{Jmuff2}):
\begin{equation} \label{Qff}
\begin{split}
\langle\theta+\epsilon,a|Q|\theta,b\rangle\!=\!\int \!\! dx  \, \langle\theta+\epsilon,a|J_0|\theta,b\rangle=
 \frac{2\pi}{{\cal M}\, \cosh \theta}\, \delta(\epsilon) \, F^{J_0}_{\bar{a}b}(\theta+i  \pi+\epsilon,\theta).
\end{split}
\end{equation}
Comparing the results of the two different computations one obtains the symmetric diagonal limit 
of the 2-particle form-factors of $J_0:$
\begin{equation} \label{J0ffsy2}
\begin{split}
F^{J_0}_{ab}(\theta+i \, \pi,\theta)={\cal M} \cosh \theta \, q_{b\bar{a}},
\end{split}
\end{equation}
with $q_{ab}$ given in (\ref{Qact1}). Formula (\ref{J0ffsy2}) and (\ref{Jmuff2}) allows one to compute the 
near diagonal limit of the 2-particle scalarized form-factor $F^{\psi}_{ab}:$
\begin{equation} \label{Fpsi2}
\begin{split}
F^{\psi}_{ab}(\theta+i \, \pi+\epsilon,\theta)=\frac{1}{\epsilon} \, q_{\bar{b}a}+O(\epsilon),
\end{split}
\end{equation}
which together with (\ref{Jmuff2}) gives the 2-particle symmetric diagonal form-factor of $J_1$ as well:
\begin{equation} \label{J1ffsy2}
\begin{split}
F^{J_1}_{ab}(\theta+i \, \pi,\theta)={\cal M} \sinh \theta \, q_{\bar{b}a}.
\end{split}
\end{equation}
We note that the pure comparison of (\ref{J0ffsy2}) and (\ref{Jmuff2}) would imply that in (\ref{Fpsi2}) 
there are $O(1)$ terms in $\epsilon$ as well. 
However, the Lorentz invariance (\ref{ax1}), the cyclic axiom (\ref{ax3}) and the charge conjugation negativity 
of $J_\mu,$ implies that the form-factor $F^{\psi}_{ab}(\theta+i \, \pi+\epsilon,\theta)$ is independent 
of $\theta$ and is an odd function of $\epsilon.$  This oddity forbids the appearance of constant in $\epsilon$ terms 
in the right hand side of (\ref{Fpsi2}).

%However, if one takes into account the Lorentz invariance, which 
%implies that our form-factors depend on only the differences of the rapidities and applies the cyclic axiom 
%to $F^{\psi}_{ab}(\theta+i \, \pi+\epsilon,\theta),$ then the charge conjugation negativity of $J_\mu$ 
%tells us that the form-factor $F^{\psi}_{ab}(\theta+i \, \pi+\epsilon,\theta)$ is independent of $\theta$ and is an 
%odd function of $\epsilon.$  This oddity forbids the appearance of constant in $\epsilon$ terms in the right hand 
%side of (\ref{Fpsi2}). 

\subsection{4-particle symmetric diagonal form-factors}

The next step in solving the form-factor axioms (\ref{ax1})-(\ref{ax4}) in the near diagonal limit 
is the determination of the 4-particle form-factors. To obtain them we need to determine the 
singular-parts of the near diagonal 4-particle  form-factors of the scalar fields $\phi$ and $\psi$ of (\ref{Tmn}) and (\ref{Jmuff1}).

To analyse the near diagonal limit of 4-particle form-factors, the following two useful  formulas can be derived from the appropriate 
combination of the axioms (\ref{ax2})-(\ref{ax4}): 
%\begin{equation} \label{AXe1}
%\begin{split}
%F_{a_2 a_1 b_1 b_2}(\hat{\theta}_2,\hat{\theta}_1,\theta_1,\theta_2)\!=\!\frac{i}{\epsilon_1} \!
%\left\{ C_{a_1 b_1}  F_{a_2 b_2}(\hat{\theta}_2,\theta_2)\!-\!{\cal T}^{\bar{a}_1}_{b_1}(\theta_1|\theta_2,\theta_2\!-\!i \pi\!+\!\epsilon_2)_{b_2 a_2}^{v_1 v_2} 
%\, F_{v_2 v_1}(\hat{\theta}_2,\theta_2)
%\right\},
%\end{split}
%\end{equation}
\begin{equation} \label{AXe1}
\begin{split}
F_{a_2 a_1 b_1 b_2}(\hat{\theta}_2,\hat{\theta}_1,\theta_1,\theta_2)\!=\!\frac{i}{\epsilon_1} \!
\left\{ C_{a_1 b_1}  F_{a_2 b_2}(\hat{\theta}_2,\theta_2)\!-\!{\cal T}^{\bar{a}_1}_{b_1}(\theta_1|\theta_2,\tilde{\theta}'_2)_{b_2 a_2}^{v_1 v_2} 
\, F_{v_2 v_1}(\hat{\theta}_2,\theta_2)
\right\}+O(1)_{\epsilon_1},
\end{split}
\end{equation}
\begin{equation} \label{AXe2}
\begin{split}
F_{a_2 a_1 b_1 b_2}(\hat{\theta}_2,\hat{\theta}_1,\theta_1,\theta_2)\!=\!-\frac{i}{\epsilon_2} \!
\left\{ C_{b_2 a_2}  F_{a_1 b_1}(\hat{\theta}_1,\theta_1)\!-\!{\cal T}^{\bar{b}_2}_{a_2}(\hat{\theta}_2|\hat{\theta}_1,{\theta}_1)_{a_1 b_1}^{v_1 v_2} 
\, F_{v_1 v_2}(\hat{\theta}_2,\theta_2)
\right\}+O(1)_{\epsilon_2},
\end{split}
\end{equation}
where we introduced the short notation   $\tilde{\theta}'_{j}=\theta_j\!-\!i \pi\!+\!\epsilon_j$ for any value of the index $j.$ 
The symbol $O(1)_{\epsilon_1}$ denotes terms which are of order one in $\epsilon_1.$

The application of formulas (\ref{AXe1}) and (\ref{AXe2}) to the 4-particle form factors of the scalar field 
$\phi,$ one obtains the result as follows: 
%For any subscript settings introduce the notation:
\begin{equation} \label{aphi}
F^{\phi}_{\alpha \beta \gamma \delta}(\hat{\theta}_2,\hat{\theta}_1,\theta_1,\theta_2)=
\frac{1}{\epsilon_1 \epsilon_2} a^\phi_{\alpha \beta \gamma \delta}(\theta_1,\theta_2)+O(\tfrac{1}{\epsilon}), 
\qquad \alpha,\beta,\gamma,\delta=\pm,
\end{equation}
where the nonzero elements of the tensor $a^\phi(\theta_1,\theta_2)$ are as follows:
\begin{equation} \label{ATHmmpp}
\begin{split}
a^{\phi}_{--++}(\theta_1,\theta_2)=a^{\phi}_{++--}(\theta_1,\theta_2)=
- G(\theta_1-\theta_2),
\end{split}
\end{equation}
\begin{equation} \label{ATHpmpm}
\begin{split}
a^{\phi}_{+-+-}(\theta_1,\theta_2)=a^{\phi}_{-+-+}(\theta_1,\theta_2)=
- \varphi(\theta_1-\theta_2),
\end{split}
\end{equation}
\begin{equation} \label{ATHpmmp}
\begin{split}
a^{\phi}_{+--+}(\theta_1,\theta_2)=a^{\phi}_{-++-}(\theta_1,\theta_2)=
- \Omega(\theta_1-\theta_2).
\end{split}
\end{equation}
The functions $\varphi$ and $\Omega$ are given by the formulas:
\begin{equation} \label{kernelsTH2}
\begin{split}
%\sigma(\theta)&=\frac{1}{2 \pi i} (\log S_0)'(\theta)=\frac{1}{2 \pi } G(\theta), \\
\varphi(\theta)&=-i \big(C_0(\theta)\, B_0'(-\theta)+B_0(\theta)\, C_0'(-\theta)\big), \\
\Omega(\theta)&=-i \big(C_0(\theta)\, C_0'(-\theta)+B_0(\theta)\, B_0'(-\theta)\big)+G(\theta), 
\end{split}
\end{equation}
where $G,$ $B_0$ and $C_0$ are defined in (\ref{G}), (\ref{B0}) and (\ref{C0}) respectively.
As a consequence of the unitarity of the S-matrix (\ref{Unit}), all the functions 
of (\ref{kernelsTH2}) are even in $\theta.$
Inserting (\ref{aphi}) with (\ref{ATHmmpp}), (\ref{ATHpmpm}) and (\ref{ATHpmmp}) into (\ref{THff1}) and taking the uniform $\epsilon_1=\epsilon_2=\epsilon \to 0$  limit, 
one obtains the symmetric diagonal 4-particle form-factors of $\Theta:$
\begin{equation} \label{FTHmmpp}
\begin{split}
F^{\Theta,symm}_{--++}(\theta_1,\theta_2)=F^{\Theta,symm}_{++--}(\theta_1,\theta_2)=
 2 \,{\cal M}^2\,  \left( 1+\cosh(\theta_1-\theta_2)\right) \, G(\theta_1-\theta_2),
\end{split}
\end{equation}
\begin{equation} \label{FTHpmpm}
\begin{split}
F^{\Theta,symm}_{+-+-}(\theta_1,\theta_2)=F^{\Theta,symm}_{-+-+}(\theta_1,\theta_2)=
 2 \,{\cal M}^2\, \big( 1+\cosh(\theta_1-\theta_2)\big) \, \Omega(\theta_1-\theta_2),
\end{split}
\end{equation}
\begin{equation} \label{FTHpmmp}
\begin{split}
F^{\Theta,symm}_{+--+}(\theta_1,\theta_2)=F^{\Theta,symm}_{-++-}(\theta_1,\theta_2)=
 2 \,{\cal M}^2\, \big( 1+\cosh(\theta_1-\theta_2)\big) \, \varphi(\theta_1-\theta_2).
\end{split}
\end{equation}
All functions entering these formulas are even, thus these form-factors are really symmetric with respect to the exchange of 
the two rapidities $\theta_1 \leftrightarrow \theta_2$.

The very same procedure can be repeated for the $U(1)$ current and for the scalar operator $\psi$ associated to it by (\ref{Jmuff1}). 
We just write down the final results below. In the near diagonal limit the 4-particle form factors of the scalar $\psi$ take the form: 
\begin{equation} \label{Fpsimmpp}
\begin{split}
F^{\psi}_{--++}(\hat{\theta}_2,\hat{\theta}_1,\theta_1,\theta_2)=-F^{\psi}_{++--}(\hat{\theta}_2,\hat{\theta}_1,\theta_1,\theta_2)=-
2 \pi \sigma(\theta_{12}) \, \left( \frac{1}{\epsilon_1}+\frac{1}{\epsilon_2}\right)+O(1)_\epsilon,
\end{split}
\end{equation}
\begin{equation} \label{Fpsipmmp}
\begin{split}
F^{\psi}_{-++-}(\hat{\theta}_2,\hat{\theta}_1,\theta_1,\theta_2)\!=\!-F^{\psi}_{+--+}(\hat{\theta}_2,\hat{\theta}_1,\theta_1,\theta_2)\!=\!
\frac{{\cal G}_0(\theta_{12})}{\epsilon_1 \, \epsilon_2}\!+\!\frac{{\cal G}_1(\theta_{12})}{\epsilon_1 }\!+\!
\frac{{\cal G}_2(\theta_{12})}{ \epsilon_2}\!+\!O(1)_\epsilon,
\end{split}
\end{equation}
\begin{equation} \label{Fpsipmpm}
\begin{split}
F^{\psi}_{-+-+}(\hat{\theta}_2,\hat{\theta}_1,\theta_1,\theta_2)\!=\!-F^{\psi}_{+-+-}(\hat{\theta}_2,\hat{\theta}_1,\theta_1,\theta_2)\!=\!
\frac{{\cal H}_0(\theta_{12})}{\epsilon_1 \, \epsilon_2}\!+\!\frac{{\cal H}_1(\theta_{12})}{\epsilon_1 }\!+\!
\frac{{\cal H}_2(\theta_{12})}{ \epsilon_2}\!+\!O(1)_\epsilon,
\end{split}
\end{equation}
where $\theta_{12}=\theta_1-\theta_2$ and
\begin{equation} \label{G0}
\begin{split}
{\cal G}_0(\theta)=-i \, \big(B_0(\theta) \, C_0(-\theta)-C_0(\theta) \, B_0(-\theta)\big),
\end{split}
\end{equation}
\begin{equation} \label{H0}
\begin{split}
{\cal H}_0(\theta)=-i \, \big(1+C_0(\theta) \, C_0(-\theta)-B_0(\theta) \, B_0(-\theta)\big),
\end{split}
\end{equation}
\begin{equation} \label{GH12}
\begin{split}
{\cal G}_j(\theta)=g_j(\theta)+G(\theta)\, \hat{g}_j(\theta), \quad 
\quad {\cal H}_j(\theta)=h_j(\theta)+G(\theta)\, \hat{h}_j(\theta), \qquad j=1,2,
\end{split}
\end{equation}
with 
\begin{equation} \label{g1}
\begin{split}
g_1(\theta)&=-i \big( B_0(\theta) \, C_0'(-\theta)-C_0(\theta) \, B_0'(-\theta)\big), \quad g_2(\theta)=-g_1(\theta), \\
\hat{g}_1(\theta)&=  B_0(\theta) \, C_0(-\theta)-C_0(\theta) \, B_0(-\theta), \qquad \quad \hat{g}_2(\theta)=-\hat{g}_1(\theta),
\end{split}
\end{equation}
\begin{equation} \label{h1}
\begin{split}
h_1(\theta)&=-i \big( C_0(\theta) \, C_0'(-\theta)-B_0(\theta) \, B_0'(-\theta)\big), \quad h_2(\theta)=-h_1(\theta), \\
\hat{h}_1(\theta)&=  C_0(\theta) \, C_0(-\theta)-B_0(\theta) \, B_0(-\theta), \qquad \quad \hat{h}_2(\theta)=-\hat{h}_1(\theta).
\end{split}
\end{equation}
Then using (\ref{Jmuff2}) the symmetric diagonal 4-particle form-factors of 
$J_\mu$ can be computed. It turns out that only the ones which correspond 
to the expectation values in pure soliton or pure antisoliton states have finite uniform $\epsilon_1=\epsilon_2=\epsilon \to 0$ limit:
\begin{equation} \label{FJmu2sym}
\begin{split}
F^{J_0,symm}_{--++}(\theta_1,\theta_2)=-F^{J_0,symm}_{++--}(\theta_1,\theta_2)=2 \,{\cal M}\, \left(\cosh \theta_1+\cosh \theta_2\right) \, G(\theta_{12}), \\
F^{J_1,symm}_{--++}(\theta_1,\theta_2)=-F^{J_1,symm}_{++--}(\theta_1,\theta_2)=2 \,{\cal M}\, \left(\sinh \theta_1+\sinh \theta_2\right) \, G(\theta_{12}).
\end{split}
\end{equation}
The other form-factors will diverge as $\frac{1}{\epsilon}$ when the symmetric diagonal limit is taken. Nevertheless it can be shown, that these divergences cancel, when 
according to (\ref{polform}) 
the symmetric diagonal{\footnote{Here the word diagonal means diagonality in the Bethe eigenstates, as well. }} 
limit is taken  between Bethe eigenvectors of the soliton transfer matrix. Simple application of the charge conjugation negativity of 
$J_\mu$ shows that these non-pure solitonic 4-particle symmetric diagonal form-factors are actually zero. 

%In the next section we will show, that 
%this divergence of the diagonal limit is a consequence of the charge conjugation negativity of the operator.

\subsection{6-particle symmetric diagonal form-factors}

If one would like to compute the symmetric diagonal limit of the 6-particle form-factors of the operators of our interest, after some computations 
it becomes obvious, that with fixed subscripts in general this diagonal limit does not exist. Namely, the $\epsilon \to 0$ limit becomes divergent. 
Nevertheless, in the P\'almai-Tak\'acs conjecture summarized in section \ref{PTsejt}, the symmetric diagonal limit of form-factors polarized with respect to 
eigenvectors of the soliton transfer matrix (\ref{polform}) should be determined. 
To do this computation, first we rewrite the necessary form-factor axioms in the language of the eigenvectors of the soliton transfer matrix (\ref{trans}).
For our computations we need the appropriate versions of two axioms, the exchange (\ref{ax2}) and the kinematical singularity (\ref{ax4}) ones. 

The kinematical pole axiom for a near diagonal settings of the rapidities can be written as follows:
\begin{equation} \label{ax3ind}
\begin{split}
F_{a_n...a_1 b_1...b_n}(\hat{\theta}_n,...,\hat{\theta}_1,\theta_1,...,\theta_n)\!=&\frac{i}{\epsilon_1} \!
\left\{ \delta_{b_1}^{\bar{a}_1}  \prod\limits_{k=2}^n  \delta_{a_k}^{\beta_k} \delta_{b_k}^{\alpha_k}\!-\!
\tau(\theta_1|\vec{\theta})_{b_1 b_2...b_n}^{l\alpha_2...\alpha_n} \, \tau^{-1}(\theta_1^\epsilon|\vec{\theta^\epsilon})_{l \bar{\beta}_2...\bar{\beta}_n}^{\bar{a}_1 \bar{a}_2...\bar{a}_n}
\right\}\!\!\times \\
& F_{\beta_n...\beta_2 \alpha_2...\alpha_n}(\hat{\theta}_n,...,\hat{\theta}_2,\theta_2,...,\theta_n)+O(1)_{\epsilon_1},
\end{split}
\end{equation}
where we introduced the notations $\theta_j^\epsilon=\theta_j+\epsilon_j$ and $\vec{\theta^\epsilon}=\{\theta_1^\epsilon,...,\theta_n^\epsilon\}.$
Now, analogously to the definition (\ref{polform}), one can sandwich this axiom with two color wave functions 
$\Psi$ and $\Psi^{(\epsilon)},$ such that they become complex conjugate to each other in the $\epsilon \to 0$ 
diagonal limit:
%an two arbitrary wave vector: $\Psi^{i_1...i_n}.$ Define:
\begin{equation} \label{Fsand1}
\begin{split}
F_{\Psi}(\hat{\theta}_n,...,\hat{\theta}_1,\theta_1,...,\theta_n)\!=\!\!\!\!\!\!\! \sum\limits_{i_1,...,i_n=\pm}  \sum\limits_{j_1,...,j_n=\pm} \!
\Psi_{j_1...j_n}^{(\epsilon)*} F_{\bar{j}_n...\bar{j}_1 i_1...i_n}(\hat{\theta}_n,...,\hat{\theta}_1,\theta_1,...,\theta_n)
\Psi^{i_1...i_n}.
\end{split}
\end{equation}
Then this form-factor satisfies the kinematical pole equation as follows:
\begin{equation} \label{ax3Fsand}
\begin{split}
F_{\! \Psi}(\hat{\theta}_n,.,\hat{\theta}_1,\theta_1,.,\theta_n)\!\!&=\!\!\frac{i}{\epsilon_1}\!\!\left\{ \Psi^{(\epsilon)*}_{k \bar{\beta}_2...\bar{\beta}_n} \! \Psi^{k \alpha_2...\alpha_n}\!\!-\!
\Psi^{i_1...i_n }\tau(\theta_1|\vec{\theta})_{i_1 i_2...i_n}^{l\alpha_2...\alpha_n} 
\tau^{-1}(\theta_1^\epsilon|\vec{\theta^\epsilon})_{l \bar{\beta}_2...\bar{\beta}_n}^{j_1 j_2...j_n} \Psi^{(\epsilon)*}_{j_1..j_n}
\right\}\!\times \\
& F_{\beta_n...\beta_2 \alpha_2...\alpha_n}(\hat{\theta}_n,...,\hat{\theta}_2,\theta_2,...,\theta_n)+O(1)_{\epsilon_1}.
\end{split}
\end{equation}
It follows, that this equation can be diagonalized, if $\Psi$ is chosen to be a left eigenvector of $\tau(\theta_1|\vec{\theta})$ and $\Psi^{(\epsilon)*}$ 
is chosen to be a right eigenvector of $\tau(\theta_1^\epsilon|\vec{\theta^\epsilon}):$
\begin{equation} \label{PPe}
\begin{split}
\Psi^{i_1...i_n }\tau(\theta_1|\vec{\theta})_{i_1 i_2...i_n}^{l\alpha_2...\alpha_n}&=\Lambda(\theta_1|\vec{\theta}) \Psi^{l\alpha_2...\alpha_n}, \\
\tau(\theta_1^\epsilon|\vec{\theta^\epsilon})_{l \bar{\beta}_2...\bar{\beta}_n}^{j_1 j_2...j_n} \Psi^{(\epsilon)*}_{j_1..j_n}&=\Lambda(\theta_1^\epsilon|\vec{\theta^\epsilon})
\Psi^{(\epsilon)*}_{l \bar{\beta}_2...\bar{\beta}_n}.
\end{split}
\end{equation}
With such sandwiching states the kinematical singularity axiom in the near diagonal limit takes the form: 
\begin{equation} \label{Fax3diag}
\begin{split}
F_{\! \Psi}(\hat{\theta}_n,.,\hat{\theta}_1,\theta_1,.,\theta_n)\!\!&=\!\frac{i}{\epsilon_1}\!\left(\!1\!-\!\frac{\Lambda(\theta_1|\vec{\theta})}{\Lambda(\theta_1^\epsilon|\vec{\theta^\epsilon})}\!\right) \Psi^{(\epsilon)*}_{k \bar{\beta}_2...\bar{\beta}_n} \! \Psi^{k \alpha_2...\alpha_n}\,
F_{\beta_n...\beta_2 \alpha_2...\alpha_n}(\hat{\theta}_n,...,\hat{\theta}_2,\theta_2,...,\theta_n)
\\&+O(1)_{\epsilon_1}
\end{split}
\end{equation}
A few important comments are in order. First, we pay the attention that $\Psi^{(\epsilon)*}$ is not the complex conjugate vector of $\Psi,$ because it is an eigenvector of a transfer matrix whose 
inhomogeneities are shifted with $\epsilon$s with respect to those of $\tau.$ They form a conjugate pair only 
in the $\epsilon_j \to 0$ limit.
On the other hand in \cite{Palmai13} the symmetric diagonal form-factors are defined by a sandwich (\ref{polform}), where $\Psi$ must be a {\bf right} eigenvector of $\tau$ (\ref{trans}).
Nevertheless, the near diagonal limit formulation of the kinematical singularity axiom (\ref{Fax3diag}) suggest, that the diagonal limit, should be taken such that in (\ref{polform}) the vector $\Psi$ 
must be the {\bf left} eigenvector of the transfer matrix (\ref{trans}). 
Actually this was the reason why we redefined the original definition of polarized form-factors (\ref{polform}) 
by the formula (\ref{polformen}). Nevertheless, in the sequel we keep the defining formula (\ref{polform}), but 
based on the implications of formulas (\ref{PPe}) and (\ref{Fax3diag}), 
\emph{we require $\Psi$ to be a left eigenvector of $\tau(\theta_1|\vec{\theta})$ and $\Psi^{(\epsilon)*}$ to be right eigenvector of $\tau(\theta_1^\epsilon|\vec{\theta^\epsilon}).$ }   

%Thus from now on we will define the polarized form-factors of a operator by the definition given in (\ref{polform}), 
%but as implied by the formulas (\ref{PPe}) \emph{we require $\Psi$ to be a left eigenvector of $\tau(\theta_1|\vec{\theta})$ and $\Psi^{(\epsilon)*}$ to be right eigenvector of $\tau(\theta_1^\epsilon|\vec{\theta^\epsilon}).$ } 

Now an important remark is in order. It is worth to analyse, what the form-factor equation (\ref{Fax3diag}) 
tells about the symmetric diagonal limit, when  $\epsilon_j$ tends to zero uniformly. 
The term  $\tfrac{i}{\epsilon_1}\!\left(\!1\!-\!\tfrac{\Lambda(\theta_1|\vec{\theta})}{\Lambda(\theta_1^\epsilon|\vec{\theta^\epsilon})}\!\right)$%two terms
  on the 
right hand side have a finite limiting value.  The sum $ \Psi^*_{k \bar{\beta}_2...\bar{\beta}_n} \! \Psi^{k \alpha_2...\alpha_n}\,
F_{\beta_n...\beta_2 \alpha_2...\alpha_n}(\hat{\theta}_n,...,\hat{\theta}_2,\theta_2,...,\theta_n)$ contains 
the sum of near diagonal form-factors with all possible indexes. In the previous section we saw, that not all 
of them have finite $\epsilon \to 0$ limit. This implies that the existence of the symmetric diagonal limit 
of a form-factor is not obvious, and if eventually  it exists, it must be a consequence of non-trivial 
cancellations between divergent terms. We will discuss this point in more detail in section \ref{sect9}.

We continue with writing the exchange axiom (\ref{ax2}) applied to the near diagonal limit  
in terms of the eigenvectors of the transfer matrix. These eigenvectors can be given as actions of the  
off diagonal elements of the monodromy matrix (\ref{monodr}) on the trivial vacuum (\ref{trivvac}).  
Using the representations (\ref{Psi1R}) and (\ref{Psi*1R}) for the Bethe-eigenvectors:
\begin{equation}
\begin{split} \label{PsiPsi}
\Psi^{a_1...a_n}\equiv \Psi^{a_1...a_n }(\{\lambda_j\}|\vec{\theta})\sim 
{}^{a_1...a_n}(\langle 0|\prod\limits_{j=1}^r {\cal C}(\lambda_j|\vec{\theta})), \\
%\sim (\prod\limits_{j=1}^r {\cal B}(\lambda_j|\vec{\theta})|0\rangle )^{a_1..a_n}, \\
\Psi^{(\epsilon)*}_{b_1...b_n}\equiv\Psi(\{\lambda_j^\epsilon\}|\vec{\theta^\epsilon})_{b_1...b_n}^*
%\sim {}_{b_1...b_n}(\langle 0|\prod\limits_{j=1}^r {\cal C}(\lambda_j^\epsilon|\vec{\theta}^\epsilon))
\sim (\prod\limits_{j=1}^r {\cal B}(\lambda_j^\epsilon|\vec{\theta^\epsilon})|0\rangle )_{b_1..b_n},
\end{split}
\end{equation}
the exchange axiom 
in the near diagonal limit can be written as follows:
\begin{equation} \label{Fax2diag}
\begin{split}
\Psi(\{\lambda_j^\epsilon\}|\vec{\theta^\epsilon})_{b_1...b_n}^* 
F_{\bar{b}_n...\bar{b}_1 a_1...a_n}(...,\hat{\theta}_{s+1},\hat{\theta}_s,...,\theta_s,\theta_{s+1},...) 
\Psi^{a_1...a_n }(\{\lambda_j\}|\vec{\theta})\!=\! S_0(\theta_{s+1}^\epsilon-\theta_s^\epsilon) \! \times\\
 S_0(\theta_{s}-\theta_{s+1}) \,
\Psi(\{\lambda_j^\epsilon\}|\vec{\theta^\epsilon}_{\!\!ex})_{b_1...b_n}^* 
F_{\bar{b}_n...\bar{b}_1 a_1...a_n}(...,\hat{\theta}_{s},\hat{\theta}_{s+1},...,\theta_{s+1},\theta_{s},...) 
\Psi^{a_1...a_n }(\{\lambda_j\}|\vec{\theta}_{ex}),
\end{split}
\end{equation}
where the set $\{\lambda_j\}_{j=1}^r$ is the solution of the Bethe-equations (\ref{ABAE}) and the set 
$\{\lambda_j^\epsilon\}_{j=1}^r$
also solves (\ref{ABAE}) but with $\theta_j \to \theta_j^\epsilon=\theta_j+\epsilon_j$ 
replacement{\footnote{We just note that (\ref{Fax2diag}) remains valid if the sets $\{\lambda_j\}_{j=1}^r$ and 
 $\{\lambda_j^\epsilon\}_{j=1}^r$ are not solutions of the Bethe-equations, but are arbitrary sets. 
Here we require them to be solutions of (\ref{ABAE}) for later convenience.}}.
The most important details of the formula are the vectors $\vec{\theta}$ and $\vec{\theta}_{ex}.$ 
In these vectors the order of the rapidity matters! The difference between them is the order of 
the exchanged rapidities $\theta_s$ and 
$\theta_{s+1}.$ Namely, 
\begin{equation} \label{thetavec}
\begin{split}
\vec{\theta}&=\{\theta_1,...,\theta_s,\theta_{s+1},..,\theta_n\}, \qquad \vec{\theta^\epsilon}\!=\!\{\theta_1\!+\!\epsilon_1,..,\theta_s\!+\!\epsilon_s,\theta_{s+1}\!+\!\epsilon_{s+1},..,\theta_n\!+\!\epsilon_n\},\\
\vec{\theta}_{ex}&=\{\theta_1,...,\theta_{s+1},\theta_{s},..,\theta_n\}, \qquad \vec{\theta^\epsilon}_{\!\!ex}\!=\!\{\theta_1\!+\!\epsilon_1,.,\theta_{s+1}\!+\!\epsilon_{s+1},
\theta_{s}\!+\!\epsilon_{s},..,\theta_n\!+\!\epsilon_n\}.
\end{split}
\end{equation}
This means that the Bethe-vectors on the left and right hand sides of the equation (\ref{Fax2diag}) are 
different, since they are eigenvectors of different transfer matrices! We would like to explain this in a bit 
more detail. The rapidities $\theta_j$ are inhomogeneities of the transfer matrix. The transfer matrix is 
not invariant under the permutation of the inhomogeneities among the $n$ lattice sites. Nevertheless, the Bethe-equations (\ref{ABAE}) and the eigenvalue expression are also invariant under the permutation of the 
rapidities. Thus the transfer matrices $\tau(\theta|\vec{\theta})$ and $\tau(\theta|\vec{\theta_{ex}})$ 
are only isospectral, but have different eigenvectors connected by a unitary transformation. This recognition 
has also some implication on the P\'almai-Tak\'acs conjecture (section \ref{PTsejt}), 
since there in the wave-function decomposition (\ref{decomp}) the orders of rapidities 
in the arguments of the wave functions matter!

\subsubsection{Solitonic 6-particle symmetric diagonal form-factors}

If one starts to analyse the 3-particle Bethe-equations (\ref{ABAE}), it becomes immediately obvious that 
the relevant solutions are the zero and 1-root solutions, since they account for all states 
in the Q=3 and Q=1 sectors. The missing Q=-3 and Q=-1 sectors can be obtained from the previous ones by the 
charge conjugation symmetry. The Q=3 sector is the pure soliton sector with no Bethe-root in (\ref{ABAE}). 
In this case the complicated sum in the right hand side of the kinematical pole equation (\ref{Fax3diag}) 
applied to the scalar operators $\phi$ and $\psi$ 
will contain only a single term, which includes only the pure solitonic near diagonal form-factors (\ref{ATHmmpp}) and (\ref{Fpsimmpp}). Since in this limit the diagonal pure solitonic matrix elements does not mix with other 
states, the computation of their symmetric and connected limits can be computed in exactly the same way 
as in a purely elastic scattering theory \cite{Mussbook}.Their explicit form  
for the operators $\Theta$ and $J_\mu$ can be found in references 
\cite{En1} and \cite{En}, respectiveley.
In these papers analytical formulas describing  
the Bethe-Yang limit of pure solitonic expectation values of the 
operators $J_\mu$ and $\Theta$ can also be found. This made it possible  
to verify the conjecture of \cite{Palmai13} in this sector for any number of solitons.
 The pure solitonic sector is very similar to the case of a purely elastic scattering theory. 
As a consequence the actual form of conjecture of \cite{Palmai13} goes through remarkable 
simplifications in this sector and becomes identical with the formula conjectured for 
diagonally scattering theories \cite{Pozsg13,PST14}.
 Our purpose is to check the general form of the conjecture of \cite{Palmai13}. Thus we will test it  
in a sector, where there is mixing between the states with different polarizations. This simplest such nontrivial 
sector is the $Q=1$ sector of the 3-particle space. In the language of the Bethe-equations (\ref{ABAE}) 
it is described by a single Bethe-root.

\subsubsection{6-particle symmetric diagonal form-factors in the $Q=1$ sector}

The first step to compute the symmetric diagonal form-factors of the operators of our interest 
in the $Q=1$ sector, is to write down the actual form of the wave functions which should sandwich 
our form-factors according to (\ref{polform}). 
Here we denote their matrix elements as follows:
\begin{equation} \label{vektorok}
\begin{split}
\Psi^{i_1 i_2 i_3 \,}=\frac{C^{i_1 i_2 i_3}}{N_\Psi}, \\
\Psi_{i_1 i_2 i_3}^{(\epsilon)*}=\frac{B^\epsilon_{i_1 i_2 i_3}}{N_\Psi},
\end{split}
\end{equation}
where the nonzero coefficients in the $Q=1$ sector can be read off from the 
formulas (\ref{Psi1R}), (\ref{Psi*1R}) coming from the Algebraic Bethe-Ansatz 
diagonalization of the soliton-transfer matrix: 
\begin{equation} \label{BC123}
\begin{split}
C^{+--}=C_1, \qquad C^{-+-}=B_1 \, C_2, \qquad C^{--+}=B_1 \, B_2 \, C_3, \\
B^\epsilon_{+--}=C_1^\epsilon \, B_2^\epsilon \, B_3^\epsilon, \qquad B^\epsilon_{-+-}= C_2^\epsilon \, B_3^\epsilon, \qquad B^\epsilon_{--+}= C_3^\epsilon,
\end{split}
\end{equation}
where for later convenience we introduced the short notations as follows:
\begin{equation} \label{BjCj}
\begin{split}
B_j=&B_0(\lambda_1-\theta_j), \qquad C_j=C_0(\lambda_1-\theta_j), \qquad 
B_j^\epsilon=B_0(\lambda_1^\epsilon-\theta_j^\epsilon), \qquad C_j^\epsilon=C_0(\lambda_1^\epsilon-\theta_j^\epsilon), \\
  &\text{with} \qquad \theta_j^\epsilon=\theta_j+\epsilon_j, \qquad \text{for} \quad j=1,2,3,
\end{split}
\end{equation}
such that the single Bethe-roots $\lambda_1$ and $\lambda_1^\epsilon$ are solutions of the Bethe-equations (\ref{3BAE}): 
\begin{equation} \label{BBE}
\begin{split}
B_1 \, B_2 \, B_3=1, \qquad B_1^\epsilon \, B_2^\epsilon\, B_3^\epsilon=1.
\end{split}
\end{equation}
The normalization factor $N_\Psi$ is chosen to be the Gaudin-norm (\ref{3NPsi}) 
of the vector{\footnote{Namely, the $\epsilon \to 0 $ limit of the scalar product $C^{i_1 i_2 i_3} B^\epsilon_{i_1 i_2 i_3}.$}} $\Psi.$ We note that this normalization factor is invariant under any permutations of the 
three rapidities $\{\theta_j\}_{j=1}^3.$

%\subsection{6-particle symmetric diagonal form-factor of $\Theta$ in the $Q=1$ subsector}
\begin{equation}\nonumber
\end{equation}

{\bf The case of $\Theta$:}
\newline

Now we are in the position to compute the 6-particle symmetric diagonal form-factors of $\Theta$ in the $Q=1$ subsector. This subsector is characterized by a single Bethe-root solving the equation (\ref{BBE}). 

Looking at the formula (\ref{THff1}) it turns out that to get the required limit of our 6-particle form-factor 
one needs to know the $\tfrac{1}{\epsilon^2}$ order part of the $\Psi$-sandwiched matrix element of the 
scalar operator $\phi$ defined in (\ref{Tmn}). To compute this part, one needs to use only the equations 
(\ref{Fax3diag}) and (\ref{Fax2diag}). These equations together with the concrete forms (\ref{aphi})-(\ref{ATHpmmp})
of the near diagonal 4-particle form factors imply the following small $\epsilon$ series for the 
required form-factor of $\phi:$
\begin{equation} \label{Wphi}
\begin{split}
W^\phi(\theta_1,\epsilon_1;\theta_2,\epsilon_2;\theta_3,\epsilon_3)=
\frac{1}{N_\Psi^2} B^\epsilon_{j_1 j_2 j_3} 
F^\phi_{\bar{j}_3 \bar{j}_2 \bar{j}_1 i_1 i_2 i_3}(\hat{\theta}_3,\hat{\theta}_2,\hat{\theta}_1, \theta_1,\theta_2,\theta_3) \, C^{i_1 i_2 i_3},
\end{split}
\end{equation}
\begin{equation} \label{Wexp}
\begin{split}
W^\phi(\theta_1,\epsilon_1;\theta_2,\epsilon_2;\theta_3,\epsilon_3)\!=\!
\frac{A_{12}(\theta_1,\theta_2,\theta_3)}{\epsilon_1 \epsilon_2}+
\frac{A_{13}(\theta_1,\theta_2,\theta_3)}{\epsilon_1 \epsilon_3}+
\frac{A_{23}(\theta_1,\theta_2,\theta_3)}{\epsilon_2 \epsilon_3}+
O(\tfrac{1}{\epsilon}).
\end{split}
\end{equation}
Then equation (\ref{Fax2diag}) tells us how $W^\phi$ of (\ref{Wphi}) changes when exchanging the pairs $(\theta_j,\epsilon_j) \leftrightarrow (\theta_k,\epsilon_k)$ in the argument. This gives the following 
relations among the $A_{ij}$ functions in (\ref{Wexp}):
\begin{equation} \label{Aijrel1}
\begin{split}
A_{13}(\theta_1,\theta_2,\theta_3)=A_{12}(\theta_3,\theta_1,\theta_2), \qquad
A_{23}(\theta_1,\theta_2,\theta_3)=A_{13}(\theta_3,\theta_1,\theta_2),%=A_{12}(\theta_2,\theta_3,\theta_1),
\end{split}
\end{equation}
and in addition $A_{ij}$ is invariant under the exchange of its $i$th and $j$th arguments.

The functions $A_{12}$ and $A_{13}$ can be directly computed from the $\tfrac{1}{\epsilon_1}$ 
pole given by equation (\ref{Fax3diag}). Then $A_{23}$ can be determined from them by using (\ref{Aijrel1}). 
Straightforward application of (\ref{Fax3diag}) leads to the following expressions for $A_{12}$ and $A_{13}:$
\begin{equation} \label{A12}
\begin{split}
A_{12}(\theta_1,\theta_2,\theta_3)\!=\!i\, \partial_3 \ln \Lambda(\theta_1|\vec{\theta}) \, 
T^\phi(\theta_1,\theta_2,\theta_3), \\
A_{13}(\theta_1,\theta_2,\theta_3)\!=\!i\, \partial_2 \ln \Lambda(\theta_1|\vec{\theta}) \, 
T^\phi(\theta_1,\theta_2,\theta_3),
\end{split}
\end{equation}
where $T^\phi$ is the singularity eliminated tensorial sum part of (\ref{Fax3diag}):
\begin{equation} \label{Tphi}
\begin{split}
T^\phi(\theta_1,\theta_2,\theta_3)\!&=\!\lim\limits_{\epsilon \to 0}
\frac{1}{N_\Psi^2} B^\epsilon_{k \bar{\beta}_2 \bar{\beta}_3} C^{k \alpha_2 \alpha_3} 
a^\phi_{\beta_3 \beta_2 \alpha_2 \alpha_3}(\theta_2,\theta_3)=\\
&-\frac{1}{N_\Psi^2}\left[ \frac{C_1^2}{B_1}G(\theta_{23})
+\left(\frac{C_2^2}{B_2}+\frac{C_3^2}{B_3}\right)\,\Omega(\theta_{23})+(1+B_1)\, C_2 \, C_3 \, \varphi(\theta_{23})
\right],
\end{split}
\end{equation}
with the constituent functions given in (\ref{kernelsTH2}) and (\ref{BjCj}).
Having the explicit expression for $A_{12}$ and $A_{13},$ with the help of the exchange relation (\ref{Aijrel1}) 
$A_{23}$ can also be obtained from them. Finally using (\ref{THff1}) and (\ref{Wphi}), the symmetric diagonal 
limit of the form-factors of $\Theta$ in a 3-particle state described by the 
Bethe-root $\lambda_1$ can be given by the 
formula as follows: 
\begin{equation} \label{FSDTH1}
\begin{split}
F^{\Theta, (\Psi)}_{6,symm}(\theta_1,\theta_2,\theta_3)=-\frac{{\cal M}^2}{N_\Psi^2} 
\left[ A_{12}(\theta_1,\theta_2,\theta_3)+A_{13}(\theta_1,\theta_2,\theta_3)+A_{23}(\theta_1,\theta_2,\theta_3)\right]
\times \\
\, \left[3+2 \cosh(\theta_{12}) +2 \cosh(\theta_{13})+2 \cosh(\theta_{23})\right].
\end{split}
\end{equation}
We note that the $\lambda_1$ dependence is implicit in this expression. It is hidden in the 
expression of $T^\Phi$ in (\ref{Tphi}) and in the derivative of the eigenvalue in (\ref{A12}). 
A useful formula for the latter is given in (\ref{Lsq}).

\begin{equation}\nonumber
\end{equation}

%\subsection{6-particle symmetric diagonal form-factor of $J_\mu$ in the $Q=1$ subsector}

{\bf The case of $J_\mu$:}
\newline

The computation of the 6-particle symmetric diagonal form-factors of the $U(1)$ current is a bit more 
subtle than that of the trace of the stress energy tensor. The method described in the previous paragraphs  
is the same, but one should be much more careful in the small $\epsilon_j$ expansion. In this case  
the linear in $\epsilon_j$ terms of the Bethe-vector $B^\epsilon_{i_1 i_2 i_3}$ (\ref{BC123}) will 
also give relevant contributions to the symmetric form-factors.
The first step is to compute the 6-particle form-factor of the scalar field $\psi$ in the near 
diagonal limit. Thus the quantity we compute is defined by:
\begin{equation} \label{Wpsi}
\begin{split}
W^\psi(\lambda^\epsilon_1,\lambda_1|1^\epsilon,2^\epsilon,3^\epsilon)=
\frac{1}{N_\Psi^2} B^\epsilon_{j_1 j_2 j_3} 
F^\psi_{\bar{j}_3 \bar{j}_2 \bar{j}_1 i_1 i_2 i_3}(\hat{\theta}_3,\hat{\theta}_2,\hat{\theta}_1, \theta_1,\theta_2,\theta_3) \, C^{i_1 i_2 i_3},
\end{split}
\end{equation}
where for short we introduced the symbolic notation for a pair: $\theta_j,\epsilon_j \to j^\epsilon,$ 
and we also indicated in the list of arguments the Bethe-root dependence of this form-factor. 
Using the kinematical pole equation for the $\tfrac{1}{\epsilon_1}$ singularity, the terms proportional to 
$\tfrac{1}{\epsilon_1}$ in the small $\epsilon$ expansion of $W^\psi$ can be computed.
To facilitate this task first we do the computations in some smaller building blocks of $W^\psi.$   
Let $Y$ denote the eigenvalue part of (\ref{Fax3diag}):  
\begin{equation} \label{Ydef}
\begin{split}
Y(\lambda^\epsilon_1,\lambda_1|1^\epsilon,2^\epsilon,3^\epsilon)=1-\frac{\Lambda(\theta_1|\vec{\theta})}{\Lambda(\theta^\epsilon_1|\vec{\theta^\epsilon})}=\sum\limits_{j=1}^3 \epsilon_j \, \partial_j \ln \Lambda(\theta_1|\vec{\theta})+
O(\epsilon^2),
\end{split}
\end{equation}
and let denote $T^\psi$ the tensorial sum part of (\ref{Fax3diag}): 
\begin{equation} \label{Tpsi}
\begin{split}
T^\psi(\lambda_1^\epsilon,\lambda|1^\epsilon,2^\epsilon,3^\epsilon)=B^\epsilon_{k \bar{\beta}_2 \bar{\beta}_3} 
C^{\, k \alpha_2 \alpha_3}
F^\psi_{\beta_3 \beta_2 \alpha_2 \alpha_3}(\hat{\theta}_3,\hat{\theta}_2,\theta_2,\theta_3) .
\end{split}
\end{equation}
Taking the near diagonal limit of the 4-particle form-factors of $\psi$ given in (\ref{Fpsimmpp})-(\ref{h1}),
one obtains the following small $\epsilon$ expansion for $T^\psi:$
\begin{equation} \label{Tpsiexp}
\begin{split}
T^\psi(\lambda_1^\epsilon,\lambda_1|1^\epsilon,2^\epsilon,3^\epsilon)=
\frac{T_{23}(\vec{\theta})}{\epsilon_2 \epsilon_3}+
\frac{T_2(\vec{\theta})}{\epsilon_2}+\frac{T_3(\vec{\theta})}{\epsilon_3}+O(1),
\end{split}
\end{equation}
where the functions $T_{23}(\vec{\theta}),\, T_{2}(\vec{\theta}), \,T_{3}(\vec{\theta})$ take the form:
\begin{equation} \label{T23}
\begin{split}
T_{23}(\vec{\theta})={\cal H}_0(\theta_{23})\left( \frac{C_3^2}{B_3}-\frac{C_2^2}{B_2}\right)\!+
{\cal G}_0(\theta_{23}) \, C_2 \, C_3 \, (B_1-1),
\end{split}
\end{equation}
\begin{equation} \label{T2}
\begin{split}
T_{2}(\vec{\theta})\!=&{\cal G}_0(\theta_{23})\!\!\left[\!B^{(3)}_3\!(\vec{\theta})C^{(2)}\!(\vec{\theta})\!-\!
B^{(2)}_3\!(\vec{\theta})C^{(3)}\!(\vec{\theta}) \!\right]
\!\!+\!
{\cal G}_1(\theta_{23})\!\left[\!B^{(3)}_0\!(\vec{\theta})C^{(2)}\!(\vec{\theta})\!-\!
B^{(2)}_0\!(\vec{\theta})C^{(3)}\!(\vec{\theta}) \!\right]\!+\\
&{\cal H}_0(\theta_{23})\!\!\left[\!B^{(3)}_3\!(\vec{\theta})C^{(3)}\!(\vec{\theta})\!-\!
B^{(2)}_3\!(\vec{\theta})C^{(2)}\!(\vec{\theta}) \!\right]
\!\!+\!
{\cal H}_1(\theta_{23})\!\left[\!B^{(3)}_0\!(\vec{\theta})C^{(3)}\!(\vec{\theta})\!-\!
B^{(2)}_0\!(\vec{\theta})C^{(2)}\!(\vec{\theta}) \!\right]\!+\\
& G(\theta_{23}) B^{(1)}_0\!(\vec{\theta}) C^{(1)}(\vec{\theta}),
\end{split}
\end{equation}
\begin{equation} \label{T3}
\begin{split}
T_{3}(\vec{\theta})\!=&{\cal G}_0(\theta_{23})\!\!\left[\!B^{(3)}_2\!(\vec{\theta})C^{(2)}\!(\vec{\theta})\!-\!
B^{(2)}_2\!(\vec{\theta})C^{(3)}\!(\vec{\theta}) \!\right]
\!\!+\!
{\cal G}_2(\theta_{23})\!\left[\!B^{(3)}_0\!(\vec{\theta})C^{(2)}\!(\vec{\theta})\!-\!
B^{(2)}_0\!(\vec{\theta})C^{(3)}\!(\vec{\theta}) \!\right]\!+\\
&{\cal H}_0(\theta_{23})\!\!\left[\!B^{(3)}_2\!(\vec{\theta})C^{(3)}\!(\vec{\theta})\!-\!
B^{(2)}_2\!(\vec{\theta})C^{(2)}\!(\vec{\theta}) \!\right]
\!\!+\!
{\cal H}_2(\theta_{23})\!\left[\!B^{(3)}_0\!(\vec{\theta})C^{(3)}\!(\vec{\theta})\!-\!
B^{(2)}_0\!(\vec{\theta})C^{(2)}\!(\vec{\theta}) \!\right]\!+\\
& G(\theta_{23}) B^{(1)}_0\!(\vec{\theta}) C^{(1)}(\vec{\theta}),
\end{split}
\end{equation}
where the functions $B^{(j)}_k$ and $C^{(k)}$ are coming from the small $\epsilon$ expansion of the 
components of the Bethe-eigenvectors (\ref{BC123}) in the following way:
\begin{eqnarray} %\label{BCeps}
B_{+--}^\epsilon&=&B^{(1)}_0(\vec{\theta})+\sum\limits_{j=1}^3 B^{(1)}_j(\theta) \epsilon_j+O(\epsilon^2), 
\qquad C_{+--}=C^{(1)}(\vec{\theta}),
\nonumber \\
B_{-+-}^\epsilon&=&B^{(2)}_0(\vec{\theta})+\sum\limits_{j=1}^3 B^{(2)}_j(\theta) \epsilon_j+O(\epsilon^2), 
\qquad C_{-+-}=C^{(2)}(\vec{\theta}), \label{BCeps} \\
B_{--+}^\epsilon&=&B^{(3)}_0(\vec{\theta})+\sum\limits_{j=1}^3 B^{(3)}_j(\theta) \epsilon_j+O(\epsilon^2), 
\qquad C_{--+}=C^{(3)}(\vec{\theta}).
\nonumber
\end{eqnarray}
Their actual form can be computed from (\ref{BC123}) and  (\ref{BjCj}). Here we give only the ones entering 
(\ref{T2}) and (\ref{T3}): 
\begin{eqnarray} 
B^{(1)}_0(\theta)\!\!&=&\!\!C_1 \, B_2 \, B_3, \qquad \qquad C^{(1)}(\vec{\theta})=C_1, \nonumber \\
B^{(2)}_0(\theta)&=&C_2 \, B_3,  \qquad \qquad C^{(2)}(\vec{\theta})=B_1 \, C_2, \label{BC1ek} \\
B^{(3)}_0(\theta)&=&C_3, \qquad \qquad C^{(3)}(\vec{\theta})=B_1 \, B_2 \, C_3, \nonumber
\end{eqnarray}
\begin{equation} \label{BB1}
\begin{split}
B^{(1)}_2(\vec{\theta})&=\partial_{\lambda_1} B^{(1)}_0(\vec{\theta})\cdot \frac{\partial \lambda_1}{\partial \theta_2}-
C_1 \, B'_2 \, B_3, \\
B^{(1)}_3(\vec{\theta})&=\partial_{\lambda_1} B^{(1)}_0(\vec{\theta})\cdot \frac{\partial \lambda_1}{\partial \theta_3}-
C_1 \, B_2 \, B'_3, 
\end{split}
\end{equation}
\begin{equation} \label{BB2}
\begin{split}
B^{(2)}_2(\vec{\theta})&=\partial_{\lambda_1} B^{(2)}_0(\vec{\theta})\cdot \frac{\partial \lambda_1}{\partial \theta_2}- C'_2 \, B_3, \\
B^{(2)}_3(\vec{\theta})&=\partial_{\lambda_1} B^{(2)}_0(\vec{\theta})\cdot \frac{\partial \lambda_1}{\partial \theta_3}- C_2 \, B'_3, 
\end{split}
\end{equation}
\begin{equation} \label{BB3}
\begin{split}
B^{(3)}_2(\vec{\theta})&=\partial_{\lambda_1} B^{(3)}_0(\vec{\theta})\cdot \frac{\partial \lambda_1}{\partial \theta_2}, \\
B^{(3)}_3(\vec{\theta})&=\partial_{\lambda_1} B^{(3)}_0(\vec{\theta})\cdot \frac{\partial \lambda_1}{\partial \theta_3}- C'_3, 
\end{split}
\end{equation}
where introduced the notations:
\begin{equation} \label{short}
\begin{split}
B'_j=B'_0(\lambda_1-\theta_j), \qquad C'_j=C'_0(\lambda_1-\theta_j), \qquad j=1,2,3. 
\end{split}
\end{equation}
In the above formulas we did not write down explicitely the $\lambda_1$ dependence of the functions. Nevertheless, it 
is important because of the  $\partial_{\lambda_1}$ partial derivatives. Here the $\lambda_1$ dependence is simply meant 
by the $\lambda_1$ dependence of the objects $B_j$ and $C_j$ given by (\ref{BjCj}).

Now we have all ingredients to compute the 6-particle symmetric diagonal form-factors of $J_\mu$ in the $Q=1$ 
sector of the 3-particle subspace. Looking at the formula (\ref{Jmuff2}) one can see that the symmetric diagonal 
limit is finite only if $F^\psi$ or equivalently $T^\psi$ has only $\tfrac{1}{\epsilon_j}$ order divergences. 
However, the order $\tfrac{1}{\epsilon^2}$ term in the expansion (\ref{Tpsiexp}) of $T^\psi$ implies, that 
 the symmetric diagonal limit is divergent in this case, provided the coefficient function $T_{23}$ is nonzero. 
Looking at its explicit form (\ref{T23}) it does not seem to be zero. Nevertheless, with some work, exploiting the 
Yang-Baxter equations (\ref{YBE}) and the Bethe-equations (\ref{BBE}) for $\lambda_1$ one can show that:  
\begin{equation} \label{T23=0}
\begin{split}
T_{23}(\vec{\theta})=0.
\end{split}
\end{equation}
This nontrivial for the first sight result ensures, that the symmetric diagonal limit of the 6-particle form-factors 
of $J_\mu$ in the $Q=1$ sector will be well defined. Nevertheless this computation sheds light on the fact that 
the higher and higher $\tfrac{1}{\epsilon}$ divergences of the nondiagonal form-factors could make the symmetric 
diagonal limit divergent{\footnote{Since their sum enter the right hand of the kinematical pole axiom (\ref{Fax3diag}). See section \ref{sect9} for a more detailed discussion.}}, too. On the other hand this computation might also imply that the special properties of 
integrability might ensure the cancellation of these (would be?) divergences. 

Due to the cancellation of the $\tfrac{1}{\epsilon^2}$ divergent term in $T^\psi,$ $W^\psi$ (\ref{Wpsi}) admits the 
following small $\epsilon$ expansion:
 \begin{equation} \label{Wpsiexp}
\begin{split}
W^\psi(\lambda^\epsilon_1,\lambda_1|1^\epsilon,2^\epsilon,3^\epsilon)&=
\frac{W_1(\vec{\theta})}{\epsilon_1}+\frac{W_2(\vec{\theta})}{\epsilon_2}+
\frac{W_3(\vec{\theta})}{\epsilon_3}+
W^{(1)}(\vec{\theta})\frac{\epsilon_1}{\epsilon_2 \epsilon_3}
+
W^{(2)}(\vec{\theta})\frac{\epsilon_2}{\epsilon_1 \epsilon_3}
+\\
&W^{(3)}(\vec{\theta})\frac{\epsilon_3}{\epsilon_1 \epsilon_2}
+O(1),
\end{split}
\end{equation}
such that the coefficient functions $W_1, \, W^{(2)}$ and $W^{(3)}$ can be computed from the kinematical 
pole equation (\ref{Fax3diag}) by using the formulas (\ref{Tpsiexp}) and (\ref{Ydef}):
\begin{equation} \label{W1}
\begin{split}
W_1(\vec{\theta})&=\frac{i}{N_\Psi^2} \, \left[T_2(\vec{\theta}) \, \partial_2 \ln \Lambda(\theta_1|\vec{\theta}) +
T_3(\vec{\theta}) \, \partial_3 \ln \Lambda(\theta_1|\vec{\theta})\right], \\
W^{(2)}(\vec{\theta})&=\frac{i}{N_\Psi^2} \, T_3(\vec{\theta}) \, \partial_2 \ln \Lambda(\theta_1|\vec{\theta}), \\ 
W^{(3)}(\vec{\theta})&=\frac{i}{N_\Psi^2} \, T_2(\vec{\theta}) \, \partial_3 \ln \Lambda(\theta_1|\vec{\theta}).
\end{split}
\end{equation}
The exchange equation (\ref{Fax2diag}) allows one to compute from (\ref{W1}) the other still unknown $W$-functions 
of the expansion (\ref{Wpsiexp}), since (\ref{Fax2diag}) implies that they are related by argument exchanges: 
\begin{equation} \label{Wujak}
\begin{split}
W_2(\theta_1,\theta_2,\theta_3)&=W_1(\theta_2,\theta_1,\theta_3), \\
W_3(\theta_1,\theta_2,\theta_3)&=W_1(\theta_3,\theta_2,\theta_1), \\
W^{(1)}(\theta_1,\theta_2,\theta_3)&=W^{(2)}(\theta_2,\theta_1,\theta_3).
\end{split}
\end{equation}
Nevertheless, (\ref{Fax2diag}) gives further relations among these functions, which can be used to test 
the obtained result. These are as follows. The functions $W_j(\theta_1,\theta_2,\theta_3)$ and 
$W^{(j)}(\theta_1,\theta_2,\theta_3)$ are symmetric with respect to the exchange of the  rapidities 
$\theta_s$ and $\theta_q$ with $s,q \neq j.$ According to (\ref{Fax2diag}), $W^{(2)}$ and $W^{(3)}$ are also not 
 independent:
\begin{equation} \label{W^23}
\begin{split}
W^{(2)}(\theta_1,\theta_2,\theta_3)&=W^{(3)}(\theta_3,\theta_1,\theta_2).
\end{split}
\end{equation}
It can be checked that our formulas in (\ref{W1}) satisfy this requirement.

With the help of (\ref{Jmuff2}) the symmetric diagonal 6-particle form-factors of the current can be expressed in 
terms of the previously computed $W$-functions as follows:
\begin{equation} \label{FJmu6}
\begin{split}
F^{J_\mu,(\Psi)}_{6,symm}(\theta_1,\theta_2,\theta_3)=(-1)^{\mu+1} {\cal M} 
\left(\sum\limits_{j=1}^3 v_j^{(\mu)}\right)
\sum\limits_{j=1}^3 \left[ W_j(\vec{\theta})+W^{(j)}(\vec{\theta})\right],
\end{split}
\end{equation}
with the vector 
\begin{equation} \label{vmu}
\begin{split}
v^{(\mu)}_j=\left\{\begin{array}{r} \cosh(\theta_j), \qquad \text{for} \quad \mu=0, \\
\sinh(\theta_j), \qquad \text{for} \quad \mu=1. 
\end{array} \right.
\end{split}
\end{equation}
The formula (\ref{FJmu6}) can also be rephrased in an equivalent way, which reflects manifestly the 
invariance of the symmetric form-factor with respect to the permutations of the rapidities:
\begin{equation} \label{FJmu6equiv}
\begin{split}
F^{J_\mu,(\Psi)}_{6,symm}(\theta_1,\theta_2,\theta_3)=(-1)^{\mu+1} \frac{\cal M}{2} 
\left(\sum\limits_{j=1}^3 v_j^{(\mu)}\right) \times \\
\sum\limits_{\sigma \in S^3} \left[ W_1(\theta_{\sigma(1)},\theta_{\sigma(2)},\theta_{\sigma(3)})
+W^{(2)}(\theta_{\sigma(1)},\theta_{\sigma(2)},\theta_{\sigma(3)})\right],
\end{split}
\end{equation}
where the second sum runs for the six possible permutations of the indexes $\{1,2,3\}.$

%\partial_{\lambda_1}

\section{Checking the P\'almai-Tak\'acs conjecture} \label{sect8}

In the previous sections we computed the symmetric diagonal form-factors of the operators $\Theta$  and $J_\mu$ 
upto 6-particles. This makes it possible to check the conjecture of P\'almai and Tak\'acs 
for the diagonal matrix elements of local operators \cite{Palmai13} 
(summarized in section \ref{PTsejt}.) against the exact results given in (\ref{THrest8})-(\ref{J18}) upto 3-particle 
expectation values. In the pure soliton sector{\footnote{For arbitrary number of solitons and not only upto 3.}} 
the validity of this conjecture have been already verified for the operators $\Theta$ and $J_\mu$ in references 
\cite{En1} and \cite{En}, respectively. This is why in our work we will only focus on states in which soliton and 
antisoliton states are mixed.

As implied by (\ref{bazis}), in the conjecture the eigenvectors of the multisoliton transfer matrix (\ref{trans}),  
play an important 
role. To test the conjecture upto 3-particle states, one needs the complete Bethe-basis on the space of 1- and 
2-particle states and one also needs the Bethe-eigenvector corresponding to the sandwiching 3-particle state.
Thus, as a first step we write down these Bethe-eigenvectors. 

For the one particle states the eigenvectors are simple: %and independent of the particle's rapidity.
\begin{equation} \label{1wave}
\begin{split}
\varphi^{(a)}_{i_1}=\delta_{i_1,a}, \qquad a,i_1=\pm.
\end{split}
\end{equation}
In the space of 1-particle states the basis is two dimensional corresponding to the soliton and the antisoliton. The  
 index $a$ distinguishes the two basis vectors of this space and $i_1$ 
is the index of the vector.
% of this two dimensional vector space. 
Here we pay the attention to two trivial, but for later considerations important 
properties of this basis. First of all the vector components are independent of the particle's rapidities. 
Second of all these vectors are real.
The 2-particle basis is also very simple \cite{Palmai13}:
\begin{equation} \label{2wave}
\begin{split}
\Psi^{(1)}_{i_1 i_2}&=\delta_{i_1 -} \delta_{i_2 -}, \qquad \qquad \qquad \qquad \qquad 
\Psi^{(2)}_{i_1 i_2}=\tfrac{1}{\sqrt{2}}\left(\delta_{i_1 +} \delta_{i_2 -}+\delta_{i_1 -} \delta_{i_2 +}\right),
\\
\Psi^{(3)}_{i_1 i_2}&=\tfrac{1}{\sqrt{2}}\left(\delta_{i_1 +} \delta_{i_2 -}-\delta_{i_1 -} \delta_{i_2 +}\right),
\qquad \quad \Psi^{(4)}_{i_1 i_2}=\delta_{i_1 +} \delta_{i_2 +}.
\end{split}
\end{equation}
Here again the superscript indexes the basis vectors and the subscripts $i_1,i_2=\pm$ denotes the vector indexes in 
the 2-particle vector space. Here we also emphasize that this 2-particle basis is real and rapidity independent.
With this remark we would like to pay the attention, that the first numerical checks of the 
P\'almai-Tak\'acs conjecture in \cite{Palmai13}, which were performed upto 2-particle states, were not 
sensible to the difference between the two definitions (\ref{polform}) and (\ref{polformen}). 

It is worth to discuss a bit more on the meaning of the basis vectors of (\ref{2wave}). 
The vectors $\Psi^{(4)}$ and $\Psi^{(1)}$ correspond to the two antisoliton and two soliton states, respectively. 
The vector $\Psi^{(2)}$ and $\Psi^{(3)}$ describe the symmetric and antisymmetric soliton-antisoliton states, 
respectively. At the level of the magnonic Bethe-equations (\ref{ABAE}),  
$\Psi^{(4)}$ and $\Psi^{(1)}$ are states without Bethe-roots, while $\Psi^{(2)}$ and $\Psi^{(3)}$ are described by
a single Bethe-root. Using the terminology of appendix \ref{appABA2} $\Psi^{(2)}$ is described by a real Bethe-root: 
$\lambda^{(2)}=\tfrac{\theta_1+\theta_2}{2}+i \tfrac{\pi}{2}$ and $\Psi^{(3)}$ is given by a self-conjugate root:
$\lambda^{(3)}=\tfrac{\theta_1+\theta_2}{2}+i \tfrac{(1+p) \pi}{2},$ provided we are in the repulsive $1<p$ regime of the 
theory. 

In the space of 3-particle states, we need the Bethe-eigenvectors only in the $Q=1$ sector. It has also a simple form:
\begin{equation} \label{3wave}
\begin{split}
\Psi_{i_1 i_2 i_3}=\Psi_{+--}\delta_{i_1 +} \delta_{i_2 -} \delta_{i_3 -}+
\Psi_{-+-}\delta_{i_1 -} \delta_{i_2 +} \delta_{i_3 -}+
\Psi_{+--}\delta_{i_1 -} \delta_{i_2 -} \delta_{i_3 +},
\end{split}
\end{equation}
where
\begin{equation} \label{Psiii}
\begin{split}
\Psi_{i_1 i_2 i_3}=\frac{C^{i_1 i_2 i_3}}{N_\Psi},
\end{split}
\end{equation}
such that $C^{i_1 i_2 i_3}$ is given by (\ref{BC123}) with (\ref{BBE}) and $N_\Psi$ is given by (\ref{3NPsi}).
Actually this vector stands for 3 eigenvectors, since it depends on a Bethe-root $\lambda_1,$ and the 
Bethe-equation (\ref{BBE}) have 3 independent solutions in this sector: a real one and two self-conjugate ones.
One can recognize that the vector $\Psi$ in (\ref{3wave}) differs by a complex conjugation with respect to the 
one enters the conjecture of \cite{Palmai13}. The reason is that at the study of the diagonal limit of the 
kinematical pole axiom (\ref{Fax3diag}), we recognized that one should sandwich the form-factor with the left 
 eigenvector of the transfer-matrix instead of its right eigenvector proposed earlier in \cite{Palmai13}.
Due to the hermiticity properties of the soliton transfer matrix (\ref{tauH}), 
this is just a complex conjugation at the level of the eigenvectors. Here the 3-particle wave vector $\Psi$ 
is complex and rapidity dependent, thus all these affairs matter. In the states upto 2-particles, 
which was studied in \cite{Palmai13} to check the conjecture, this complex conjugation problem didnot arise. 

Now, we know how the Bethe-eigenvectors we need are described by the roots of the magnonic Bethe-equations (\ref{ABAE}). 
This makes possible to write down the densities of the states (\ref{Rhot}) corresponding to the eigenstates under consideration. 
They can be simply read off from the formulas (\ref{dldh})-(\ref{tildeG}), which give the Bethe-Yang limit of the 
Gaudin-matrix.  We will specialize these formulas for the zero and one root cases
 of the 1-, 2- and 3-particle eigenstates. 
 First, we rewrite the matrix $\Phi$ (\ref{Phi}) for the zero and one-root states by emphasizing  the Bethe-root and particle number dependence better:
 %The matrix $\Phi$ for zero and one-root states takes the form: 
\begin{equation} \label{Phin}
\begin{split}
\Phi_{j,k}^{(n)}(\lambda)&=\left(\ell \cosh \theta_j+\sum\limits_{s=1}^n \tilde{G}_{j s}(\lambda) \right) \, \delta_{j k}-\tilde{G}_{j k}(\lambda), \\ 
\tilde{G}_{jk}(\lambda)&=G(\theta_j-\theta_k)+\frac{1}{i} \frac{V_j(\lambda) \, V_{k}(\lambda)}{\psi^{(n)}(\lambda)},
\quad \psi^{(n)}(\lambda)=\sum\limits_{j=1}^n V_j(\lambda), \\
V_j(\lambda)&=(\ln B_0)'(\lambda-\theta_j).
\end{split}
\end{equation}
For the zero root case the $\sim \frac{V_j(\lambda) \, V_{k}(\lambda)}{\psi^{(n)}(\lambda)}$ term must be skipped{\footnote{We denote this case by writing symbolically 
$\emptyset$ instead of $\lambda$ into the argument.}}. 

Then the densities of the states given by the Bethe-vectors (\ref{1wave}), (\ref{2wave}) and (\ref{3wave}) can be given by determinants of this matrix. 
Now, we list the necessary densities below. 
The densities (\ref{Rhot}) for the 1-particle states (\ref{1wave}) are the same as those of a free particle:
\begin{equation} \label{1partrho}
\begin{split}
\rho^{(\pm)}_1(1)\equiv\rho_1(1)=\ell \cosh \theta_1=\ell \, c_1, 
\end{split}
\end{equation}
where for short we introduce the notations $c_j=\cosh \theta_j$ and $s_j=\sinh \theta_j.$
The densities (\ref{Rhot}) for the 4-dimensional basis of 2-particle states (\ref{1wave}) are given by the determinants as follows: :
\begin{equation} \label{2rhos}
\begin{split}
\rho^{(1)}_2(1,2)&=\text{det}_{ \!\!\!\! \!\!\!\!\!\!\!{\atop 2 \times 2}} \Phi^{(2)}(\emptyset)=\ell^2 c_1 \, c_2+ \ell (c_1+c_2) G(\theta_{12}), \\
\rho^{(2)}_2(1,2)&=\text{det}_{ \!\!\!\! \!\!\!\!\!\!\!{\atop 2 \times 2}} \Phi^{(2)}(\lambda^{(2)})=\ell^2 c_1 \, c_2+ \ell (c_1+c_2) \tilde{G}_{12}(\lambda^{(2)}), \\
\rho^{(3)}_2(1,2)&=\text{det}_{ \!\!\!\! \!\!\!\!\!\!\!{\atop 2 \times 2}} \Phi^{(2)}(\lambda^{(3)})=\ell^2 c_1 \, c_2+ \ell (c_1+c_2) \tilde{G}_{12}(\lambda^{(3)}), \\
\rho^{(4)}_2(1,2)&=\text{det}_{ \!\!\!\! \!\!\!\!\!\!\!{\atop 2 \times 2}} \Phi^{(2)}(\emptyset)=\ell^2 c_1 \, c_2+ \ell (c_1+c_2) G(\theta_{12}),
\end{split}
\end{equation}
with
\begin{equation} \label{lam23}
\begin{split}
\lambda^{(2)}=\tfrac{\theta_1+\theta_2}{2}+i \tfrac{\pi}{2}, \qquad 
\lambda^{(3)}=\tfrac{\theta_1+\theta_2}{2}+i \tfrac{\pi(1+p)}{2}.
\end{split}
\end{equation}

Finally the density corresponding to the $Q=1$ sector of the 3-particle states is given by:
\begin{equation} \label{3rho}
\begin{split}
\rho_3^{(\Psi)}(1,2,3)
%(\theta_1,\theta_2,\theta_3)
&=\text{det}_{ \!\!\!\! \!\!\!\!\!\!\!{\atop 3 \times 3}} \Phi^{(3)}(\lambda_1),
\end{split}
\end{equation}
where $\lambda_1$ denotes the solution of the magnonic Bethe-equations (\ref{BBE}).
%$\text{det}_{ \!\!\!\! \!\!\!\!\!\!\!{\atop 2 \times 2}}\,\, \, \Phi$ 
Now we are in the position to check the conjecture of \cite{Palmai13} analytically in the 2-particle sector.

\subsection{Checking the conjecture for 2-particle states}

We make the test only in the $Q=0$ sector of the 2-particle space, since in the purely solitonic $Q=\pm 2$ sectors the conjecture has been verified 
for the operators $J_\mu$ and $\Theta$ and for any number of solitons in papers \cite{En} and \cite{En1}. In the sequel we compute the expectation values 
in the states described by the color wave functions $\Psi^{(2)}$ and $\Psi^{(3)}$ given in (\ref{2wave}). The first step in the computation is to determine the 
branching coefficients of these wave-functions with respect to the 1-particle color wave-functions of (\ref{1wave}). Due to the simple form of these vectors the branching 
coefficients of the decomposition (\ref{decomp}) of $\Psi^{(2)}$ and $\Psi^{(3)}$  can be read off immediately:
\begin{equation} \label{branch2}
\begin{split}\left.
\begin{array}{l}
C_{+-}^{(s)}=\tfrac{1}{\sqrt{2}}, \qquad C_{+-}^{(s)}=\tfrac{r_s}{\sqrt{2}}, 
\qquad \\
C_{--}^{(s)}=0, \qquad \quad C_{++}^{(s)}=0, \end{array} \right\}
\text{for} \quad s=2,3 \quad \text{with} \quad r_2=1, \quad r_3=-1.
\end{split}
\end{equation}
Now applying the conjectured formula (\ref{PTfv}) %of \cite{Palmai13} 
to the $Q=0$ states of the 2-particle space, one obtains:
\begin{equation} \label{FOs2}
\begin{split}
F^{{{\cal O}, (s)}}_2(\theta_1,\theta_2)=\langle {\cal O} \rangle_0+\frac{1}{\rho_2^{(s)}(1,2)}\left\{ 
F_{4,symm}^{{\cal O}, (s)}(1,2)+\frac{1}{2} F_{2,symm}^{{\cal O}, (+)}(1) \, \rho_1^{(-)}(2)+
\right.\\
\left. \frac{1}{2} F_{2,symm}^{{\cal O}, (-)}(1) \, \rho_1^{(+)}(2)+ 
\frac{1}{2} F_{2,symm}^{{\cal O}, (+)}(2) \, \rho_1^{(-)}(1)+
\frac{1}{2} F_{2,symm}^{{\cal O}, (-)}(2) \, \rho_1^{(+)}(1)
\right\}, \qquad s=2,3,
\end{split}
\end{equation}
where $\langle {\cal O} \rangle_0$ denotes the vacuum expectation value, the densities are given by (\ref{1partrho}) and (\ref{2rhos}) and 
the symmetric diagonal form-factors in the states  $\Psi^{(2)}$ and $\Psi^{(3)}$ can be determined from 
the formula (\ref{polform}) using the results (\ref{kernelsTH2})-(\ref{FJmu2sym}). 

It is easier to start testing the conjecture with the operator $J_\mu.$ Since $J_\mu$ is a charge conjugation negative operator all the symmetric form-factors 
entering (\ref{FOs2}) become zero. The vacuum expectation value is also zero because of the same reason $\langle J_\mu \rangle_0=0.$ Thus,  the conjecture of \cite{Palmai13} 
suggests that $F^{ J_\mu,(s)}_{2}(\theta_1,\theta_2)=0, \, \,$ for $s=2,3.$ Due to the charge conjugation negativity of the current it is true exactly, as well. 
Thus for $J_\mu$ the conjecture gives the expected  trivial result. 

For the trace of the stress energy tensor, due to the charge conjugation positivity{\footnote{Namely, in this case  
 form-factors are invariant with respect to conjugating their indexes.}}
 of the operator, (\ref{FOs2}) simplifies: 
\begin{equation} \label{FOs21}
\begin{split}
F_2^{\Theta,(s)}\!(\theta_1,\theta_2)\!=\!\!\langle {\Theta} \rangle_0\!+\!\!\frac{1}{\rho_2^{(s)}(1,\!2)}\!\!\left\{ 
\! F_{4,symm}^{{\Theta}\, (s)}(1,\! 2)\!+\!F_{2,symm}^{\Theta\, (+)}(1) \, \rho_1(2)
\!+ \!F_{2,symm}^{\Theta\, (+)}(2) \, \rho_1(1) \!\right\}\!,
 \, \, s\!=\!2,\!3,
\end{split}
\end{equation}
where the  symmetric diagonal 2-particle form-factor is a constant as it can be read off from (\ref{TH2F}):
\begin{equation} \label{TH2Fuj}
\begin{split}
F_{2,symm}^{\Theta, (\pm)}(1)={\cal M}^2,
\end{split}
\end{equation}
the necessary densities are listed in (\ref{1partrho}) and (\ref{2rhos}) and the 
symmetric diagonal 4-particle form-factors can be constructed from  (\ref{FTHpmpm}), (\ref{FTHpmmp})
by the prescription (\ref{polform}) and exploiting the charge conjugation positivity:
\begin{equation} \label{F4STH}
\begin{split}
F_{4,symm}^{{\Theta}, (s)}(1,2)\!=\!F^{\Theta, symm}_{+-+-}(\theta_1,\theta_2)+r_s \, F^{\Theta, symm}_{+--+}(\theta_1,\theta_2),
\end{split}
\end{equation}
with $r_s$ given in (\ref{branch2}). Using the identity 
\begin{equation} \label{azon}
\begin{split}
%\frac{1}{i} \frac{V_1(\lambda^{(s)}) \, V_2(\lambda^{(s)})}{\psi_0(\lambda^{(s)})}=
\tilde{G}_{12}(\lambda^{(s)})=
%2 \pi \left(
\Omega(\theta_{12})+r_s \,
\varphi(\theta_{12}) 
%\right)
, \qquad s=2,3,
\end{split}
\end{equation}
and the formulas (\ref{FTHpmpm}), (\ref{FTHpmmp}) 
the concrete form of these form-factors can be written in the form as follows:
\begin{equation} \label{F4THconc}
\begin{split}
F_{4,symm}^{{\Theta}, (s)}(1,2)\!=\!2 {\cal M}^2 (1+c_1 \, c_2-s_1 \, s_2) \,\tilde{G}_{12}(\lambda^{(s)}), \qquad s=2,3.
\end{split}
\end{equation}
Putting everything together one ends up with the final formula for the expectation value as follows: 
\begin{equation} \label{FOs2TH}
\begin{split}
F_2^{\Theta,(s)}\!(\theta_1,\theta_2)\!=\!\!\langle {\Theta} \rangle_0\!+\!\!\frac{1}{\rho_2^{(s)}(1,2)}\!\!\left\{ 
2 {\cal M}^2 (1\!+\!c_1 \, c_2\!-\!s_1 \, s_2) \,\tilde{G}_{12}(\lambda^{(s)})\!+\!
{\cal M}^2 \ell (c_1\!+\!c_2) \right\}, \quad s=2,3,
\end{split}
\end{equation}
which is exactly the same as the formula (\ref{THrest8}) coming from the exact result and specified to 
the single Bethe-root configurations $\lambda^{(2)}$ and $\lambda^{(3)}.$

The next step is to check the conjecture of \cite{Palmai13} in the $Q=1$ sector of the space of 
3-particle states. Here the computations are much more involved, this is why we will write down only the main steps 
and list the ingredients of the necessary computations.

\subsection{Checking the conjecture for 3-particle states}

The first step in the computation is the decomposition (\ref{decomp}) of the 3-particle wave function (\ref{3wave}) in terms of 1- (\ref{1wave}) 
and 2-particle (\ref{2wave}) wave functions. For a subset $A=\{A_1\}\subset\{1,..,3\}$ with a single element,  
the branching coefficients can be computed by the scalar product as follows:
\begin{equation} \label{branchc1}
\begin{split}
C_{st}(A)=\sum\limits_{i_1,i_2,i_3=\pm} \varphi_{i_{A_1}}^{(s) \, *} \psi_{i_{\bar{A}_1} i_{\bar{A}_2}}^{(t) *} \, \Psi_{i_1 i_2 i_3}, \qquad 
s\in\{+,-\}, \quad t\in\{1,2,3,4\}.
\end{split}
\end{equation}
Writing the analogous formula for the case, when $A\subset\{1,..,3\}$  has two elements, 
one obtains the relation:
\begin{equation} \label{Cstrel}
\begin{split}
C_{s t}(A)=C_{t s }(\bar{A}), \qquad s\in\{+,-\}, \quad t\in\{1,2,3,4\}, \qquad A\subset\{1,..,3\}.
\end{split}
\end{equation}
Thus from (\ref{branchc1}) and from the "color" wave functions (\ref{1wave}), (\ref{2wave}) and (\ref{3wave}), 
all the necessary branching coefficients can be determined:
\begin{equation} \label{Clist}
\begin{split}
C_{1+}(\{1,2\})&=C_{+1}(\{3\})=\Psi_{--+}, \quad C_{2-}(\{1,2\})=C_{-2}(\{3\})=\frac{\Psi_{+--}+\Psi_{-+-}}{\sqrt{2}}, \\
C_{3-}(\{1,2\})&=C_{-3}(\{3\})=\frac{\Psi_{+--}-\Psi_{-+-}}{\sqrt{2}}, \quad C_{1+}(\{1,3\})=C_{+1}(\{2\})=\Psi_{-+-}, \\
C_{2-}(\{1,3\})&=C_{-2}(\{2\})=\frac{\Psi_{--+}+\Psi_{+--}}{\sqrt{2}}, \quad C_{3-}(\{1,3\})=C_{-3}(\{2\})=\frac{\Psi_{+--}-\Psi_{--+}}{\sqrt{2}}, \\
C_{1+}(\{2,3\})&=C_{+1}(\{1\})=\Psi_{+--}, \quad C_{2-}(\{2,3\})=C_{-2}(\{1\})=\frac{\Psi_{--+}+\Psi_{-+-}}{\sqrt{2}}, \\
C_{3-}(\{2,3\})&=C_{-3}(\{1\})=\frac{\Psi_{-+-}-\Psi_{--+}}{\sqrt{2}},
\end{split}
\end{equation}
where $\Psi_{i_1 i_2 i_3}$ is given in (\ref{Psiii}). In the main formula (\ref{PTfv}) for the diagonal matrix elements, the absolute value squared of these 
coefficients arise. They can be expressed in terms of the elements of a Hermitian matrix $M$:  
\begin{equation} \label{Mform}
\begin{split}
M=\begin{pmatrix} \tfrac{V_1}{V_1+V_2+V_3} & \tfrac{C_1 C_2 B_3}{N_\Psi} & \tfrac{C_1 C_3}{N_\Psi} \\
\tfrac{C_1 C_2 }{N_\Psi} & \tfrac{V_2}{V_1+V_2+V_3} & \tfrac{B_1 C_2 C_3}{N_\Psi} \\
\tfrac{C_1 C_3 B_2 }{N_\Psi} & \tfrac{C_2 C_3 }{N_\Psi} & \tfrac{V_3}{V_1+V_2+V_3}
\end{pmatrix},
\end{split}
\end{equation}
in the following way{\footnote{We just recall: $B_j=B_0(\lambda_1-\theta_j), \quad C_j=C_0(\lambda_1-\theta_j), \quad V_j=(\ln B_0)'(\lambda_1-\theta_j).$}}:
\begin{equation} \label{Cnormlist}
\begin{split}
|C_{1+}(\{1,2\})|^2&\!=\!|C_{+1}(\{3\})|^2\!=\!M_{33}, \quad |C_{2-}(\{1,2\})|^2\!=\!|C_{-2}(\{3\})|^2\!=\!L_{12}^{(+)}, \\
|C_{3-}(\{1,2\})|^2&\!=\!|C_{-3}(\{3\})|^2\!=\! L_{12}^{(-)}, \quad |C_{1+}(\{1,3\})|^2\!=\!|C_{+1}(\{2\})|^2\!=\!M_{22} , \\
|C_{2-}(\{1,3\})|^2&\!=\!|C_{-2}(\{2\})|^2\!=\!L_{13}^{(+)}, \quad |C_{3-}(\{1,3\})|^2\!=\!|C_{-3}(\{2\})|^2\!=\! L_{13}^{(-)},  \\
|C_{1+}(\{2,3\})|^2&\!=\!|C_{+1}(\{1\})|^2\!=\! M_{11}, \quad |C_{2-}(\{2,3\})|^2\!=\!|C_{-2}(\{1\})|^2\!\!=\!L_{23}^{(+)},  \\
|C_{3-}(\{2,3\})|^2&\!=\!|C_{-3}(\{1\})|^2\!=\!L_{23}^{(-)},
\end{split}
\end{equation}
where we introduced the short notation: $L_{ij}^{(\pm)}=\tfrac{M_{ii}+M_{jj}\pm M_{ij}\pm M_{ji}}{2}.$

Now we have all the necessary ingredients to compare the conjectured formula (\ref{PTfv}) to the exact ones (\ref{THrest8})-(\ref{J18}) for the 
operators $\Theta$ and $J_\mu.$ Now, the computations become quite involved, thus 
they were performed by the  software {\emph{Mathematica}}. 
We just write down in words the strategy of the comparison. The $\tfrac{1}{\rho^{(\Psi)}_3}$ term naturally arises in the exact formulas (\ref{THrest8})-(\ref{J18}), if 
 the inverse Gaudin-matrix is expressed by the co-factor matrix ${\cal K}$:
\begin{equation} \label{Gaudinv}
\begin{split}
\Phi^{-1}_{jk}=\frac{{\cal K}_{kj}}{\det \Phi}=\frac{{\cal K}_{kj}}{\rho^{(\Psi)}_3(1,2,3)}.
\end{split}
\end{equation}
Then only the numerator of the conjectured formula (\ref{PTfv}) remains to be checked. Inserting all previously computed form factors and 
branching coefficients, it turns out the numerator is a second order polynomial in $\ell,$  such that the coefficients are composed of elementary functions 
multiplied by $G(\theta_{ij})$ transcendental terms{\footnote{The same structure arose in the 
2-particle case. See (\ref{FOs2TH}).}}. It is easy to see that this "transcendental" structure is the same for both the conjecture (\ref{PTfv}) and 
the exact results: (\ref{THrest8})-(\ref{J18}). The coefficients of these transcendental terms are complicated combinations of elementary functions containing the 
single Bethe-root $\lambda_1$ in the argument. These coefficients  do not seem to match for the first sight, but exploiting the Bethe-equations finally it turns out that 
they agree. Thus upto 3-particle states in the sine-Gordon model, 
the P\'almai-Tak\'acs conjecture \cite{Palmai13} gives the correct result for Bethe-Yang limit of the diagonal matrix elements of the $U(1)$ current and the trace of the stress energy tensor, provided one 
modifies the definition of polarized form-factors from the original form (\ref{polform}) to (\ref{polformen}). 
This simple modification corresponds to a $\Psi \to \Psi^*$ exchange in the original definition of \cite{Palmai13}.
%makes a $\Psi \to \Psi^*$ exchange in their definition of polarized form factors (\ref{polform}). 

%\tfrac{M_{11}+M_{33}+M_{13}+M_{31}}{2}

%C_{}(\{\})

\section{Comments on some subtle points of the conjecture of \cite{Palmai13}} \label{sect9}

In the previous section for the operators $\Theta$ and $J_\mu,$ 
we checked the conjecture of \cite{Palmai13} for the Bethe-Yang limit of 
expectation values of local operators upto 3-particle states. The agreement found between the conjectured and 
the exact results seems to be a convincing evidence for the correctness of this conjecture. 
Nevertheless, in this section we would like to pay the attention to some delicate points of the conjecture, which 
require further work to be confirmed. 
 These two subtle points are the existence of 
the symmetric diagonal limit of form-factors and the order of rapidities in the Bethe-wave functions. 

\subsection{Existence of symmetric diagonal limit of form-factors in a non-diagonally scattering theory}

In this section we argue, that the existence of symmetric diagonal limit of form-factors in a non-diagonally 
scattering theory is not obvious at all. 
Apparently, we cannot prove the existence of this limit in general,  
thus it cannot be excluded, that this limit is divergent in most of the 
cases. 

We start our argument by writing down the kinematical singularity axiom (\ref{ax4}) in the limit, when only the rapidities of the sandwiching states are close to each other. Similarly to (\ref{Fax3diag}) we formulate the axiom 
on the basis of the Bethe-eigenvectors. Let $\Psi$ and $\Phi^{(\epsilon)}$ be the color wave functions corresponding to 
the two states sandwiching the operator. Thus, $\Psi$ is a left eigenvector of the soliton transfer-matrix 
(\ref{trans}) with rapidity parameters: $\vec{\theta}=\{\theta_1,..,\theta_n\}$ and 
$\Phi^{(\epsilon)*}$ is a right eigenvector of the soliton transfer-matrix 
 with rapidity parameters{\footnote{We just recall the notation used in the preceding sections: 
$\theta_j^{\epsilon}=\theta_j+\epsilon_j,\quad j=1,..,n.$}}: $\vec{\theta^{\epsilon}}=\{\theta^{\epsilon}_1,..,\theta^{\epsilon}_n\}.$
Then they satisfy the eigenvalue equations:
\begin{equation} \label{Pshe}
\begin{split}
\Psi^{i_1...i_n }\tau(\theta_1|\vec{\theta})_{i_1 i_2...i_n}^{l\alpha_2...\alpha_n}&=\Lambda_\Psi(\theta_1|\vec{\theta}) \Psi^{l\alpha_2...\alpha_n}, \\
\tau(\theta_1^\epsilon|\vec{\theta^\epsilon})_{l \bar{\beta}_2...\bar{\beta}_n}^{j_1 j_2...j_n} \Phi^{(\epsilon) *}_{j_1..j_n}&=\Lambda_\Phi(\theta_1^\epsilon|\vec{\theta^\epsilon})
\Phi^{(\epsilon) *}_{l \bar{\beta}_2...\bar{\beta}_n}.
\end{split}
\end{equation}
With these sandwiching states the kinematical singularity axiom takes the form: 
\begin{equation} \label{Fax3sh}
\begin{split}
F_{\Phi\Psi}(\hat{\theta}_n,.,\hat{\theta}_1,\theta_1,.,\theta_n)\!\!&
=\!\frac{i}{\epsilon_1}\!\!\left(\!1\!-\!\frac{\Lambda_\Psi(\theta_1|\vec{\theta})}{\Lambda_\Phi(\theta_1^\epsilon|\vec{\theta^\epsilon})}\!\right) \Phi^{(\epsilon)*}_{k \bar{\beta}_2...\bar{\beta}_n} \! \Psi^{k \alpha_2...\alpha_n}\,
\!F_{\beta_n...\beta_2 \alpha_2...\alpha_n}\!(\hat{\theta}_n,..,\hat{\theta}_2,\theta_2,..,\theta_n)
\\&+O(1)_{\epsilon_1}
\end{split}
\end{equation}
where in accordance with the definition (\ref{polform}), $F_{\Phi\Psi}$ denotes the form-factor polarized with the 
Bethe-vectors $\Phi$ and $\Psi:$ 
\begin{equation} \label{Fsh}
\begin{split}
&F_{\Phi\Psi}(\hat{\theta}_n,...,\hat{\theta}_1,\theta_1,...,\theta_n)=
\sum\limits_{b_1,..,b_n=\pm} \, \sum\limits_{a_1,..,a_n=\pm} \Phi_{b_1...b_m}^{(\epsilon)*}(\theta^\epsilon_1,..,\theta^\epsilon_n) \times \\
&F_{\bar{b}_n...\bar{b}_1 a_1...a_n}(\theta^\epsilon_n+i \, \pi,...,\theta^\epsilon_1+i \, \pi,\theta_1,...,\theta_n) \,
\Psi_{a_1...a_n}(\theta_1,..,\theta_n) .
\end{split}
\end{equation}
Again, we used the short notation: $\hat{\theta}_j=\theta_j+\epsilon_j+i \pi.$
Formula (\ref{Fax3sh}) has serious implications on the existence of the symmetric diagonal limit of 
form-factors in a non-diagonally scattering theory. 

If the theory is of purely elastic scattering, than there is no index structure in (\ref{Fax3sh}). 
Thus $\Phi=\Psi=1$ and $\Lambda_\Phi(\theta_1^\epsilon|\vec{\theta^\epsilon})\to\Lambda_\Psi(\theta_1|\vec{\theta})
$ can be written. In this case the prefactor $\frac{i}{\epsilon_1}\!\!\left(\!1\!-\!\frac{\Lambda_\Psi(\theta_1|\vec{\theta})}
{\Lambda_\Psi(\theta_1^\epsilon|\vec{\theta^\epsilon})}\!\right)$ 
becomes $O(1)$ in $\epsilon,$ which imples that the symmetric diagonal 
limit always exist (finite). Moreover the exchange axiom (\ref{ax2}) ensures, that the limiting form-factor is a 
symmetric function of the rapidities.  

If the theory is of non-diagonally scattering the prefactor  $\frac{i}{\epsilon_1}\!\!\left(\!1\!-\!\frac{\Lambda_\Psi(\theta_1|\vec{\theta})}{\Lambda_\Phi(\theta_1^\epsilon|\vec{\theta^\epsilon})}\!\right)$ in (\ref{Fax3sh}) is not always $O(1)$ in $\epsilon!$ 
Moreover it is always divergent if the vectors $\Psi$ and $\Phi^{(\epsilon)}\big|_{\epsilon=0}$ are not equal. 
This means, that the near diagonal in rapidity limit of the form-factors is divergent if the matrix element is 
nondiagonal in the color space. 
On the other hand, the prefactor  
$\frac{i}{\epsilon_1}\!\!\left(\!1\!-\!\frac{\Lambda_\Psi(\theta_1|\vec{\theta})}{\Lambda_\Phi(\theta_1^\epsilon|\vec{
\theta^\epsilon})}\!\right)$ in (\ref{Fax3sh}) has a finite value in $\epsilon \to 0$ limit, 
if{\footnote{In the $\epsilon \to 0$ limit.}}
 both sandwiching states correspond to the same eigenstate of the soliton transfer-matrix.
Nevertheless, this fact alone doesnot guarantee, that the symmetric diagonal limit of the form-factors would be finite. This is so, because apart form the prefactor we analyzed, there is another term in (\ref{Fax3sh}), a sum of 
the near diagonal in rapidity limit of form-factors with all polarizations, weighted by the color wave-functions. 
We argued in the previous lines, that the near diagonal in rapidity limit of form-factors is divergent in general, 
which means that this sum is composed of divergent terms in the $\epsilon \to 0$ 
limit. {Actually, the degree of divergence in $\epsilon$ increases with the number of sandwiching 
particles.} 
To get finite result for these matrix elements very nontrivial cancellations must occur!
Actually such nontrivial cancellations happened, when we computed the symmetric diagonal limit 
of the 3-particle form-factors of the current in section \ref{sect7}.

Nevertheless, our conclusion is that the symmetric diagonal limit of form-factors in a non-diagonally scattering 
theory is not obviously finite. Thus, to trust the conjecture of \cite{Palmai13} 
beyond 3-particle states, it would be necessary to prove 
that the symmetric diagonal limit of form-factors  exists for generic states, as well.

\subsection{The order of rapidities}

The next delicate point in the conjecture of \cite{Palmai13} is the matter of the order of rapidities. 
Here we will not state, that there might be problems with the conjecture, but rather we would like to 
shed light on the fact, that the so far achieved analytical tests of this paper are still not enough 
to confirm certain parts of the conjecture. This unconfirmed part is how to do correctly the 
color-wave function decomposition (\ref{decomp}). The issue here is that in general these wave functions 
do depend on the particle's rapidities. What's more they do depend on their orderings, as well. 
In this paper we did computations upto 3-particle matrix elements. Thus 3-particle wave functions must 
have been decomposed with respect to 1- and 2-particle color wave functions. But, as it is emphasized in section 
\ref{sect8}, incidentally the 1- and 2-particle color wave functions are independent of the rapidities. 
Consequently, our computations cannot confirm, whether the ordering of rapidities 
in the arguments of the wave-functions in the right hand side of the decomposion 
formula (\ref{decomp})  is correct if more than three particle states are considered. 

A possible reassuring solution to this problem could be, if one could prove that the conjectured formula of 
\cite{Palmai13} is invariant under any permutations of the rapidities of the sandwiching state.

%The deficiency of these tests 

\section{Summary and conclusion} \label{sect10}

In this paper we consider two important local operators of the sine-Gordon theory; 
the trace of the stress energy tensor and the $U(1)$ current. 

We showed, that the finite volume expectation values of these operators in any eigenstate of the Hamiltonian of the model, 
can be expressed in terms of solutions of sets of linear integral equations (\ref{ujset})-(\ref{bfGdef}).
The large volume solution of these equations allowed us to get analytical formulas in the repulsive regime for the Bethe-Yang limit 
of these diagonal matrix elements. These formulas are expressed in terms of the Bethe-roots characterizing the 
corresponding eigenstate of the soliton transfer matrix (\ref{trans}). This analytical formula allowed us to check a
former conjecture \cite{Palmai13} for the Bethe-Yang limit of expectation values of local operators in a non-diagonally scattering theory.
We computed all expectation values upto 3-particle states both from our analytical formulas and from the conjectured formula of \cite{Palmai13}, 
and we found perfect agreement between the results of the two different computations. To be more precise to get agreement we had to make a 
tiny modification in the conjectured formula of \cite{Palmai13}. Namely, we had to change slightly the definition of the symmetric diagonal form-factors, 
which are basic building blocks of the formula. In the conjecture of \cite{Palmai13} they are defined as appropriately (\ref{polform}) polarized sandwiches of the 
form-factors with right eigenvectors of the soliton transfer matrix. However, from our computations it turns out that they should be defined as 
polarized sandwiches of the form-factors with left eigenvectors of the soliton transfer matrix. Since upto 2-particle states the left and right eigenvectors 
are the same, this issue arises first at the level of 3-particle states, which were not tested in the original paper \cite{Palmai13}.

Despite the success of the 3-particle checks, there are still some subtle points of the conjecture, which 
could not be confirmed by our analytical computations. First of all,  
the finiteness of the symmetric diagonal limit of form-factors for a generic state  in a non-diagonally scattering theory is still unproven. Second, the hereby performed analytical tests were still not sensible to some details of the conjectured formula. Namely, upto 3-particle states 
our computations could not check the correctness of the rapidity dependence of the eigenvectors entering 
the right hand side of the decomposition rule (\ref{decomp}), 
since all 1- and 2-particle wave-functions are incidentally independent of the rapidities.

Thus our final conclusions are as follows. Our analytical checks gave very strong support for the 
validity of the conjectured formula of \cite{Palmai13} for the Bethe-Yang limit of expectation values in non-diagonally scattering theories. 
%To be more precise, 
Our computations suggest, that the conjecture is well established upto 3-particle states, 
but to firmly trust it beyond 3-particle states, two further statements should be proven.
 First, it should be proven that the symmetric diagonal limit of form-factors is finite in a non-diagonally scattering theory, as well. 
Second, to get some more confidence about whether the rapidity dependence of wave functions is correctly embedded into the conjectured formula, 
one should also prove that the conjectured formula is invariant with respect to the permutations of the rapidities of the sandwiching state. 

Nevetheless, the fact that the conjecture of \cite{Palmai13} was found to be correct at least upto 3-particle states, 
opens the door to safely apply it to compute 
finite temperature correlators \cite{PT10T}, and various
one-point functions \cite{BucTak,Bert14,CortCub17,HKT},  
by their form-factor series representations upto 3-particle contributions.

%%%%%%%%%%%%%%%%%%%%%%%%%%%%%%%%%%%%%%%%%%%%%%%%%%%%%%%%%%%%%%%%%%%%%%%%%%%%%%%%%%%%%%%%%%%%%%%%%%%%%%%%%%%%%%%%%%%%%%%%%%%%%
%%%%%%%%%%%%%%%%%%%%%%%%%%%%%%%%%%%%%%%%%%%%%%%

%%%%%%%%%%
%%%%%%
%%%%%%%%%%%
%%%%%%%
%%%%%%%
%%%%%%%%

\vspace{1cm}
{\tt Acknowledgments}

\noindent 
The author would like to thank Zolt\'an Bajnok and J\'anos Balog for useful discussions and
G\'abor Tak\'acs for his useful comments on the manuscript.  
This work was supported by the Hungarian Science Fund OTKA (under K116505) and by an MTA-Lend\"ulet Grant.

\appendix

\section{Integral equations for the derivatives of the counting-function} \label{appA}

In this appendix we write down the linear integral equations satisfied by the $\theta-$ and $\ell-$ derivatives of the 
counting-function. The equations we list below can be obtained by differentiating the NLIE (\ref{DDV})-(\ref{DDVII}).  
The equations related to the derivative of $Z(\theta|\ell)$ with respect to $\theta$ (\ref{Gd}) and $\ell$ (\ref{Gl}) can be written in an incorporated way, because the equations for the two different derivatives differ only in a single source 
term. To have a more compact representation of the equations it is useful to pack all complex roots into a single set: 
\begin{equation} \label{ujset}
\begin{split}
\{u_j\}_{j=1}^{m_K}=\{c_j\}_{j=1}^{m_C} \cup \{w_k\}_{k=1}^{m_W}, \qquad m_K=m_C+m_W, 
\end{split}
\end{equation}
such that
\begin{equation} \label{ujindex}
\begin{split}
u_j&=c_j, \qquad j=1,...,m_C, \\
u_{m_C+j}&=w_j, \qquad j=1,...,m_W.
\end{split}
\end{equation}
%It follows the same index correspondence for the $X$ variables of (\ref{Gd}) and (\ref{Gl}):
In accordance with (\ref{ujindex}), from (\ref{Gd}) and (\ref{Gl}) we define the corresponding $X$ variables as well:
\begin{equation} \label{Xu}
\begin{split}
X^{(u)}_{\nu,j}&=X^{(c)}_j, \qquad \nu\in\{d,\ell\}, \quad j=1,...,m_C, \\
X^{(u)}_{\nu,m_C+j}&=X^{(w)}_j, \qquad \nu\in\{d,\ell\}, \quad j=1,...,m_W.
\end{split}
\end{equation}
Using this notation the linear integral equations take the form:
\begin{equation} \label{Gnu}
\begin{split}
{\cal G}_\nu(\theta)\!=\!f_\nu(\theta)\!+\!\sum\limits_{j=1}^{m_H} {\bf G}(\theta,h_j) X^{(h)}_{\nu,j}\!-\!
\sum\limits_{j=1}^{m_S} \left( {\bf G}(\theta,y_j-i \eta)+{\bf G}(\theta,y_j+i \eta)\right) X^{(y)}_{\nu,j}
- \\
\sum\limits_{j=1}^{m_K} {\bf G}(\theta,u_j) X^{(u)}_{\nu,j}\!+\!
\sum\limits_{\alpha=\pm} \! \int\limits_{-\infty}^{\infty} \!\! \frac{d \theta'}{2 \pi} 
{\bf G}(\theta,\theta'-i \, \alpha \, \eta) {\cal G}_{\nu}(\theta'+i \, \alpha \, \eta) \, 
{\cal F}_{\alpha}(\theta'+i\, \alpha \, \eta),
\end{split}
\end{equation}
\begin{equation} \label{Qhk}
\begin{split}
&\sum\limits_{k=1}^{m_H} \left[ Z'(h_j) \delta_{jk}-{\bf G}(h_j,h_k)\right]X_{\nu,k}^{(h)}\!=\!
f_\nu(h_j)\!-\!
\sum\limits_{k=1}^{m_S}\left({\bf G}(h_j,y_k+i \eta)+{\bf G}(h_j,y_k-i\eta)\right) X_{\nu,k}^{(y)}
\!-\! \\
&\sum\limits_{k=1}^{m_K}\!{\bf G}(h_j,u_k) X_{\nu,k}^{(u)}\!+\!\!
\sum\limits_{\alpha=\pm} \! \int\limits_{-\infty}^{\infty} \!\!\! \frac{d \theta'}{2 \pi} 
{\bf G}(h_j,\theta'\!-\!i \, \alpha \, \eta) {\cal G}_{\nu}(\theta'\!+\!i\, \alpha \,\eta) \, 
{\cal F}_{\alpha}(\theta'\!+\!i\, \alpha \, \eta), \qquad j\!=\!1,..,m_H,
\end{split}
\end{equation}
\begin{equation} \label{Quk}
\begin{split}
&\sum\limits_{k=1}^{m_K} \left[ Z'(u_j) \delta_{jk}+{\bf G}(u_j,u_k)\right]X_{\nu,k}^{(u)}\!=\!
f_\nu(u_j)\!-\!
\sum\limits_{k=1}^{m_S}\left({\bf G}(u_j,y_k+i \eta)+{\bf G}(u_j,y_k-i\eta)\right) X_{\nu,k}^{(y)}
\!+\! \\
&\sum\limits_{k=1}^{m_H}\!{\bf G}(u_j,h_k) X_{\nu,k}^{(h)}\!+\!\!
\sum\limits_{\alpha=\pm} \! \int\limits_{-\infty}^{\infty} \!\! \frac{d \theta'}{2 \pi} 
{\bf G}(u_j,\theta'\!-\!i\, \alpha \,\eta) {\cal G}_{\nu}(\theta'\!+\!i \,\alpha \,\eta) \, 
{\cal F}_{\alpha}(\theta'\!+\!i \, \alpha\, \eta), \qquad j\!=\!1,..,m_K,
\end{split}
\end{equation}
\begin{equation} \label{Qyk}
\begin{split}
&\sum\limits_{k=1}^{m_S} \left[ Z'(y_j) \delta_{jk}+{\bf G}(y_j,y_k+i \eta)
+{\bf G}(y_j,y_k-i \eta)\right]X_{\nu,k}^{(y)}\!=\!
f_\nu(y_j)\!+\!
\sum\limits_{k=1}^{m_H}\!{\bf G}(y_j,h_k) X_{\nu,k}^{(h)}
\!-\! \\
&\sum\limits_{k=1}^{m_K}\!{\bf G}(y_j,u_k) X_{\nu,k}^{(u)}\!+\!\!
\sum\limits_{\alpha=\pm} \! \int\limits_{-\infty}^{\infty} \!\! \frac{d \theta'}{2 \pi} 
{\bf G}(y_j,\theta'\!-\!i\, \alpha \,\eta) {\cal G}_{\nu}(\theta'\!+\!i \,\alpha \,\eta) \, 
{\cal F}_{\alpha}(\theta'\!+\!i \, \alpha\, \eta), \qquad j\!=\!1,..,m_S,
\end{split}
\end{equation}
where $\eta$ is a positive contour deformation parameter{\footnote{If $m_S\neq 0$ it is preferable to consider 
$\eta$ to be a positive infinitesimal parameter.}}
 such that $\eta<\text{min} (p , p\, \pi,|\text{Im}\, u_j|),$ 
${\cal F}_\pm(\theta)$ is defined in (\ref{calF}),  
the index $\nu$ can be either $d$ or $\ell$ telling us which derivative of $Z(\theta)$ is considered.
The source term $f_\nu(\theta)$ for the two choices of the index $\nu$ is given by the formulas: 
\begin{equation} \label{fd}
\begin{split}
f_{d}(\theta)=\left\{ 
\begin{array}{r}
\ell \cosh( \theta), \qquad \qquad \qquad \quad 
\qquad |\text{Im} \theta|\leq\text{min}(\pi, p \pi), \\
\ell \cosh_{II}(\theta) \quad 
\qquad \text{min}(\pi, p \pi)<|\text{Im} \theta|\leq \tfrac{\pi}{2}(1+p),
\end{array}\right.
\end{split}
\end{equation}
\begin{equation} \label{fl}
\begin{split}
f_{\ell}(\theta)=\left\{ 
\begin{array}{r}
\sinh( \theta), \qquad \qquad \qquad \quad 
\qquad |\text{Im} \theta|\leq\text{min}(\pi, p \pi), \\
 \sinh_{II}(\theta) \quad 
\qquad \text{min}(\pi, p \pi)<|\text{Im} \theta|\leq \tfrac{\pi}{2}(1+p),
\end{array}\right.
\end{split}
\end{equation}
where the second determination of a function is defined by (\ref{fII}).
The function ${\bf G}(\theta,\theta')$ in the equations (\ref{Gnu})-(\ref{Qyk}) agrees with $G(\theta-\theta')$ 
of (\ref{G}) in the fundamental domain and it is equal to the appropriate second determination of $G(\theta)$ if 
either of its arguments goes out of the fundamental domain $|\text{Im} (\theta-\theta')|\leq\text{min}(\pi, p \pi)$.
In the sequel we give the precise prescription, how to compute ${\bf G}(\theta,\theta')$ for any pair of values 
of its arguments. In this way we can get rid of the possible errors which can be easily committed when multiple 
 second determination of a function should be done. 
The function ${\bf G}(\theta,\theta')$ will be defined as the solution of a linear integral equation.  
Let:
\begin{equation} \label{K}
\begin{split}
K(\theta)=\frac{1}{p+1} \, 
\frac{\sin\tfrac{2 \pi}{p+1}}{\sinh\tfrac{\theta-i \pi}{p+1} \, \sinh\tfrac{\theta+i \pi}{p+1}}.
\end{split}
\end{equation}
This function is the  derivative of the scattering-phase of the elementary magnon excitations 
of the 6-vertex model with anisotropy parameter $\gamma=\tfrac{\pi}{p+1}.$ 
Then ${\bf G}(\theta,\theta')$ for arbitrary values of $\theta$ and $\theta'$ can be determined by solving the 
linear integral equation as follows:
\begin{equation} \label{bfGdef}
\begin{split}
{\bf G}(\theta,\theta')+\!\!\int\limits_{-\infty}^{\infty} \!\frac{d \theta''}{2 \pi} K(\theta-\theta'')\,
{\bf G}(\theta'',\theta')=K(\theta-\theta').
\end{split}
\end{equation}
This equation can be solved  by means of Fourier transformation along any horizontal lines of the complex plane. 
When both arguments are in the fundamental domain: $\text{max} \{|\text{Im} (\theta)|,\, |\text{Im} \theta')|\}\leq\text{min}(\pi, p \pi),$ then the solution of (\ref{bfGdef}) gives the well known kernel 
 of the NLIE of the sine-Gordon theory. Namely, 
${\bf G}(\theta,\theta')=G(\theta-\theta')$ with $G(\theta)$ given in (\ref{G}).  
The linear integral equation (\ref{bfGdef}) tells us how to continuate analytically ${\bf G}(\theta,\theta')$ out of this 
fundamental regime. For example, if one continues one of the variables of ${\bf G}(\theta,\theta')$ out of the fundamental domain, then one gets the second determination of $G(\theta-\theta')$ defined by (\ref{fII}) etc. 
Thus the function ${\bf G}(\theta,\theta')$ incorporates all possible second determinations which appear in the 
NLIE (\ref{DDV})-(\ref{DDVII}) of the model. This means that one does not need to take care of the subtle 
rules of second determination, but the solution of (\ref{bfGdef})
 will automatically give the functional form of ${\bf G}$ in any regime of the complex plane.

\section{Algebraic Bethe Ansatz for the soliton transfer matrix} \label{appB}

The monodromy and transfer matrices  made 
out of the S-matrix (\ref{Smatr}) of the sine-Gordon model are of central importance in this paper. 
They enter the form-factor axiom (\ref{ax4}) and play an important role in the conjecture of \cite{Palmai13} for the 
diagonal matrix elements of local operators of the theory. 

In this appendix we summarize the most important properties of the monodromy matrix and  
recall the Algebraic Bethe Ansatz \cite{FST79} diagonalization of the transfer matrix. 

The basic object is the $n$-particle monodromy matrix built from the S-matrix of the model (\ref{Smatr}): 
\begin{equation} \label{monodr}
\begin{split}
{\cal T}_a^b(\theta|\theta_1,...,\theta_n)_{a_1 a_2 ... a_n}^{b_1 b_2 ...b_n}=
{\cal S}_{a \, a_1}^{k_1 \, b_1}(\theta-\theta_1)
{\cal S}_{k_1 \,  a_2}^{k_2 \, b_2}(\theta-\theta_2)...
{\cal S}_{k_{n-1}\,  a_n}^{b \, b_n}(\theta-\theta_n).
\end{split}
\end{equation}
For the algebraic Bethe Ansatz techniques, it is generally written as a 2 by 2 matrix in the auxiliary space:
\begin{equation} \label{ABCD}
\begin{split}
{\cal T}(\theta|\vec{\theta})=
\begin{pmatrix} {\cal T}_-^-(\theta|\vec{\theta}) & {\cal T}_-^+(\theta|\vec{\theta}) \\
{\cal T}_+^-(\theta|\vec{\theta}) & {\cal T}_+^+(\theta|\vec{\theta})\end{pmatrix}
=\begin{pmatrix} A(\theta|\vec{\theta}) & B(\theta|\vec{\theta}) \\
C(\theta|\vec{\theta}) & D(\theta|\vec{\theta})
\end{pmatrix},
\end{split}
\end{equation}
such that the entries act on the $2^n$ dimensional vector space spanned by $n$ soliton dublets
${\cal V}_n=\left({\mathbb C}^2\right)^{\otimes n} $.
Here for short we introduced the notation $\vec{\theta}=\{\theta_1,\theta_2,...,\theta_n\}.$

As a consequence of the Yang-Baxter equation (\ref{YBE}), the entries of the monodromy matrix satisfy the 
Yang-Baxter algebra relations: 
\begin{equation} \label{YBA}
\begin{split}
{\cal S}_{a_1 \, a_2}^{k_1 \,k_2}(\theta-\theta') \,
{\cal T}_{k_1}^{b_1}(\theta|\vec{\theta}) \,
{\cal T}_{k_2}^{b_2}(\theta'|\vec{\theta})=
{\cal T}_{a_2}^{k_1}(\theta'|\vec{\theta}) \, 
{\cal T}_{a_1}^{k_2}(\theta|\vec{\theta}) \,
{\cal S}_{k_1 \, k_2}^{b_2 \,b_1}(\theta-\theta').
\end{split}
\end{equation}
The transfer matrix is defined as the trace of the monodromy matrix over the auxiliary space:
\begin{equation} \label{trans}
\begin{split}
\tau(\theta|\vec{\theta})=\sum\limits_{a=\pm} {\cal T}_a^a(\theta|\vec{\theta}). 
\end{split}
\end{equation}
As a consequence of (\ref{YBA}) the transfer matrices form a commuting family of operators on ${\cal V}_n$:
\begin{equation} \label{transcom}
\begin{split}
\tau(\theta|\vec{\theta}) \, \tau(\theta'|\vec{\theta})=\tau(\theta'|\vec{\theta})\, \tau(\theta|\vec{\theta}).
\end{split}
\end{equation}
This means that the eigenvectors of the transfer matrices are independent of the spectral parameter $\theta,$ 
but the they do depend  on the inhomogeneity vector $\vec{\theta},$ such that the order of rapidities within 
this vector matters, as well! 
The transfer matrix commutes with the solitonic charge ${\cal Q}$ and the charge parity ${\cal C}$ operators, 
which act on a vector $V\in{\cal V}_n$ as follows:
\begin{equation} \label{calQ}
\begin{split}
({\cal Q} V)_{i_1 \, i_2 ....i_n}=Q \, V_{i_1 \, i_2...i_n}\, \qquad Q=\sum\limits_{k=1}^n i_k, \qquad 
\end{split}
\end{equation}
\begin{equation} \label{calC}
\begin{split}
({\cal  C}V)_{i_1\, i_2...i_n}=V_{\bar{i}_1 \, \bar{i}_2...\bar{i}_n}, \qquad \text{with} \quad \bar{i}_k=-i_k, \qquad 
k=1,...,n.
\end{split}
\end{equation}
The $B(\theta|\vec{\theta})$ and $C(\theta|\vec{\theta})$ elements of the monodromy matrix act as 
charge raising and lowering operators: 
\begin{equation} \label{QB}
\begin{split}
[{\cal Q},B(\theta|\vec{\theta})]=2 \, B(\theta|\vec{\theta}),
\end{split}
\end{equation}
\begin{equation} \label{QC}
\begin{split}
[{\cal Q},C(\theta|\vec{\theta})]=-2 \, C(\theta|\vec{\theta}).
\end{split}
\end{equation}
The diagonalization of the transfer matrix can be done using the usual procedure of the 
Algebraic Bethe Ansatz \cite{FST79}. 
There exist a trivial eigenvector of $\tau(\lambda|\vec{\theta}),$ the pure antisoliton state:
\begin{equation} \label{trivvac}
\begin{split}
|0\rangle_{a_1 \, a_2...a_n}=\prod\limits_{j=1}^n \delta_{a_j}^-.
\end{split}
\end{equation}
Then the eigenvectors of the transfer matrix are given by acting a sequence of $B$-operators on this 
trivial eigenstate:
\begin{equation} \label{eivec1}
\begin{split}
\Psi(\{\lambda_j\}|\vec{\theta})=\frac{1}{{\cal N}_\Psi}\, B(\lambda_1|\vec{\theta})\, B(\lambda_2|\vec{\theta})...
B(\lambda_r|\vec{\theta})|0\rangle,
\end{split}
\end{equation}
such that the $\lambda_j$ spectral parameters of the $B$-operators satisfy the Bethe-equations as follows:
\begin{equation} \label{ABAE}
\begin{split}
\prod\limits_{k=1}^n B_0(\lambda_j-\theta_k)=\prod\limits_{k\neq j}^r \frac{B_0(\lambda_k-\lambda_j)}{B_0(\lambda_j-\lambda_k)}, \qquad j=1,..,r.
\end{split}
\end{equation}
The term ${\cal N}_{\Psi}$ in (\ref{eivec1}) is to fix the norm of the state to the required value.
In our computations the normalization condition for ${\cal N}_\Psi$ is that the norm  of $\Psi$ should be $1.$ 
%we need eigenvectors normalized to $1,$ thus we will choose the value $|\Psi|^2$ for ${\cal N}_\Psi.$ 
The eigenvalue of the transfer matrix on the state $\Psi(\{\lambda_j\}|\vec{\theta});$
\begin{equation} \label{taupsi}
\begin{split}
\tau(\lambda|\vec{\theta}) \, \Psi(\{\lambda_j\}|\vec{\theta})=\Lambda(\lambda,\{\lambda_j\}|\vec{\theta})\, \Psi(\{\lambda_j\}|\vec{\theta}),
\end{split}
\end{equation}
is given by the formula:
\begin{equation} \label{eival}
\begin{split}
\Lambda(\lambda,\{\lambda_j\}|\vec{\theta})=\prod\limits_{j=1}^n S_0(\lambda-\theta_k)\, \Lambda_0(\lambda,\{\lambda_j\}|\vec{\theta}),
\end{split}
\end{equation}
where $S_0(\theta)$ is given in (\ref{CHI}) and $\Lambda_0(\lambda,\{\lambda_j\}|\vec{\theta})$ is the eigenvalue 
of the transfer matrix made out of $S_{ab}^{cd}(\theta)$ and it is given by:
\begin{equation} \label{Lambda0}
\begin{split}
\Lambda_0(\lambda,\{\lambda_j\}|\vec{\theta})=\prod\limits_{j=1}^r \frac{1}{B_0(\lambda_j-\lambda)}+
\prod\limits_{k=1}^n B_0(\lambda-\theta_k) \,
\prod\limits_{j=1}^r \frac{1}{B_0(\lambda-\lambda_j)}.
\end{split}
\end{equation}
In the computations of the paper we need an analogous to (\ref{eivec1}) expression for the complex conjugate vector 
 of $\Psi(\{\lambda_j\}|\vec{\theta}),$ too. 
For this reason we need the properties of the monodromy and transfer matrices under hermitian conjugation. 
From the properties (\ref{Bose})-(\ref{Real}) of the S-matrix and from the definition (\ref{monodr}) one can prove 
the following hermitian conjugation rule for the monodromy matrix: 
\begin{equation} \label{monoHerm}
\begin{split}
{\cal T}_{a}^b(\lambda|\vec{\theta})^\dag={\cal T}_{\bar a}^{\bar b}(\lambda^*+i \, \pi|\vec{\theta}),
\end{split}
\end{equation}
which implies for the components the following rules:
\begin{equation} \label{ABCDH}
\begin{split}
A^\dag(\lambda|\vec{\theta})=D(\lambda^*+i \, \pi|\vec{\theta}), \qquad 
D^\dag(\lambda|\vec{\theta})=A(\lambda^*+i \, \pi|\vec{\theta}), \\
B^\dag(\lambda|\vec{\theta})=C(\lambda^*+i \, \pi|\vec{\theta}), \qquad 
C^\dag(\lambda|\vec{\theta})=B(\lambda^*+i \, \pi|\vec{\theta}).
\end{split}
\end{equation}
It follows for the transfer matrix that:
\begin{equation} \label{tauH}
\begin{split}
\tau^\dag(\lambda|\vec{\theta})=\tau(\lambda^*+i \, \pi|\vec{\theta}).
\end{split}
\end{equation}
Thus, the transfer matrix is a hermitian operator along the line: $\lambda=\rho+i \tfrac{\pi}{2},$ with 
$\rho \in {\mathbb R}.$
The hermitian conjugation relations (\ref{ABCDH}) imply that the complex conjugate vector $\Psi^*(\{\lambda_j\}|\vec{\theta})$ can be represented as follows:
\begin{equation} \label{Psi*}
\begin{split}
\Psi^*(\{\lambda_j\}|\vec{\theta})=\langle 0|C(\lambda_1^*+i \, \pi|\vec{\theta})\, 
C(\lambda_2^*+i \, \pi|\vec{\theta})...C(\lambda_r^*+i \, \pi|\vec{\theta}) \, \frac{1}{{\cal N}_\Psi}.
\end{split}
\end{equation}
It can be seen that if a set $\{\lambda_j\}_{j=1}^r$  is a solution of the Bethe-equations (\ref{ABAE}), 
then the set  $\{\lambda^*_j+i \, \pi\}_{j=1}^r$ is also a solution of (\ref{ABAE}). 
Thus for solutions which are invariant under this transformation the complex conjugate vector can be written 
in a simpler form: 
\begin{equation} \label{Psi*1}
\begin{split}
\Psi^*(\{\lambda_j\}|\vec{\theta})=\langle 0|C(\lambda_1|\vec{\theta})\, 
C(\lambda_2|\vec{\theta})...C(\lambda_r|\vec{\theta}) \, \frac{1}{{\cal N}_\Psi}.
\end{split}
\end{equation}
Now it is easy to determine the normalization constant ${\cal N}_\Psi,$ because it is nothing but the 
Gaudin-norm \cite{Gaudin0,Gaudin1,Korepin} of the Bethe-state $B(\lambda_1|\vec{\theta})\, B(\lambda_2|\vec{\theta})...
B(\lambda_r|\vec{\theta})|0\rangle:$
\begin{equation} \label{calN}
\begin{split}
{\cal N}_\Psi^2=\langle 0|C(\lambda_1|\vec{\theta})\, 
C(\lambda_2|\vec{\theta})...C(\lambda_n|\vec{\theta}) \, B(\lambda_1|\vec{\theta})\, B(\lambda_2|\vec{\theta})...
B(\lambda_r|\vec{\theta})|0\rangle.
\end{split}
\end{equation}
If one would like to apply the Algebraic Bethe Ansatz technique directly to $\tau(\lambda|\vec{\theta})$, 
one should  carry unnecessarily a lot of $S_0(\theta)$ factors. This can be avoided, if one diagonalizes 
the transfer matrix constructed out of the $S_0$ removed part of the S-matrix. To be more concrete analogously to 
(\ref{monodr}) one should define the "reduced" monodromy matrix by the formula:  
\begin{equation} \label{monodrR}
\begin{split}
{\sc T}_a^b(\theta|\theta_1,...,\theta_n)_{a_1 a_2 ... a_n}^{b_1 b_2 ...b_n}=
{ S}_{a \, a_1}^{k_1 \, b_1}(\theta-\theta_1)
{ S}_{k_1 \,  a_2}^{k_2 \, b_2}(\theta-\theta_2)...
{ S}_{k_{n-1}\,  a_n}^{b \, b_n}(\theta-\theta_n),
\end{split}
\end{equation}
where $S_{ab}^{cd}(\theta)$ is the matrix part of the S-matrix (\ref{Smatr}) given by (\ref{Selem})-(\ref{C0}).
Analogously to (\ref{ABCD}) it can be written as a 2 by 2 matrix in the auxiliary space:
\begin{equation} \label{ABCDR}
\begin{split}
{\sc T}(\lambda|\vec{\theta})=
\begin{pmatrix} {\sc T}_-^-(\lambda|\vec{\theta}) & {\sc T}_-^+(\lambda|\vec{\theta}) \\
{\sc T}_+^-(\lambda|\vec{\theta}) & {\sc T}_+^+(\lambda|\vec{\theta})\end{pmatrix}
=\begin{pmatrix} {\cal A}(\lambda|\vec{\theta}) & {\cal B}(\lambda|\vec{\theta}) \\
{\cal C}(\lambda|\vec{\theta}) & {\cal D}(\lambda|\vec{\theta})
\end{pmatrix}.
\end{split}
\end{equation}
Its matrix elements satisfy the same Yang-Baxter algebra (\ref{YBA}) as those of ${\cal T}(\lambda|\vec{\theta}).$ 
The "reduced" transfer matrix ${\sc t}(\lambda|\vec{\theta})$ is defined 
by taking the trace in the auxiliary space:
\begin{equation} \label{transR}
\begin{split}
{\sc t}(\lambda|\vec{\theta})=\sum\limits_{a=\pm} {\sc T}_a^a(\lambda|\vec{\theta}).
\end{split}
\end{equation}
It differs from $\tau(\lambda|\vec{\theta})$ in only a trivial scalar factor:
\begin{equation} \label{tautau}
\begin{split}
\tau(\lambda|\vec{\theta})=\prod\limits_{k=1}^n S_0(\lambda-\theta_k)\, {\sc t}(\lambda|\vec{\theta}).
\end{split}
\end{equation}
Thus their common eigenvector $\Psi(\lambda,\{\lambda_j\}|\vec{\theta})$ (\ref{eivec1}) and its complex conjugate 
(\ref{Psi*1}) can be expressed in terms of the elements of the "reduced" monodromy matrix completely analogously 
to the formulas (\ref{eivec1}) and (\ref{Psi*1}): 
\begin{equation} \label{Psi1R}
\begin{split}
\Psi(\{\lambda_j\}|\vec{\theta})=\frac{1}{N_\Psi}\, {\cal B}(\lambda_1|\vec{\theta})\, {\cal B}(\lambda_2|\vec{\theta})...
{\cal B}(\lambda_r|\vec{\theta})|0\rangle,
\end{split}
\end{equation}
\begin{equation} \label{Psi*1R}
\begin{split}
\Psi^*(\{\lambda_j\}|\vec{\theta})=\langle 0|{\cal C}(\lambda_1|\vec{\theta})\, 
{\cal C}(\lambda_2|\vec{\theta})...{\cal C}(\lambda_n|\vec{\theta}) \, \frac{1}{ N_\Psi}.
\end{split}
\end{equation}
Certainly the normalization factor is also changed compared to (\ref{eivec1}) and (\ref{Psi*1}): 
\begin{equation} \label{NPsi}
\begin{split}
{N}_\Psi^2=\langle 0|{\cal C}(\lambda_1|\vec{\theta})\, 
{\cal C}(\lambda_2|\vec{\theta})...{\cal C}(\lambda_n|\vec{\theta}) \, {\cal B}(\lambda_1|\vec{\theta})\, {\cal B}(\lambda_2|\vec{\theta})...
{\cal B}(\lambda_r|\vec{\theta})|0\rangle,
\end{split}
\end{equation}
which can be written as a Slavnov-determinant \cite{Slavnov}.
The eigenvalue of ${\sc t}(\lambda|\vec{\theta})$ on $\Psi(\{\lambda_j\}|\vec{\theta})$ is exactly 
$\Lambda_0(\lambda,\{\lambda_j\}|\vec{\theta})$ given in (\ref{Lambda0}).

We continue this appendix by specializing the main formulas of this appendix to the 
3-particle case.

\subsection{Formulas for the 3-particle case}

Due to the charge conjugation symmetry, in the 3-particle case  the number of Bethe-roots in (\ref{ABAE}) 
can be either zero or one. The zero root case corresponds to the trivial pure solitonic eigenvector (\ref{trivvac}). 
Here we do not deal with this trivial case, but we are interested in the state described by a single Bethe-root.
In this case the Bethe-equations take the simple form:
\begin{equation} \label{3BAE}
\begin{split}
\prod\limits_{j=1}^3 B_0(\lambda_1-\theta_k)=1.
\end{split}
\end{equation}
The eigenvalue of the soliton transfer matrix (\ref{eival}), when its spectral parameter 
takes the value of one of the rapidities,   
is given{\footnote{Since here we have only one 
Bethe-root, for short we skipped it from the list of arguments of the eigenvalue.}} by:
\begin{equation} \label{3eival}
\begin{split}
\Lambda(\theta_j|\vec{\theta})=\prod\limits_{k=1}^3 S_0(\theta_j-\theta_k) \frac{1}{B_0(\lambda_1-\theta_j)}, \qquad j=1,2,3.
\end{split}
\end{equation}
In this one-root case the normalization factor $N_\Psi$ in (\ref{NPsi}) takes the form:
\begin{equation} \label{3NPsi}
\begin{split}
{N}_\Psi^2=\langle 0|{\cal C}(\lambda_1|\vec{\theta})\, 
 {\cal B}(\lambda_1|\vec{\theta})|0\rangle=p \, \sinh\left( \tfrac{i \, \pi}{p}\right) \, \sum\limits_{j=1}^3 (\ln B_0)'(\lambda_1-\theta_j).
\end{split}
\end{equation}
In the computation of the symmetric form-factors some derivatives with respect to the particle's rapidities will 
be important. Differentiating (\ref{3eival}) with respect to $\theta_q$ one obtains:
\begin{equation} \label{Lsq}
\begin{split}
\partial_q \log \Lambda(\theta_s|\vec{\theta})=-i \,G(\theta_s-\theta_q)-(\ln B_0)'(\lambda_1-\theta_s) \,
\frac{\partial \lambda_1}{\partial \theta_q}, \qquad %\sigma_{sq}=\frac{1}{2 \pi } G(\theta_s-\theta_q), 
\qquad s\neq q,
\end{split}
\end{equation}
with $G(\theta)$ given in (\ref{G}). The derivative $\frac{\partial \lambda_1}{\partial \theta_q}$ can be obtained by differentiating the Bethe-equation (\ref{3BAE}): 
\begin{equation} \label{3lamder}
\begin{split}
\frac{\partial \lambda_1}{\partial \theta_q}=
\frac{(\ln B_0)'(\lambda_1-\theta_q)}{\sum\limits_{k=1}^3 (\ln B_0)'(\lambda_1-\theta_k) }, \qquad q=1,2,3.
\end{split}
\end{equation}
If we have one single root then due to the $\lambda \to \lambda^*+i \, \pi$ symmetry of the Bethe-equation the 
single root of the equation can be either "real" or "self-conjugate". 
A solution $\lambda_1$ is called real, if it is a fixed point of the symmetry $\lambda \to \lambda^*+i \, \pi,$
 i.e.  $\lambda_1= \lambda_1^*+i \, \pi.$ Here we use the term "real", because in a more convenient 
 parameterization this type of roots would be actually a real numbers. Namely, 
if it is parameterized as $\lambda_1=\rho_1+i \tfrac{\pi}{2}$ then the fix point equation restricts $\rho_1$ to be real. 

Due to the $i \,p \,  \pi $ symmetry of the functions entering the 
the Bethe-equations (\ref{ABAE}), they have another symmetry, as well. 
Namely if $\lambda_j$ is a solution of the equations 
then $\lambda_j+i \, \pi \, p$ is also a solution. This means that the solutions can be resticted to a 
fundamental domain given by the strip of width $i \, p \, \pi.$ 
By definition a "self-conjugate" root satisfies the combination of symmetries: 
$\lambda \to \lambda^*+i \, \pi$ and $\lambda \to \lambda\pm i \,p \, \pi,$ namely 
\begin{equation} \label{lsc}
\begin{split}
\lambda_1=\lambda_1^*+i\, \pi\pm i \, p \pi.
\end{split}
\end{equation}
If it is parameterized again as $\lambda_1=\rho_1+i \, \tfrac{\pi}{2},$ then $\rho_1$ has a fixed imaginary part:
$\text{Im} \rho_1=\tfrac{p \, \pi}{2}.$

The numerical solution of the equation 
(\ref{3BAE}) shows that in the repulsive regime $(1<p)$
from the 3 different solutions,  two ones are self-conjugated and one is real.

\subsection{Classification of the magnonic Bethe-roots} \label{appABA2}

As the simple discussion at the end of the previous subsection shows, there are two symmetries of the magonic  
Bethe-equations (\ref{ABAE}):
\begin{equation} \label{sym12}
\begin{split}
&\bullet \qquad \qquad \{\lambda_j\}_{j=1}^r= \{\lambda_j^*+i \, \pi\}_{j=1}^r, \\
&\bullet \qquad \qquad \{\lambda_j\}_{j=1}^r= \{\lambda_j+i \,p\,  \pi\}_{j=1}^r.
\end{split}
\end{equation}
They imply the following classification of the roots. 
\begin{eqnarray}
&\bullet& \, \text{Real-roots:} \qquad \, \text{Im}\, (\lambda_j-i \, \tfrac{\pi}{2})=0, \qquad 
\qquad \qquad \qquad \qquad j=1,..,n_r, \nonumber \\
&\bullet& \, \text{Close-roots:} \qquad |\text{Im}\, (\lambda_j-i \, \tfrac{\pi}{2})|\leq 
\text{min} (\tfrac{\pi}{2},\tfrac{ (2 p-1)\, \pi}{2}) , \qquad  \quad \,  j=1,..,n_c, \label{rootclassR} \\
&\bullet& \, \text{Wide-roots:} \qquad \text{min} (\tfrac{\pi}{2},\tfrac{ (2 p-1)\, \pi}{2}) < |\text{Im}\,
 (\lambda_j-i \, \tfrac{\pi}{2})|\leq \tfrac{p \, \pi}{2} , \quad j=1,..,n_w. \nonumber
\end{eqnarray}
A special type of wide-root is the self-conjugate root, whose imaginary part is exactly $i \tfrac{(1+p) \,\pi}{2}.$
From the symmerties (\ref{sym12}) of the asymptotic Bethe-equations it also follows that all roots, which are neither 
real nor self-conjugate appear in pairs being symmetric to the line $\text{Im} z=\tfrac{\pi}{2}.$ 
In this way we can speak about close-and wide-pairs similarly to the Bethe-roots 
entering the NLIE (\ref{DDV}), which describes  
the exact finite volume spectrum of the sine-Gordon model. 

%Closing this subsection we just note that in the attractive regime, when $0<p<1,$

%\begin{equation} \label{}
%\begin{split}
%\end{split}
%\end{equation}

%\begin{eqnarray}
%\end{eqnarray}

%\begin{itemize}\item
%\end{itemize}

%%%%%%%%%%%%%%%%%%%%%%%%%%%%%%%%%%%%%%%%%%%%%%%%%%%%%%%%%%%%%%%%%%%%%%%%%%%%%%%%%%%%%%%%%%%
%%%%%%%%%%%%%%%%%%%%%%%%%%%%%%%%%%%%%%%%%%%%%%%%%%%%%%%%%%%%%%%%%%%%%%%%%%%%%%%%%%%%%%%%%%%
%%%%%%%%%%%%%%%%%%%%%%%%%%%%%%%%%%%%%%%%%%%%%%%%%%%%%%%%%%%%%%%%%%%%%%%%%%%%%%%%%%%%%%%%%%%

\end{document}